\documentclass[iop,numberedappendix]{emulateapj}
\usepackage{url}
\usepackage{enumerate}
\usepackage{graphics}
\usepackage{amsmath}
\usepackage{graphicx}
\usepackage{hyperref}
\usepackage{epstopdf}
\usepackage{natbib}

\newcommand{\cmd}{\mbox{$\mbox{cm}^{-2}$}}
\newcommand{\kms}{\mbox{\,km$\,$s$^{-1}$}}
\newcommand{\kpc}{\mbox{\,kpc}}

\newcommand{\lsun}{\mbox{$L_{\sun}$}}
\newcommand{\msun}{\mbox{$M_{\sun}$}}

\newcommand{\Nhtwo}{$N_{\rm H_2}$}

\newcommand{\K}{\mbox{K}}

\newcommand{\vlsr}{\mbox{$V_{\mbox{\tiny LSR}}\,$}}

\newcommand{\tdust}{\mbox{$T_{\tiny{\rm dust}}$}}

\newcommand{\hii}{H \scriptsize{II}\normalsize\,}

\newcommand{\cmc}{\mbox{\,cm$^{-3}$}\,}

\newcommand{\av}{A$_\mathrm{V}$}

\shorttitle{Low-velocity shocks in W43}
\shortauthors{Quang Nguyen Luong et al.}

\begin{document}

\title{Low-velocity shocks traced by extended SiO emission along the W43 ridges: witnessing the formation of young massive clusters
}

\author{Q. Nguy$\tilde{\hat{\rm e}}$n Lu{\hskip-0.65mm\small'{}\hskip-0.5mm}o{\hskip-0.65mm\small'{}\hskip-0.5mm}ng\altaffilmark{1},
		F. Motte\altaffilmark{2},
		P. Carlhoff\altaffilmark{3},
		F. Louvet \altaffilmark{2},
		P. Lesaffre \altaffilmark{4},
	    P. Schilke\altaffilmark{3},
		T. Hill \altaffilmark{2}
		M. Hennemann \altaffilmark{2},	
		A. Gusdorf \altaffilmark{4},
		P. Didelon \altaffilmark{2},		
		N.~Schneider\altaffilmark{5a,5b},
		S.~Bontemps\altaffilmark{5a,5b},
		A.~Duarte-Cabral\altaffilmark{5a,5b},
		K. M. Menten \altaffilmark{6}, 
		P.~G.~Martin\altaffilmark{1}, 
 	    F.~Wyrowski \altaffilmark{6},
 	    G.~Bendo\altaffilmark{7},
		H. Roussel \altaffilmark{8},   
	    J-P. Bernard  \altaffilmark{9}, 	
	    L. Bronfman \altaffilmark{10},	     	
	    T. Henning  \altaffilmark{11},	    
	    C. Kramer  \altaffilmark{12},
	    F. Heitsch \altaffilmark{13}
							    				                                                        }

\altaffiltext{1}{Canadian Institute for Theoretical Astrophysics, University of Toronto, 60 St. George Street, Toronto, ON M5S~3H8, Canada}
\altaffiltext2{Laboratoire AIM Paris-Saclay, CEA/IRFU - CNRS/INSU - Universit\'e Paris Diderot, Service d'Astrophysique, B\^at. 709, CEA-Saclay, F-91191, Gif-sur-Yvette Cedex, France}
\altaffiltext{3}{I. Physikalisches Institut, Universit\"at zu K\"oln, Z\"ulpicher Str. 77, 50937 K\"oln, Germany}
\altaffiltext{4}{ENS, LERMA, UMR 8112, CNRS, Observatoire de Paris, 24 rue Lhomond 75005 Paris, France LRA/ENS, France}
\altaffiltext{5}{Universit\'e de Bordeaux, LAB, CNRS, UMR~5804, 33270 Floirac, France}
\altaffiltext{5a}{Univ. Bordeaux, LAB, UMR 5804, F-33270, Floirac, France}
\altaffiltext{5b}{CNRS, LAB, UMR 5804, F-33270, Floirac, France}
\altaffiltext{6}{Max-Planck-Institut f\"ur Radioastronomie, Auf dem H\"ugel 69, D-53121 Bonn, Germany}
\altaffiltext{7}{ UK ALMA Regional Centre Node, Jodrell Bank Centre for Astrophysics, School of Physics and Astronomy, University of Manchester, Oxford Road, Manchester M13 9PL, United Kingdom}
\altaffiltext{8}{Institut d'Astrophysique de Paris, UMR 7095 CNRS,Universit\'e Pierre \& Marie Curie, 98~bis boulevard Arago, 75014, Paris, France}
\altaffiltext{9}{Universit\'e de Toulouse, UPS, CESR, 9 avenue du colonel Roche, 31028 Toulouse Cedex 4; CNRS, UMR 5187, 31028 Toulouse, France}
\altaffiltext{10}{Departamento de Astronomía, Universidad de Chile, Santiago, Chile}
\altaffiltext{11}{Max Planck Institute for Astronomy, Koenigstuhl 17, D-69117 Heidelberg, Germany}
\altaffiltext{12}{Instituto Radioastronom\'{i}a Milim\'etrica (IRAM), Av. Divina Pastora 7, N\'ucleo Central, E-18012 Granada, Spain}
\altaffiltext{13}{(Department of Physics and Astronomy, University of North Carolina Chapel Hill, Phillips Hall, Chapel Hill, NC, 27599-3255, USA}

\altaffiltext{\dag}{\url{qnguyen@cita.utoronto.ca}}

\begin{abstract}
The formation of high-mass stars is tightly linked to that of their parental clouds. We here focus on the high-density parts of W43, a molecular cloud undergoing an efficient event of star formation.
Using a column density image derived from
\emph{Herschel}\footnote{
\emph{Herschel} is an ESA space observatory with science instruments provided by European-led Principal Investigator consortia and with important participation from NASA.} 
submm continuum maps, we identify two high-density filamentary clouds, named the W43-MM1 and W43-MM2 ridges. Both have gas masses $>$2$\times10^{4}~\msun$ above $>$10$^{23}~\cmd$ and within areas of $\sim$6 and $\sim$14~pc$^2$, respectively. 
The W43-MM1 and W43-MM2 ridges are structures coherent in velocity and gravitationally bound despite the large velocity dispersion as shown by the N$_2$H$^+$~(1--0) lines of the W43-HERO IRAM\footnote{The W43-HERO (W43 Hera/EmiR Observation) project is an IRAM~30m Large Program, named ``Origins of molecular clouds and star formation in W43'', led by Fr\'ed\'erique Motte and Peter Schilke. The project description and data can be accessed here: \url{http://www.astro.uni-koeln.de/projects/schilke/IRAMLargeProject/MainPage}. IRAM is supported by INSU/CNRS (France), MPG (Germany) and IGN (Spain)} Large Program. 
Another intriguing result is that these ridges harbour wide-spread ($\sim$10~pc$^{2}$) bright SiO~(2--1) emission which 
we interpret as arising from low-velocity shocks ($\le$10\,\kms).
We measure a steep relation between the SiO~(2--1) luminosity and velocity extent and propose it to distinguish our observations from the high-velocity shocks associated with outflows.
We use state-of-the-art shock models to demonstrate that low-velocity shocks with a small amount (10\%) of Si atoms initially in gas phase or in grain mantles can explain the observed SiO column density in the W43 ridges. 
The spatial and velocity overlaps between the ridges high-density gas and the shocked SiO gas suggest that ridges could be forming via colliding flows driven by gravity and accompanied by low-velocity shocks. This mechanism may be the initial conditions for the formation of young massive clusters.  
\end{abstract}

\section{Introduction}
Though being an important entity in the energy budget 
and the cloud structure of galaxies, the origin of high-mass stars (OB-type, $>$8~$M_\odot$) are uncertain. An essential question is how they can gain such relatively large masses within their short protostellar lifetime (10$^{5}$-10$^{6}$~yr). Two main theoretical explanations were proposed to answer this question. In the quasi-static view, a high degree of turbulence allows the formation of a massive dense core inside which a powerful accretion develops to build up the stellar mass (e.g. \citealt{
krumholz09,
hosokawa09}). In the dynamical picture, competitive accretion and/or cloud formation generate colliding flows which then funnel gas to a large potential well  (e.g. \citealt{bonnell06,heitsch08,inoue09,hartmann12}). Dynamical signatures such as gravitational streamers and shearing motions at molecular-cloud scales (10--100~pc; 
\citealt{nguyenluong11}), filamentary scales (1--10~pc; \citealt{schneider10,hennemann12}) and protostellar core scales ($0.1-1$~pc; \citealt{csengeri11,csengeri11b}),
have been observed. All of these studies convey the idea that the formation of high-mass stars is tightly linked to the density, kinematics, and formation of their parental cloud.

HOBYS\footnote{The \emph{Herschel} imaging survey of OB Young Stellar objects (HOBYS) is a \emph{Herschel} key program. See \url{http://hobys-herschel.cea.fr}}, a key mapping survey with \emph{Herschel} is dedicated to the formation of OB-type stars within molecular complexes \citep{motte10,motte12}. 
It has shown that molecular clouds are dominated by networks of filaments and that clusters of high-mass protostars are forming in ``cloud ridges" \citep{hill11,nguyenluong11b,hennemann12}.
These ``ridges" are high-density (equivalent \Nhtwo$>10^{23}~\cmd$), elongated cloud structures, which dominate and shape their surroundings.
Clusters of massive stars are forming within these ridges, making them excellent candidate sites to host mini-starbursts, i.e. miniature and instant model of intense star formation activity \citep{hill11,nguyenluong11b}. \cite{schneider12} also showed that intermediate-mass stellar clusters generally form at the junction of filaments.
The density structure and star formation efficiency of ridges suggest that they have formed through dynamical processes such as colliding flows and/or filaments merging. 
In the converging flows theory, ridges can be the result of these large-scale processes that sweep up and accumulate gas mass \citep{ballesterosparedes99,hartmann01,inoue09}. Alternatively, they could be formed by the gravitational focusing of gas toward large potential wells \citep{hartmann07}. The massive dense cores are then formed within shock-compressed ridges, which were created by shock waves from cloud-cloud collision \citep{inoue13}.

The W43 complex lies at 6~kpc from the Sun, at the meeting point of the Scutum-Centaurus (or Scutum-Crux) Galactic arm and the bar, a dynamically complex region where high-velocity streams could easily collide \citep{nguyenluong11}. The peculiar properties of W43 has been confirmed by the high-angular resolution observations of the ${13}$CO~(2--1) and C$^{18}$O~(2--1) lines of the W43-HERO IRAM Large Program (W43 Hera/EmiR Observations) (see Carlhoff et al. submitted).
Close to its center, W43-Main is undergoing a remarkably efficient episode of star formation and qualifies as a mini-starburst (\emph{SFE}~$\sim25\%$, $\Sigma_\text{SFR}~\sim 1000~\msun$yr$^{-1}$kpc$^{-2}$; \citealt{motte03}). 
According to the criteria of \cite{bressert12}, the W43-MM1 and W43-MM2 clouds are indeed good candidates to form the future young massive clusters (see \citealt{ginsburg12}).
Among the ~15 dense cores of W43-Main (0.2~pc \textit{FWHM} size and $>$5$\times 10^{5}~\cmc$ density) potentially forming high-mass stars are the three extremely massive, dense cores W43-MM1 ($M=3600$~\msun), W43-MM2 ($M=1600$~\msun), and W43-MM3 ($M=1000$~\msun) (\citealt{motte03}, confirmed by \citealt{bally10a}). 
Adjacent to W43-Main is a giant \hii region, illuminated by a cluster of Wolf-Rayet (WR) and OB stars, emitting $\sim10^{51}$ Lyman continuum photons per second and a far-infrared continuum luminosity of $\sim3.5\times10^{6}~\lsun$ \citep{smith78,blum99,bik05}. 
It is not yet clear what is the impact of this starburst cluster on the W43-Main cloud located 2--10~pc away \citep{motte03,bally10a}. With its special characteristics, W43 represent for a type of molecular cloud complex which host high luminosity embedded clusters. Other examples can be found, such as W49 (Galv\'an-Madrid in prep.) or W51 \citep{bieging10}.

Theoretically, a series of shocks are expected to emerge at the swept-up edges of turbulent colliding flows as a consequence of compression and shears of material \citep{koyama00,heitsch08b}. Detecting these shocks observationally will generate the firmest evidence for the collision and merging of several gas flows.
Classically, SiO is the diagnostic of high-velocity shocks ($v_{\rm shock}\sim20-50~\kms$) in protostellar outflows \citep{schilke97,gueth98,gusdorf08b}. It has also been proven to trace medium- to high-velocity irradiated shocks ($v_{\rm shock}\sim10-20~\kms$ to $v_{\rm shock}\sim50~\kms$) from hot cores/hot corinos \citep{hatchell01,jorgensen11}, and photon dominated regions \citep[e.g.][]{schilke01}, or very high velocity shocks ($v_{\rm shock}\sim100~\kms$) in  the 
Central Molecular Zone of our Galaxy
\citep[e.g.][]{martin-pintado97,jones12}.
SiO emission is also detected in the locations where protostellar outflows interact with local gas condensations \citep{lefloch98}.
Most of the surveys performed toward high-mass protostars interpreted the SiO emission as solely tracing high-velocity protostellar outflows \citep[e.g.][]{beuther07c,lopez-sepulcre11}.
When analyzing the SiO spectra of the Cygnus~X massive dense cores, \cite{motte07} however felt the necessity to decompose the line in two shocked gas components: a broad one tracing the protostellar outflow and a narrow one interpreted as arising from the hot core. More recently, Duarte-Cabral et al. (2013, in prep) revisited the question of the SiO line profiles in Cygnus X and proposed that the narrow component is most probably associated with the shocks expected at the locations of the velocity shears observed by \cite{csengeri11,csengeri11b} at the scale of massive dense cores (0.1~pc).
\cite{jimenez-serra10} have also proposed that part of the extended SiO emission observed along IRDC G035.39--00.33 is associated with low-velocity shocks from colliding flows \citep[see also][]{henshaw13}. 
This statement was confirmed, at one location of the filament, by the lack of intermediate- to high-mass protostars that could explain the SiO emission by shocks from a protostellar outflow or a hot core \citep[see][]{nguyenluong11b}.

In the present paper, we report the discovery of bright and wide-spread SiO emission along the W43-MM1 and W43-MM2 ridges, whose physical properties are constrained through \emph{Herschel} maps, N$_2$H$^+$~(1--0) and SiO~(2--1) spectral cubes. The paper is organized such that Sect.~\ref{sect:observation} describes the data employed, Sect.~\ref{sect:ridgeresult} characterizes the W43-MM1 and W43-MM2 cloud ridges, and Sect.~\ref{sect:sioresult} characterize their SiO emission. In Sect.~\ref{sect:discussion}, we showthat state-of-the-art models can create strong SiO emission with low-velocity shocks, and we propose that this is the origin of SiO emission we observe in W43 ridges. We also propose in Sect.~\ref{sect:discussion} that the extended SiO emission directly relates to the formation of ridges and we discuss the implications for the subsequent high-mass star formation. We conclude and summarize our findings in Sect.~\ref{sect:conclusion}.

\section{Observations}
\label{sect:observation}
 To study the connection between the density and the velocity structures of W43-Main and the shock properties within this region, we used  both continuum data from \emph{Herschel} and SiO, N$_2$H$^+$ emission lines from the IRAM 30~m telescope.

\subsection{Dust continuum imaging with Herschel}
\label{sect:obsherschel}
We used the \emph{Herschel} images obtained from the Hi-GAL survey (\emph{Herschel} Infrared Galactic Plane Survey, OBSIDs:1342186275 and 1342186276; \citealt{molinari10}) at $70/160\,\micron$ with PACS \citep{poglitsch10} and at $250/350/500\,\micron$ with SPIRE \citep{griffin10}.
We complemented the Hi-GAL nominal-mode dataset with SPIRE $250/350/500\,\micron$ bright-mode observations from the HOBYS key program (OBSIDs: 1342239977 and 1342239978; \citealt{motte10}). The Hi-GAL data were taken in parallel mode with a fast scanning speed of 60$\arcsec {\rm s}^{-1}$ and the HOBYS data in SPIRE-only mode with a 30$\arcsec {\rm s}^{-1}$ scanning speed. 
The raw (level-0) data of each individual scan from both PACS and SPIRE were calibrated and deglitched using HIPE\footnote{HIPE is a joint development software by the \emph{Herschel} Science Ground Segment Consortium, consisting of ESA, the NASA \emph{Herschel} Science Center, and the HIFI, PACS, and SPIRE consortia.} version 10.0. The SPIRE and PACS level-1 data were then fed to version 18 of the Scanamorphos software package\footnote{http://www2.iap.fr/users/roussel/herschel/} \citep{roussel12}, which subtracts brightness drifts by exploiting the redundancy of observed points on the sky, masks remaining glitches, and produces maps. The observational parameters of the \emph{Herschel} images as well as the conversion factor from the original intensity units to MJy/sr are listed in Table~\ref{table:obs}. 

\begin{table}[htbp]
\begin{center}
\caption{Observational parameters of \emph{Herschel} images}
\label{table:obs}
\begin{tabular}{lcrlll}
 \hline
 \hline
 {$\lambda$ (\micron)}  &  \textit{HPBW} &   {f$^{a}$} &  \multicolumn2{c}{1$\sigma$ rms  (MJy/sr)} & Pixel \\
 {/Camera} & (\arcsec)  &  -   & {Hi-GAL} & {HOBYS} & size (\arcsec)\\
\hline
70/PACS   & ~~5.8$\times12.1$    & 21706 &  0.02 & - & 1.40\\
160/PACS   & 11.4$\times13.4$  & 5237  & 0.08 & -& 2.85 \\
250/SPIRE  & 18.1       & 115   &  1.00 & 0.1 & 4.50 \\
350/SPIRE  & 25.2       & 60    &  1.10  & 0.2 & 6.25\\
500/SPIRE  & 36.9       & 27    &  1.20  & 0.4 & 9.00 \\
\hline
\end{tabular}
\end{center}
Note: ~$^a$ Conversion factor from original units, Jy/pixel for PACS data and Jy/beam for SPIRE data, to MJy/sr. 
\end{table}

\begin{figure}[!tbhp]
\centering
$\begin{array}{c}
\hspace{-0.1cm}
 \includegraphics[scale =0.4]{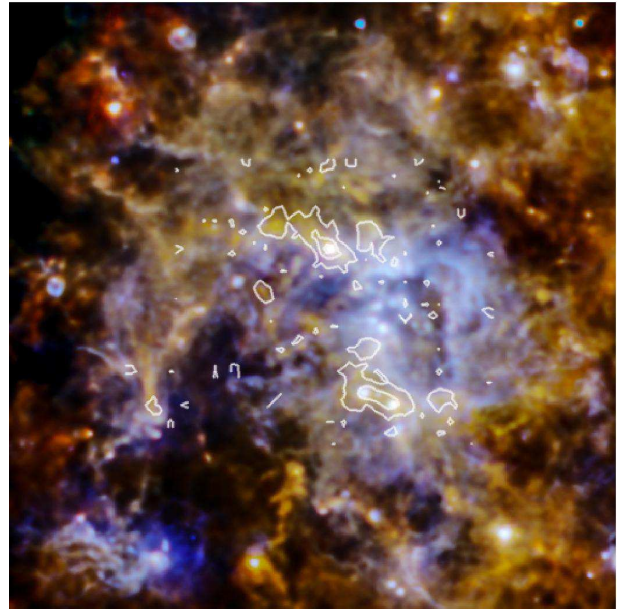}
\\
\end{array}$
\caption{Composite 3-color \emph{Herschel} image of the W43-Main mini-starburst ($70\,\micron$: blue, $160\,\micron$:  green, $250\,\micron$: red).  The blue component traces H{ \scriptsize II} and photon-dominated regions while earlier stage star-forming sites such as cores and filaments are traced by the red component. The contours are the integrated intensity of SiO in levels of $3\sigma=0.8$~K\kms and $6\sigma=1.6$~K\kms. We note that the SiO map does not cover the full extent of the 3 colour image. The figure is oriented in RA-Dec and the area is equal to that covered in Fig.~\ref{fig:w43coldens}.}
\label{fig:w43main_3colorsb}
\end{figure}

The Hi-GAL dataset covers the entire W43 molecular complex but saturation occurs around the bright ($\gtrsim$200~Jy/beam) structures in the 250 and $350\,\micron$ SPIRE images. The Hi-GAL saturated areas were corrected by the bright-mode HOBYS dataset. 
For each wavelength, we used the same grid for the HOBYS and Hi-GAL maps built by Scanamorphos. The pixels of the Hi-GAL 250 and $350\,\micron$ maps with $\ge$150~Jy/beam are then replaced by those of the HOBYS map, after the addition of offsets measured by cross-correlating unsaturated Hi-GAL and HOBYS pixels (see Appendix~\ref{sect:appendixC}). 
The resulting three-color (RGB=$250\,\micron$/$160\,\micron$/$70\,\micron$) image is presented in Figs.~\ref{fig:w43main_3colorsb} and \ref{fig:w43_3colors}.

\begin{figure*}[!htbp]
$\begin{array}{cc}
\hspace{-0.1cm}
\includegraphics[angle=0,height=10.5cm]{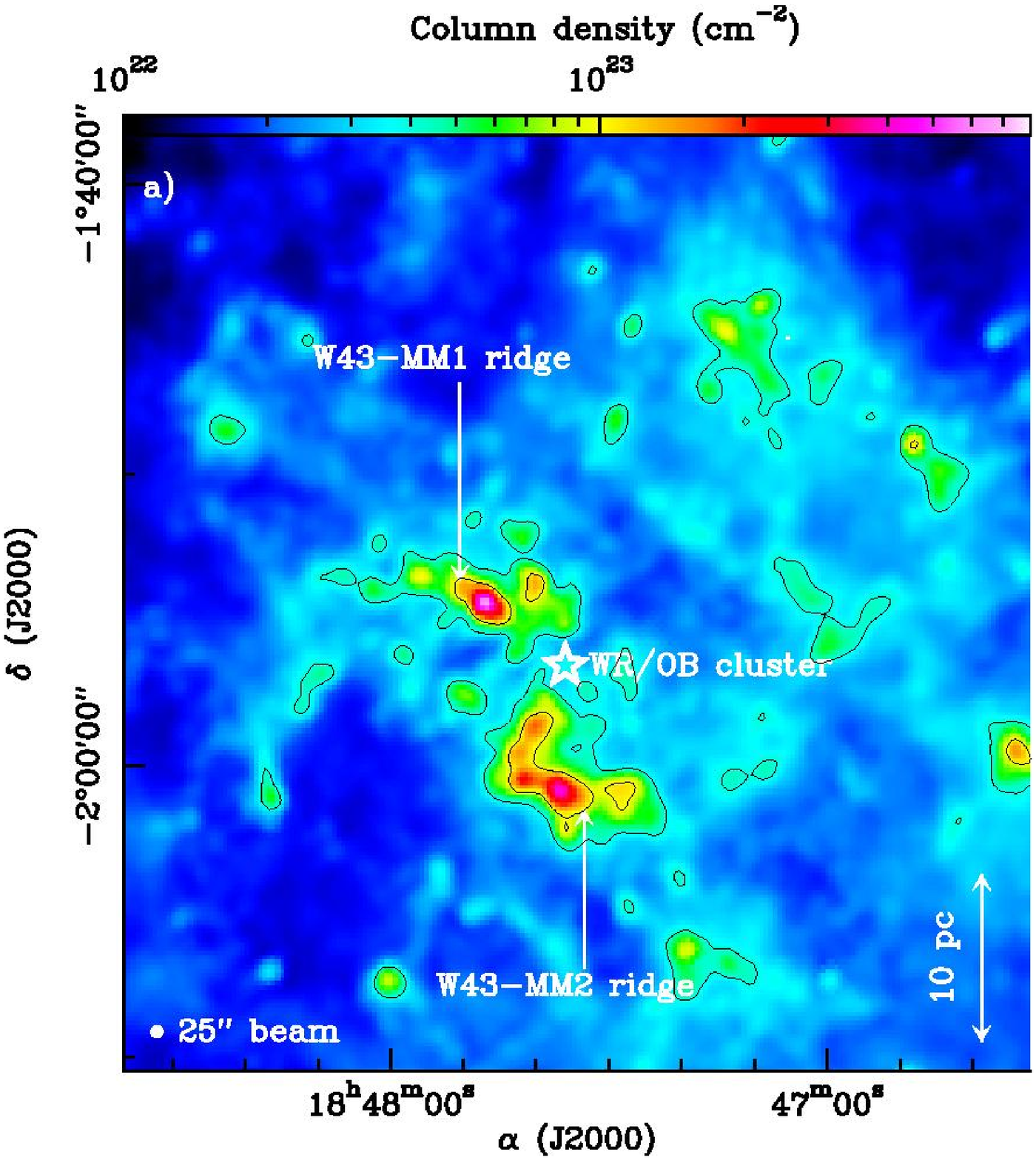} &
\includegraphics[angle=0,height=10.5cm]{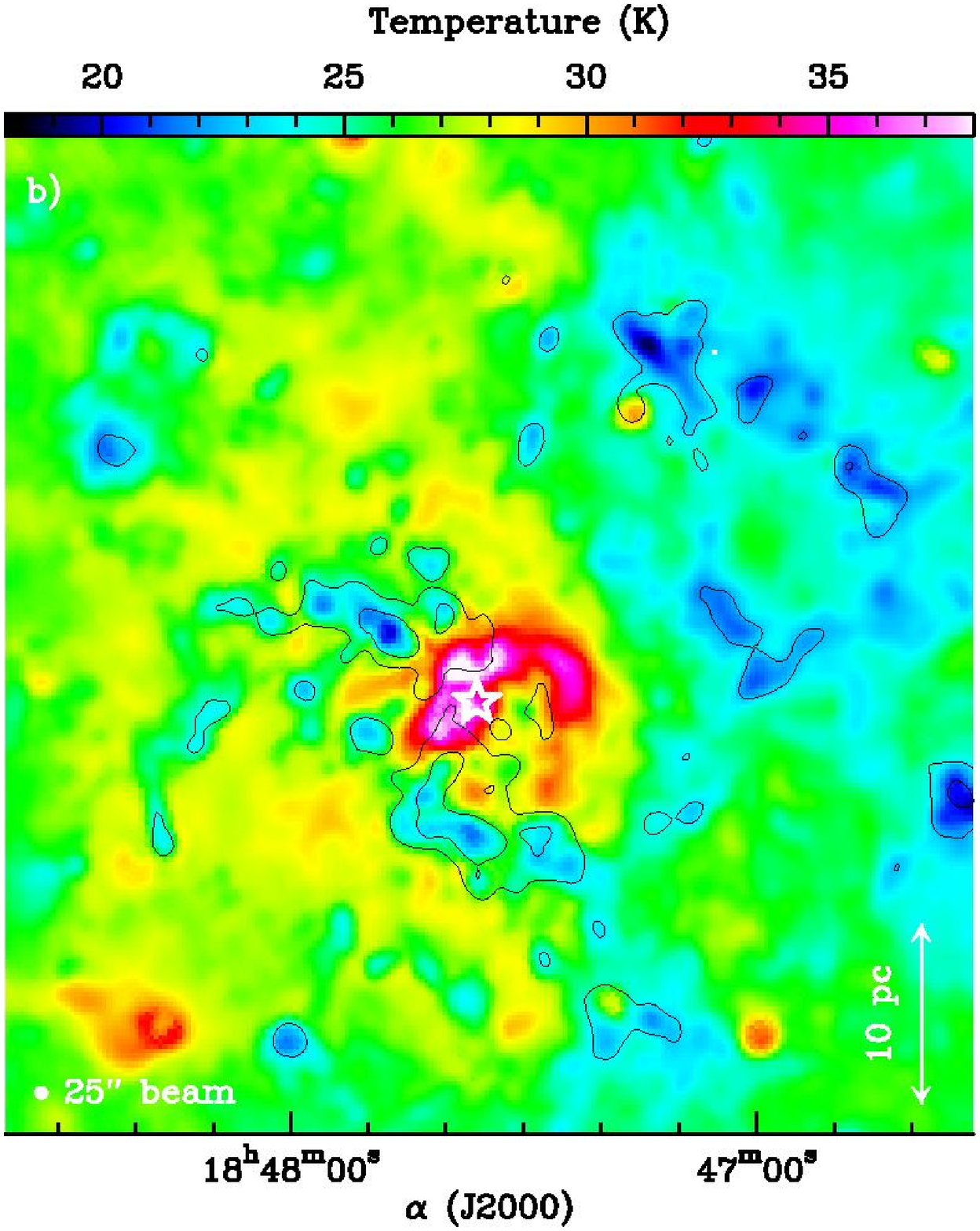}
\\
\end{array}$
\caption{The W43-Main mini-starburst in {\bf (a)} column density (color and contours), and {\bf (b)} dust temperature (color) and column density (contours) derived from \emph{Herschel} images. The black contours are the $4\times10^{22}~\cmd$ and $1\times10^{23}~\cmd$ levels, outlining the W43-MM1 and W43-MM2 ridges and their immediate surroundings. The star symbol pinpoints the WR/OB stars cluster responsible for a giant {\hii} region seen  in {\bf b} as a heated bubble. 
}
\label{fig:w43coldens}
\end{figure*}

\subsection{SiO and N$_2$H$^+$ mappings with the IRAM 30~m}
\label{mappingsurvey}
We extracted the SiO~(2--1) and N$_2$H$^+$~(1--0) emission lines from a $\sim$160$\arcmin^2$ mapping survey of the W43-Main mini-starburst region made at 3~mm (85--93~GHz). These observations were performed as part of the W43-HERO IRAM 30~m Large Program\footnote{see \url{http://www.astro.uni-koeln.de/projects/schilke/IRAMLargeProject/MainPage}}. Its first results are jointly presented in present paper and in Carlhoff et al. (submitted).
SiO~(2--1) and N$_2$H$^+$~(1--0) emission lines were observed with two different frequency setups of the Eight MIxer Receiver (EMIR) in December 2010, January 2011, and March 2011.
We used the low spectral resolution (0.67\,\kms) but large bandwidth (8~GHz) Fast Fourier Transform Spectrometer (FTS) backend for SiO lines and the high spectral resolution (0.13\,\kms) but narrow bandwidth (80~MHz) Versatile SPectrometer Array (VESPA) for N$_2$H$^+$~(1--0) lines. The observational parameters of the SiO and N$_2$H$^+$ mappings are summarized in Table~\ref{table:SiOobs}.

\begin{table}[htbp]
\centering
\caption{Observational parameters for the IRAM 30~m lines}
\label{table:SiOobs}
\begin{tabular}{lrrr}
\hline
\hline
Parameter	 	  & Unit	    &   SiO~(2--1)  & N$_2$H$^+$~(1--0) \\
\hline
Frequency  & (GHz)	    &   86.846      &  93.173 \\
Angular  \textit{HPBW} & (\arcsec)       &   31   &28\\
Linear  \textit{HPBW} & (pc)        &   0.90   &0.81\\
$\Delta$V$_{\mbox{\tiny res}}$ & (\kms) & 0.67 & 0.13\\
3$\sigma$~rms  & (K\kms)      &    0.05 & 0.01 \\
$B_{\rm eff}$  & - &  0.81  & 0.81 \\
$F_{\rm eff}$  & - &  0.95  & 0.95 \\
\hline
\end{tabular}
\flushleft
\end{table}

The maps were done using the on-the-fly mode in two perpendicular scanning directions (RA and Dec) and a full sampling which provides the best angular resolutions (see Table~\ref{table:SiOobs}). Pointing, focus, and  calibration observations were repeated frequently. Pointing error is less than 3\arcsec\, and calibration accuracy is within 10\%.
The basic data reduction steps such as pointing correction, focus correction, and flux calibration were done using  
MIRA which is part of the GILDAS 
package\footnote{GILDAS (Grenoble Image and Line Data Analysis Software) is a software package mainly developed and maintained by IRAM to reduce and analyze data obtained with the 30~m telescope and Plateau de Bure interferometer. See {\tt www.iram.fr/IRAMFR/GILDAS}}. 
We removed a baseline from each spectrum by masking the main velocity range of W43-Main (80--100\,\kms) and fitting a first-order polynomial to the spectrum observed over the remaining velocity ranges.
We converted the units of the data cubes from antenna temperature, $T_{\rm  A^*}$, to main beam temperature, $T_{\rm MB}$, following the usual conversion: $T_{\rm MB}= T_{\rm A^*}\times\frac{F_{\rm eff}}{B_{\rm eff}}$ with main beam and forward efficiencies given in Table~\ref{table:SiOobs}.
In the remaining of the paper, we used $T_{\rm MB}$ as the intensity unit.
Finally, we combined the reduced spectra into 
gridded data cubes with a $10\arcsec$ Gaussian kernel, 
leading to a slightly degraded resolution (see Table~\ref{table:SiOobs}).

\begin{table*}[htbp]
\setlength{\tabcolsep}{5pt}
\caption{Basic characteristics of the W43 ridges derived from \emph{Herschel} images, N$_2$H$^+$~(1--0), and SiO (2--1) and  line mappings}
\begin{tabular}{lrrccccrrrrr
c}
\hline
\hline
Cloud structure  & $A$ & $l$ & $r\,^{a}$ & $V\,^{a}$ & $T$ & {$\langle N_{\rm H_2}\rangle\,^{b}$} & $M\,^{b}$ &  $n_{\rm H_2}\,^{a,b}$  & $\Sigma_{S}\,^{a,b}$ & $M_{\rm vir}\,^{c}$ & $\frac{M}{M_{\rm vir}}\,^{c}$ & $L^{\rm total}_{\rm SiO}$\\
 & {(pc$^2$)} & {(pc)} & {(pc)} & {(pc$^{3}$)} & {(K)} & {(\cmd)} & {(\msun)} &  {(\cmc)} & {($\msun\,$pc$^{-2}$)} & (\msun) & - & 
(K\,$\kms$\\
\multicolumn{1}{c}{(1)} & (2) & (3) &(4) &(5) &(6) &(7) & \multicolumn{1}{c}{(8)} & \multicolumn{1}{c}{(9)} & \multicolumn{1}{c}{(10)} & \multicolumn{1}{c}{(11)} & \multicolumn{1}{c}{(12)} & \,kpc$^{2}$)(13) 
\\
\hline
\multicolumn{10}{l}{\bf The ridges ($N_{\rm H_2}>$ 10$^{\bf 23}~\cmd$)}\\
\hline
W43-MM1 ridge     & 6 & 4.3    & 0.7 & 6.6 & $19-27$ & 1.9$\times10^{23}$ & 21000    &  6.0$\times10^{4}$ & 3500 & 3700 & 5.5 & 1.0$\times 10^4$\\
W43-MM2 ridge    & 14   & 8.8   &0.8 & 17.6 & $21-28$ & 1.2$\times10^{23}$ & 35000   & 3.4$\times10^{4}$ & 2300 & 11000 & 2.9 &  1.0$\times 10^4$\\
DR21$^{d}$ ridge & 2.3  & 4.0 & 0.3 & 1.1 & $14-25$ & 2.3$\times10^{23}$ & 9800   & 1.7$\times10^{5}$ & 4300 & - & 
-\\
\hline
\multicolumn{10}{l}{\bf The ridge and its immediate surrounding ($N_{\rm H_2} >$ 4${\bf \times10^{22}~\cmd}$)}\\
\hline
W43-MM1 sur     & 43  & 11.9   & 1.8 & 121.5 & $19-40$ & 4.4$\times10^{22}$ & 32000  & 5.5$\times10^{3}$ & 830 & 58000 & 0.6 & 2.0$\times 10^4$ \\
W43-MM2 sur    & 62   & 13.5  	& 2.3 & 223.9 & $21-38$ & 4.6$\times10^{22}$ & 53000  & 4.5$\times10^{3}$ & 900  & 120000 & 0.5 & 2.0$\times 10^4$ \\
DR21$^{d}$ sur & 6.0    & 5.8    & 0.5 & 4.6 & $14-25$ & 1.1$\times10^{23}$ & 12200  & 5.1$\times10^{4}$ & 2000 & - & - & -\\
\hline
\end{tabular}

$^{a}$ The radii and thus volumes are not deconvolved from the $N_{\rm H_2}$ beam ($25\arcsec$ in W43 and $37\arcsec$ in DR21). The resulting volume number densities quoted here may be underestimated by a factor up to $\sim$2: 
e.g. $1.0\times10^{5}~\cmc$ instead of $6.0\times10^{4}~\cmc$ for the W43-MM1 ridge and $5.8\times10^{4}~\cmc$ instead of $3.4\times10^{4}~\cmc$ for the W43-MM2 ridge . 

$^{b}$ Properties measured after subtracting the common surrounding column density background of $N_{\rm H_2 \,, sur }^{\rm background} = 4\times10^{22}~\cmd$.

$^{c}$ Mass estimated from Col.~2 and the median velocity dispersion of N$_2$H$^+$~(1--0) lines given in Table~\ref{tab:n2h+fitting}.

$^{d}$ All properties of the DR21 ridge are drawn from \cite{hennemann12}. The temperature range given for DR21 excludes the area toward DR21 itself due to saturation, and where we expect higher dust temperatures.

\label{tab:ridgeproperties}
\end{table*}

\section{PROPERTIES of Cloud ridges}
\label{sect:ridgeresult}

In this section, we present the properties of the W43-MM1 and W43-MM2 ridges using column density and temperature images derived from \emph{Herschel} maps (Sect.~\ref{sect:coltem}), and the velocity structure traced by N$_2$H$^+$ emission lines (Sect.~\ref{sect:vel}).

\subsection{Density structure}
\label{sect:coltem}
Using three of the four longest wavelengths of \emph{Herschel} (160--350\,\micron), we derived the total (gas+dust) column density ($N_{\rm H_2}$) and average dust temperature maps of W43-Main with an angular resolution of $25\arcsec$, which is that of the $350\,\micron$ map. 
Following the procedure fully described in  \citet{hill11,hill12}, 
we fitted pixel-by-pixel spectral energy distributions (SEDs) with modified blackbody models. We used a dust opacity law similar to that of \citet{hildebrand83} but with $\beta = 2$ instead of $\beta = 1$ and assumed a gas-to-dust ratio of 100: $\kappa_{\nu} =  0.1 \times (300\,{\rm \mu m}/\lambda)^{2}$~cm$^2$\,g$^{-1}$.
Given the high quality of the \emph{Herschel} data, a robust fit can be done even when dropping the 500\,\micron\, data point. The column density values are only changed by a factor of less than 10\% in region with $N_{\rm H_2}>10^{22}~\cmd$ 
\citep[see][]{hill12}. 
Changing $\beta$ from $2$ to $1.5$ would globally lower the column density values by $\sim$20\%. Carlhoff et al. (subm.) made  a detailed comparison of \emph{Herschel} versus CO column density maps and showed that they do agree with each other to within a factor of $50\%$ for intermediate-density clouds.

The column density and temperature maps of W43-Main and the entire W43 molecular complex are plotted in Fig.~\ref{fig:w43coldens} and~\ref{fig:w43coldenslarge}. 
The noise level, estimated over areas with minimum intensity variations, are  $\sigma\sim1.5\,\times10^{21}~\cmd$ and $\sim$1.5\,\K.
The average column density and temperature found toward W43-Main are $\sim$2.5 $\times\,10^{22}~\cmd$ (\av~$\sim$ 25\,mag) and  $\sim26\,\K$. 
The dominating structure of Fig.~\ref{fig:w43coldens} is the cold and dense Z-shaped filament discussed in \cite{motte03}, which reaches column densities up to $\sim$6.6\,$\times10^{23}~\cmd$ and temperatures down to $\sim$19\,\K. 
It contains two clouds with \Nhtwo\,$> 10^{23}~\cmd$, i.e. above the column density requirement of ridges as defined by \cite{hill11}. They host the W43-MM1 and W43-MM2 massive dense cores \citep{motte03} and are hereafter denoted as the W43-MM1 and W43-MM2 ridges. 
These ridges stand out from their environment at a column density level of $\sim$4\,$\times 10^{22}~\cmd$, inside which we defined their immediate surroundings (see Fig.~\ref{fig:w43coldens}a). 
Between these ridges lies a region with higher temperature ($T>22~\K$) and lower column density \Nhtwo$\,\sim 3-7\times10^{22}~\cmd$ (see Fig.~\ref{fig:w43coldens}). This hot bubble of gas is produced by the strong UV radiation from the massive stars within the WR/OB cluster associated with W43-Main (star symbol on Fig.~\ref{fig:w43coldens}).

We used the \Nhtwo\,$= 10^{23}~\cmd$ and  $4 \times 10^{22}~\cmd$ contours of Fig.~\ref{fig:w43coldens}a to measure the areas, $A$, and estimate the rough lengths, $l$, of the W43-MM1 and W43-MM2 ridges and their immediate surroundings. We assumed they are well represented by a cylindrical geometry to calculate their radii, $r=A/(2\,l)$, and volumes, $V=\pi\,l\,r^2$. We used the \Nhtwo\,$=4 \times 10^{22}~\cmd$ value as background emission in measuring the column density averaged over the W43-MM1 and W43-MM2 ridges, and their immediate surroundings.
The mean column density after background subtraction, $\langle N_{\rm H_2}\rangle$, translates into mass, volume number density, and mass surface density ($M$, $n_{\rm H_2}$, $\Sigma_S$) by the relations:
\begin{eqnarray}
\label{eq:nh22masssurfden}
M & = & \mu\, m_{\rm H} \, \langle N_{\rm H_2}\rangle A \simeq 1853\,\msun \times \frac{\langle N_{\rm H_2}\rangle}{10^{23}\, {\rm cm^{-2}}} \times \frac{A}{1~{\rm pc^2}}\\
n_{\rm H_2} & = &  \frac{M}{V} \simeq
22450~{\rm cm^{-3}}
\times \frac{\langle N_{\rm H_2}\rangle}{10^{23}\, {\rm cm^{-2}}}  \times \left( \frac{r}{1~{\rm pc}} \right)^{-1}\\
\Sigma_S & = & \mu\, m_{\rm H} \, \langle N_{\rm H_2}\rangle \simeq
1853\,\msun\,{\rm pc^{-2}} \times \frac{\langle N_{\rm H_2}\rangle}{10^{23}\, {\rm cm^{-2}}}
\end{eqnarray}
where $\mu=2.33$  is the mean molecular weight and $m_{\rm H}$ is the hydrogen mass. 
Table~\ref{tab:ridgeproperties} lists all the characteristics derived from the \emph{Herschel} column density and temperature images.
We also list the properties of the DR21 ridge, drawn from \cite{hennemann12} with the same background subtractions, to allow a comparison with this prototypical ridge of the HOBYS survey.

Both the W43-MM1 and W43-MM2 ridges have higher temperatures than the DR21 ridge, due to their proximity to the WR/OB cluster and a higher background of low-density, high-temperature cloud impacting the SED fits.
The mean column densities of the W43-MM1 and W43-MM2 ridges are similar and are also similar to that of the DR21. 
For both the cloud ridges and their immediate surroundings, the masses measured for W43-MM1 and W43-MM2 are 2-4 times higher than the corresponding ones of DR21.
 This is a natural consequence of the mass being integrated over a larger area since the W43-MM1 and W43-MM2 ridges and surroundings are a few times larger than the DR21 ridge. 
This also explains the lower values of volume number density and mass surface density, by factors of 3--10 and 1--2, measured in the W43-MM1 and the W43-MM2 ridges when compared with the DR21 ridge (see Table~\ref{tab:ridgeproperties}). 
One should however remember that DR21 is characterized with three times higher linear spatial resolution than W43 (0.25~pc vs. 0.7~pc). With a similar resolution, the $\langle N_{\rm H_2}\rangle$ value measured for W43-MM1 would increase and the cloud volume would decrease, thus increasing by a factor of a few the estimated volume number density (see Table~\ref{tab:ridgeproperties} for a tentative deconvolution of $n_{\rm H_2}$).

\begin{figure*}[bhpt]
\includegraphics[angle=0,width=17cm,height=18.5cm]{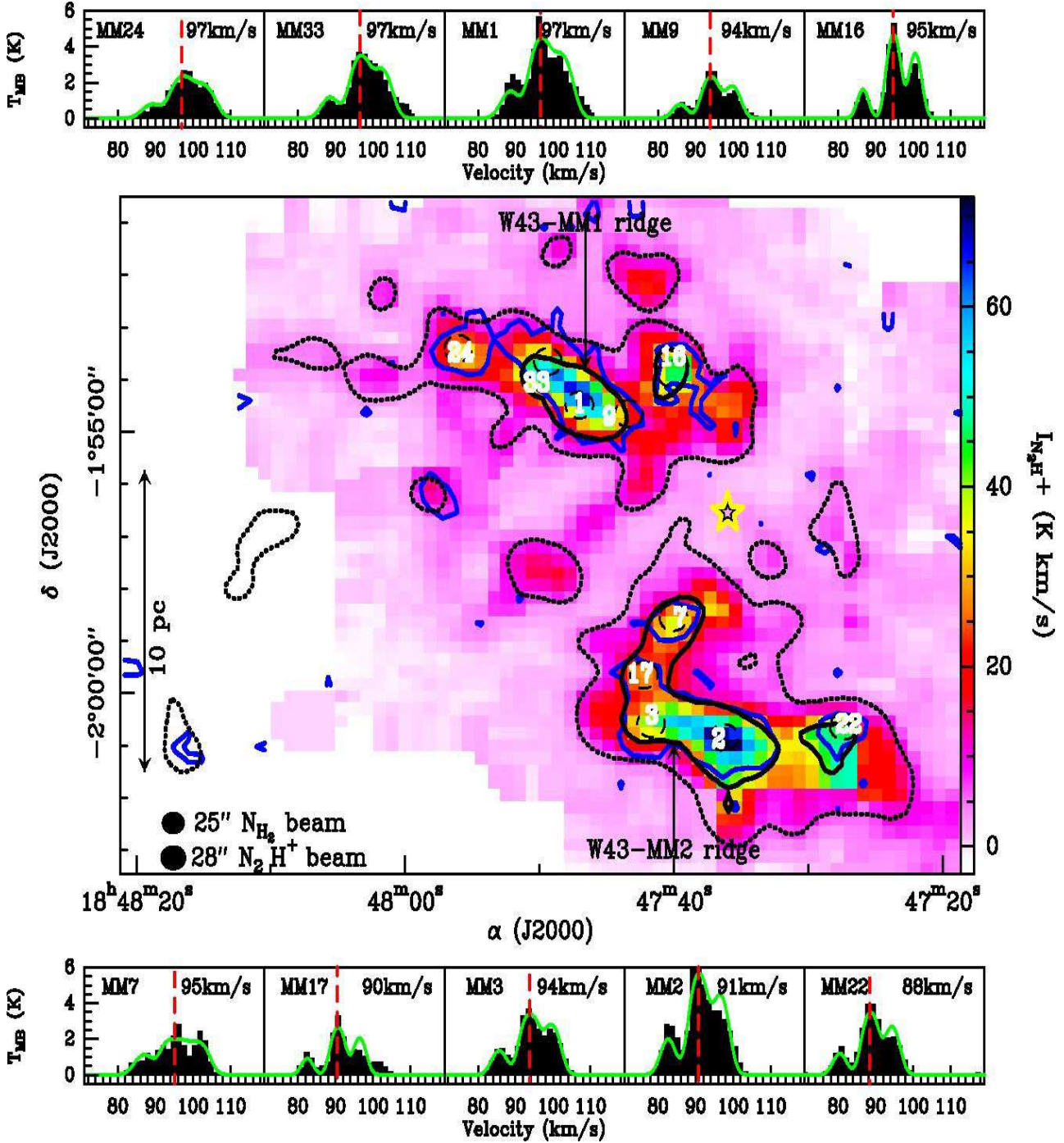} \\ \\
\caption{The N$_2$H$^+$~(1--0) integrated intensity (color image), \emph{Herschel} column density (black contours), and SiO~(2--1) integrated intensity (blue contour) maps of W43-Main.  The range of integration for both lines is $80-110~\kms$ and the 3$\sigma$ level of the N$_2$H$^+$~(1--0) integrated map is $\sim\,$4~K$\,\kms$. The dotted and solid contours are the $4\times10^{22}~\cmd$ and $1\times10^{23}~\cmd$ \emph{Herschel} column density levels, outlining the ridges boundary and their surroundings. The blue contours indicate the regions with SiO emission detected above the 3$\sigma=0.8~$K$\,\kms$ level (see Fig.~\ref{fig:w43ridgeSiO}).  The star symbol pinpoints the WR/OB stars cluster. 
N$_2$H$^+$~(1--0) spectra, measured at beam locations outlined by dashed circles. and labelled in white numbers (according to the massive dense cores numbering of \citealt{motte03}), are plotted as insets where black histograms are the measured spectra. The green curves represent the hyperfine structure fits to the spectra (see text), and red dashed lines mark the sources peak velocity derived from the fits.}  
\label{fig:w43ridgen2h+}
\end{figure*}

\begin{table*}[htb]
\setlength{\tabcolsep}{3pt}
\begin{center}
\scriptsize
\caption{Kinematics overview of a few positions within the W43-MM1 and W43-MM2 ridges and their mean characteristics.}
\begin{tabular}{llllll|lllr|r}
\hline
\hline
{Position/} & \multicolumn{5}{c}{N$_2$H$^+$ (1--0)} & \multicolumn{3}{c}{SiO (2--1)} &  \\
{Name} & {$T_{\rm N_2H^+}^{\,a}$} & {$V_{\rm LSR}^{\,a}$} & {{$\Delta V_\textit{FWHM}$}$^{a}$} & {\textit{$\sigma_{1D}$}$^{d}$} & {$\tau_{\rm main}$ $^{a}$} & {$\int T dv^{b}$} & {$V_{\rm LSR}$$^{b}$} & {$\Delta V_\textit{FWHM}\,^{b}$}   & {\textit{$\sigma_{1D}$}$^{d}$} &  $\Delta_{\vlsr}$$^{c}$\\ 
{} & {(K)} & {(km/s)} & {(km/s)} & {(km/s)} & {(K km/s)} & {(km/s)} & {(km/s)} & {(km/s)}& {(km/s)}\\
\multicolumn{1}{c}{(1)} & \multicolumn{1}{c}{(2)}  & \multicolumn{1}{c}{(3)} & \multicolumn{1}{c}{(4)} & \multicolumn{1}{c}{(5)} & \multicolumn{1}{c}{(6)} & \multicolumn{1}{c}{(7)} & \multicolumn{1}{c}{(8)} & \multicolumn{1}{c}{(9)} & \multicolumn{1}{c}{(10)} & \multicolumn{1}{c}{(11)}\\
\hline
\multicolumn{11}{l}{\bf W43-MM1 Ridge} \\
W43-MM24 & 0.30$\pm$0.01  &  97.1$\pm$0.03  &  5.8$\pm$0.05  & 2.5$\pm$0.02  & 0.10$\pm$0.01   &   2.04$\pm$0.15  &  98.4$\pm$0.44  &  12.2$\pm$1.08 & $5.2\pm0.46$ &-1.3$\pm$0.44 \\ 
W43-MM33 & 0.45$\pm$0.01  &  96.6$\pm$0.11  &  5.1$\pm$0.18  & 2.1$\pm$0.08  &  0.10$\pm$0.01  &  3.37$\pm$0.10  &  97.9$\pm$0.09  &  6.8$\pm$0.22  & $2.9\pm0.09$ & -1.3$\pm$0.14\\ 
W43-MM1  & 0.58$\pm$0.01  &  96.6$\pm$0.08  &  5.5$\pm$0.18  & 2.3$\pm$0.08  & 0.10$\pm$0.03  &  7.00$\pm$0.05  &  98.3$\pm$0.08  &  7.8$\pm$0.22 & $3.3\pm0.09$ & -1.7$\pm$0.11 \\ 
W43-MM9  & 0.62$\pm$0.01  &  94.6$\pm$0.01  &  3.4$\pm$0.01  & 1.4$\pm$0.01  & 0.10$\pm$0.01  &   1.20$\pm$0.11 &  94.0$\pm$0.29  &  5.9$\pm$0.83  &$2.5\pm0.35$ & 0.6$\pm$0.29 \\ 
W43-MM16 & 0.31$\pm$0.01  &  93.9$\pm$0.03  &  4.3$\pm$0.05  & 1.8$\pm$0.02  & 0.10$\pm$0.01  &   0.77$\pm$0.08  &  94.1$\pm$0.27  &  6.5$\pm$0.85 & $2.8\pm0.36$ & -0.2$\pm$0.3\\ 
{W43-MM1 ridge$^{\,e}$} &{\it 0.45$\pm$0.01}  &  {\it 96.6$\pm$0.08}  &{\it  5.1$\pm$0.18}  &{\it 2.2$\pm$0.08}  &{\it 0.10$\pm$0.01}  & {\it  2.04$\pm$0.15}  &  {\it 97.9$\pm$0.09 }  &{\it  6.8$\pm$0.22} & {\it 2.9$\pm$0.09} & {\it -1.3$\pm$0.20}\\

\hline
\multicolumn{11}{l}{\bf W43-MM2 Ridge} \\
W43-MM7  & 0.53$\pm$0.01  &  95.0$\pm$0.02  &  5.3$\pm$0.04  & 2.3$\pm$0.02  &  0.26$\pm$0.01  &  1.83$\pm$0.19  &  93.8$\pm$0.49  &  12.3$\pm$1.24 & $5.2\pm0.53$ & 1.2$\pm$0.50\\ 
W43-MM17 & 0.60$\pm$0.67  &  90.6$\pm$0.10  &  2.5$\pm$0.17  & 1.1$\pm$0.07  & 0.21$\pm$0.32  &  1.21$\pm$0.12  &  90.9$\pm$0.25  &  6.8$\pm$0.62 & $2.9\pm0.26$ & -0.3$\pm$0.27 \\ 
W43-MM3  & 0.55$\pm$0.01  &  93.7$\pm$0.01  &  4.5$\pm$0.02  & 1.9$\pm$0.01 &  0.14$\pm$0.01  &  3.00$\pm$0.13  &  93.2$\pm$0.12  &  6.9$\pm$0.29 &$2.9\pm0.12$ & 0.5$\pm$0.12\\ 
W43-MM2  & 0.78$\pm$0.01  &  90.7$\pm$0.01  &  4.5$\pm$0.02  & 1.9$\pm$0.01 & 0.11$\pm$0.01  &  4.25$\pm$0.18  &  91.0$\pm$0.12  &  8.0$\pm$0.35  & $3.4\pm0.15$ & -0.3$\pm$0.12 
\\ 
W43-MM22 & 0.46$\pm$0.01  &  88.5$\pm$0.05  &  4.1$\pm$0.09  & 1.7$\pm$0.04  & 0.10$\pm$0.01  &  1.69$\pm$0.19  &  88.3$\pm$0.26  &  6.5$\pm$0.73 & $2.8\pm0.31$ & 0.2$\pm$0.27 \\
{ W43-MM2 ridge$^{\,e}$ }& {\it 0.55$\pm$0.01 } &  {\it 90.7$\pm$0.01 } & {\it  4.5$\pm$0.02 } & {\it 1.9$\pm$0.01 } & {\it  0.14$\pm$0.01 } & {\it 1.83$\pm$0.19 } &    {\it 91.0$\pm$0.12 } &  {\it 6.9$\pm$0.29 } &{\it 2.9$\pm$0.12 } & {\it -0.3$\pm$0.12 }\\
\hline
\end{tabular}
\label{tab:n2h+fitting} 
\end{center}
$^{a}$ Results of the Gaussian hyperfine fits to the N$_2$H$^+$~(1--0) multiplet. $T_{\rm N_2H^+}$ (Col.~2) is the line temperature and $\tau_{\rm main}$ (Col.~5) the optical depth of the strongest N$_2$H$^+$ (1--0) multiplet, related by the equation $T_{\rm N_2H^+} =T_{\rm MB}\times\tau_{\rm main}$.\\
$^{b}$ Results of the Gaussian fits to the SiO~(2--1) lines.\\
$^{c}$ The shift between the peak velocities of the N$_2$H$^+$~(1--0) and SiO~(2--1) lines is measured as $\Delta_{\vlsr}=V_{\rm LSR}({\rm N^2H^+})-V_{\rm LSR}({\rm SiO})$.\\
$^{d}$ The 1D velocity dispersion is calculated as $\sigma_{1D}=\frac{\Delta V_\textit{FWHM}}{\sqrt{8 \,{\rm ln} 2}}$, using Col.~4 for $\rm N_2H^+$ Col.~10 for SiO.\\
$^{\,e}$ The values of the ridges are the mean values of all pixels contained in each ridge.

\end{table*}

\subsection{Velocity structure}
\label{sect:vel}

The integrated intensity map of the N$_2$H$^+$~(1--0) emission observed toward the W43-Main mini-starburst as well as spectra at a few selected locations are shown in Fig.~\ref{fig:w43ridgen2h+}. The N$_2$H$^+$~(1--0) lines are integrated from 80 to 110~\kms, which is the velocity range of W43-Main as defined in \cite{nguyenluong11}. 
The selected positions are paving the W43-MM1 and W43-MM2 ridges and have clear SiO peaks (see Fig.~\ref{fig:w43ridgeSiO}).  
They are coincident or close to massive dense cores identified by \cite{motte03} and whose names are used here for discussion (see Figs.~\ref{fig:w43ridgen2h+}, \ref{fig:w43ridgeSiO} and Table~\ref{tab:n2h+fitting}).
With a critical density of $\sim$4$~\times 10^{5}~\cmc$, N$_2$H$^+$~(1--0) is a good tracer of the cold dense gas in the W43-MM1 and W43-MM2 ridges as well as their massive dense cores. 

The N$_2$H$^+$~(1--0) line splits into seven hyperfine structure components. We thus fitted the observed N$_2$H$^+$~(1--0) spectra by seven Gaussians, whose 
frequency spacings and relative line strengths are theoretically determined and cataloged in 
splatalogue\footnote{The splatalogue is a catalog of spectroscopic information of astronomical spectral lines which can be accessed from \url{www.splatalogue.net}} 
(see Fig.~\ref{fig:w43ridgen2h+} and Table~\ref{tab:n2h+fitting}). 
Considering the noise level in the spectra and the high probability for $\sim$0.9~pc beams to contain several cloud components, the multiple hyperfine structure component fits to the N$_2$H$^+$ (1--0) spectra are acceptable at all locations.
The derived optical depths range from 0.1 to 0.26, meaning that N$_2$H$^+$ (1--0) emission is optically thin and traces all material along the line of sight.
The W43-MM1 and W43-MM2 ridges have their $\vlsr$ spanning ranges of $94-97~\kms$ and $88-95~\kms$, respectively. 
At the ten locations except W43-MM1, the N$_2$H$^+$ $\vlsr$ are only $\le$1~$\kms$ different from the values determined for H$^{13}$CO$^+$~3--2 lines \citep{motte03}, it is $\sim$2~$\kms$ higher for the W43-MM1 dense core. 

We used the volume $V=\frac{4}{3}\pi\,r_\text{sph}^3$ of the W43-MM1 and W43-MM2 ridges and the median velocity dispersion of their N$_2$H$^+$ lines, $\sigma_{\rm N_2H^+}$, given in Tables~\ref{tab:ridgeproperties}-\ref{tab:n2h+fitting} to calculate their virial mass, $M_{\rm vir}$,
according to:
\begin{equation}
M_{\rm vir}=3\times \frac{r_\text{sph}\, \sigma_{\rm tot}^{2}}{G},
\end{equation}
where $G$ is the gravitational constant and $\sigma_{\rm tot}$ the total (thermal + nonthermal) velocity dispersion. The latter is calculated as
\begin{eqnarray}
\sigma_{\rm tot} & = & \sqrt{\sigma_{\rm T}^{2}+\sigma_{\rm NT}^{2}} 
=  \sqrt{\sigma_{\rm N_2H^+} + \frac{k\,T}{m_{\rm H}} \left( \frac{1}{\mu} - \frac{1}{\mu_{\rm\tiny N_2H^+}}\right) }
\end{eqnarray}
where $k$ is the Boltzmann constant, $\mu=2.33$ and $\mu_{\rm N_2H^+}=29$ are the mean and N$_2$H$^+$ molecular weights. 
We assumed that $T$ is close to the dust temperature measured in the ridges to be $\sim$23~K.
This calculation only provides approximated  $M_{\rm vir}$ values for the W43-MM1 and W43-MM2 ridges since they are better represented by cylinders than spheres and are embedded within backgrounds whose N$_2$H$^+$ emission is not removed here.
The unusually large virial ratios measured, $M/M_{\rm vir}\sim5.5$ and 2.9, indicate that, despite their high level of turbulence, the W43-MM1 and W43-MM2 ridges are most probably gravitationally bound cloud structures. The immediate surroundings, however, have much smaller virial ratios, 0.6 and 0.5, indicate that they are just at the limit to be
gravitationally bound which requires $M/M_{\rm vir} > 0.5$.
Since $\vlsr$ velocities in the W43-MM1 and W43-MM2 ridges vary within $5~\kms$ ranges, which are close to the measured line widths, and since the virial ratio suggest boundedness/collapse, W43-MM1 and W43-MM2 ridges can be considered as coherent structures in both in space and velocity. 

\section{SiO emission Result \& Analysis}
\label{sect:sioresult}

\subsection{SiO spectra and integrated intensity map}
\label{sect:siointspec}
\begin{figure}[bhtp]
\centering
$\begin{array}{c}
\includegraphics[angle=0,width=8cm]{{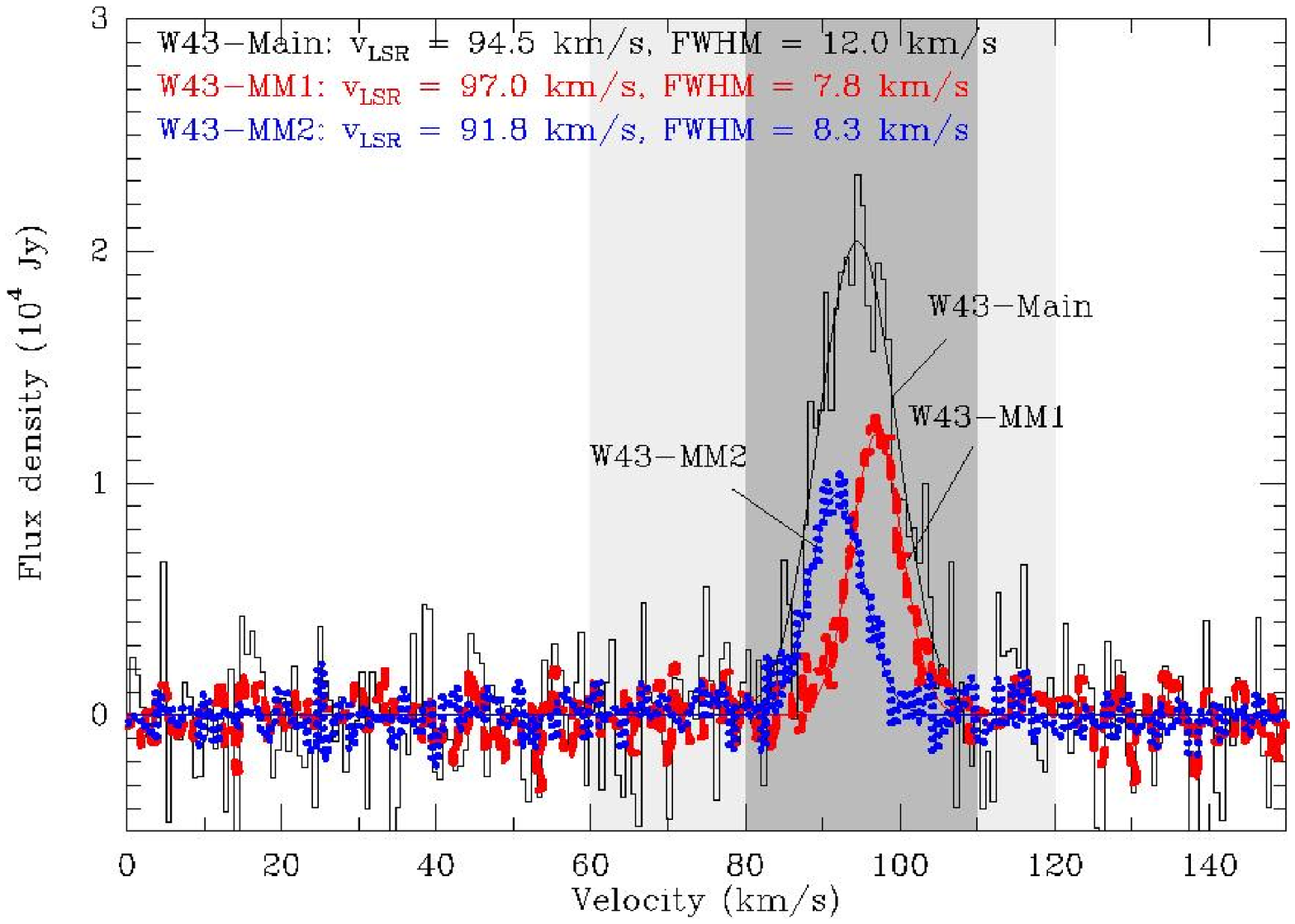}} 
\end{array}$
\caption{SiO~(2--1) accumulated line spectra resulting from the sum of all spectra observed in the entire W43--Main area (black line), compared to
those summed over the W43-MM1 ridge (red dashed line) and W43-MM2 ridge
(blue dashed-dotted line). 
The main and complete velocity ranges of W43 defined by  \cite{nguyenluong11} are indicated with light- and dark-gray filling.
}
\label{fig:siospectraw43}
\end{figure} 

In Fig.~\ref{fig:siospectraw43}, we plot the accumulated spectrum representing the SiO~(2--1) emission integrated over the entire W43-Main region, as well as the spectra integrated over the W43-MM1 and the W43-MM2 ridges defined in Sect.~\ref{sect:coltem}.
The spectrum integrated over the entire W43-Main region spans a velocity range of 80--110~\kms, in agreement with the main $^{13}$CO~(1--0) velocity range of W43 derived in \cite{nguyenluong11}. Note that Carlhoff et al. (subm.) used the $^{13}$CO~(2--1) data of our IRAM large program to 
refine the main velocity range for the W43 complex to $78-110~\kms$.
The similarity in velocities ranges between the $^{13}$CO~(1--0) and SiO~(2--1) lines is remarkable since SiO lines are generally tracing shocks from outflows and are thus expected to have very extended wings \citep[e.g.][]{schilke97}. 
Some SiO emission arising from a collection of outflows driven by protostars forming in the W43-MM1 and W43-MM2 ridges could be hidden in these broad lines.
The SiO~(2--1) lines of Fig.~\ref{fig:siospectraw43} seem however to be dominated by one line component without velocity wings.
We fitted Gaussian profiles to all three averaged spectra and obtained $\vlsr$ of 97.0\,\kms\, for the W43-MM1 ridge, 91.8\,\kms\, for the W43-MM2 ridge, and 94.5~\kms\, for the entire W43-Main. 
The \textit{FWHM} line width measured toward the W43-MM1 and W43-MM2 ridges both are $\sim$8~$\kms$ while that of the averaged spectrum in the entire W43-Main is somewhat larger, $\sim$12~\kms.   
The integrated peak SiO fluxes in the W43-MM1 and W43-MM2 ridges are of the same order: $\sim$1.2\,$\times10^{4}$~Jy (see Table~\ref{tab:ridgeproperties}).

\begin{figure*}[hbtp]
\hspace{-1cm}
\includegraphics[height=16cm,width=19cm,height=19.5cm]{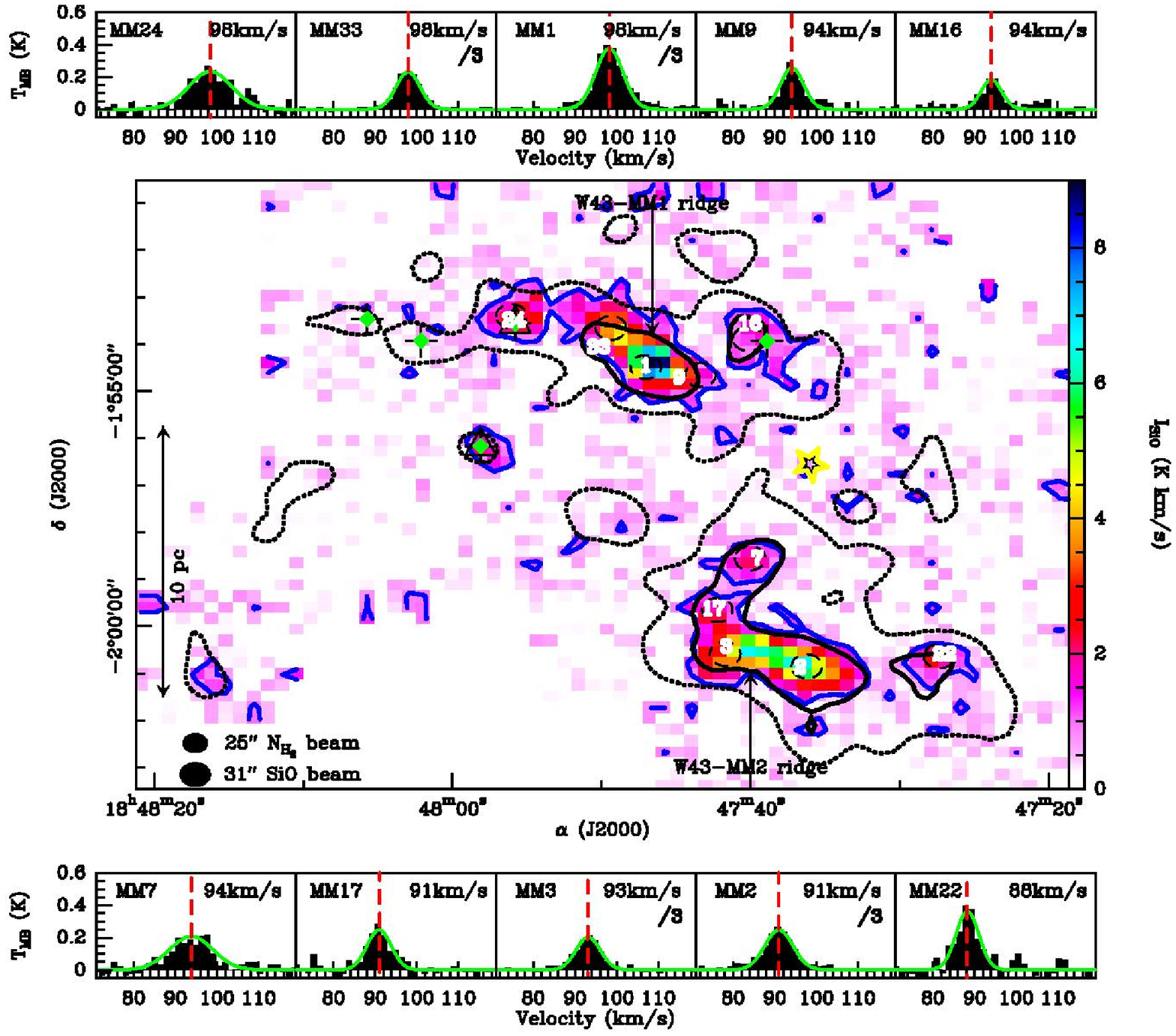}
\caption{The SiO~(2--1) integrated intensity (color image and blue contour), \emph{Herschel} column density (black contours) maps of W43-Main. The range of integration of SiO~(2--1) is $80-110~\kms$ and the blue contours indicate the regions with a  3$\sigma=0.8~$K$\,\kms$ detection or higher. The dotted and solid contours are the $4\times10^{22}~\cmd$ and $1\times10^{23}~\cmd$ \emph{Herschel} column density levels, outlining the ridges boundary and their surroundings. The star symbol pinpoints the WR/OB stars cluster. The green diamonds indicate the locations where SiO~(2--1) emission was detected by  \cite{beuther07c} and the ones with outer triangles indicate the locations where SiO~(2--1) is high (see Fig.~\ref{fig:SiOcomparison} for more details).
SiO~(2--1) spectra, measured at beam locations outlined by dashed circles and labelled in white numbers (according to the massive dense cores numbering of \citealt{motte03}), are plotted as insets where black histograms are the measured spectra. The SiO 2-1 intensities of MM33, MM1, MM3 and MM2 are divided by a factor of three for the visualization purpose. The green curves represent the Gaussian fits to the spectra and red dashed lines mark the sources peak velocity derived from the fits.
}
\label{fig:w43ridgeSiO}
\end{figure*}

The intensity map of the SiO~(2--1) emission, integrated from 80 to 110~\kms, is shown in Fig.~\ref{fig:w43ridgeSiO} for the W43-Main mini-starburst.
The SiO emission above a 3$\sigma$ level, i.e. $>$0.8~K$\,\kms$, appears as two extended features following the density structure of the two W43-MM1 and W43-MM2 ridges
(see and compare the blue and black $N_{\rm H_2}>10^{23}~\cmd$ contours in Fig.~\ref{fig:w43ridgeSiO}). 
A closer inspection shows that there is an exception north of the W43-MM33 position since the SiO emission peak is offset by $\sim$0.5~pc ($\sim$15\arcsec) from the dust column density peak.
The spatially integrated SiO fluxes in the W43-MM1 and W43-MM2 regions
are $\sim$39~K$\,\kms$\,pc$^2$ for areas of $\sim$12 and $\sim$16~pc$^2$, respectively.  
Their SiO emitting regions 
have mean dust column density of $\sim$1.5~$\times10^{23}~\cmd$\, and $\sim$5.9~$\times10^{23}~\cmd$, respectively. 
The strongest SiO emission peaks are at the position of massive dense cores in the W43-MM1 and W43-MM2 ridges, especially toward  W43-MM1 and W43-MM2, but SiO emission is also detected toward other more isolated sources. 
\cite{beuther07c} observed five locations in the W43-Main region, among which two remained undetected in our lower-sensitivity image (see Fig.~\ref{fig:w43ridgeSiO}). Their observations suggest that the SiO~(2--1) emission would be even more extended if observed with a higher sensitivity.

To probe in more detail the kinematics of shocked gas traced by SiO emission, we plotted the SiO spectra toward several peak positions in Fig.~\ref{fig:w43ridgeSiO}. These spectra are gridded at the \textit{HPBW} of $\sim$31$\arcsec$ or $\sim$0.9~pc at a distance of 6~kpc, each box thus representing a beam-separated independent spectrum. 
These spectra illustrate that the SiO emission is resolved and extended over several beams toward both the W43-MM1 and W43-MM2 ridges. 
The line parameters change from one location to another but most of the SiO spectra can be very well approximated by a single Gaussian profiles (see Table~\ref{tab:n2h+fitting}). 
The velocity dispersions of SiO (2--1) spectra are larger than those of N$_2$H$^+$~(1--0) spectra: $\sim$2.9~$\kms$ instead of  $\sim$2.2~$\kms$. 
W43-MM24 and W43-MM7, which both are at one end of the ridges, are  the locations with the broadest SiO~(2--1) and N$_2$H$^+$ lines (see Table~\ref{tab:n2h+fitting}) compared to the rest of the ridges . The comparison of the $\rm N_2H^+$~(1--0) and SiO~(2--1) peak velocities 
reveals a median offset of $\sim$1~$\kms$ in the W43-MM1 ridge 
(see Table~\ref{tab:n2h+fitting}).

\begin{table*}[htbp]
\centering
\caption{Median properties of SiO lines in W43-Main and a few comparative samples}
\footnotesize
\label{table:SiOlum}
\begin{tabular}{lccrrccccc}
\hline
\hline
\multicolumn{1}{c}{Source samples$^a$} & $\langle d \rangle$ & $3\sigma$ rms & $\langle {\rm \textit{FWZP}}\rangle$ & $\langle L_{\rm SiO (2-1)} \rangle$ & $\langle I_{\rm SiO (2-1)}^{\rm int} \rangle$ & \multicolumn{2}{c}{$L= \alpha_{\rm L}$$\times$${\rm \textit{FWZP}}$--$\beta_{\rm L}$} &  \multicolumn{2}{c}{$I= \alpha_{\rm I}$$\times$${\textit{FWZP}}$--$\beta_{\rm I}$} \\
 & {\scriptsize (kpc)} &  {\scriptsize (K/\kms)} &  {\scriptsize(\kms)} &  {\footnotesize(K$\,\kms$\,kpc$^2$)} &  {\scriptsize(K\,\kms)} &  {\footnotesize $\alpha_{\rm L}^{b}$ } &{\footnotesize $\beta_\text{L}^{b}$ } &  {\footnotesize $\alpha_\text{I}^{b}$ } &  {\footnotesize $\beta_{\rm I}^{b}$} \\
\multicolumn{1}{c}{(1)} & (2) & (3) & \multicolumn{1}{c}{(4)} & (5) & (6) & (7) & (8) & (9) & (10)\\ 
 \hline
{\footnotesize W43-Main ($>$ $3\sigma$) }     & 6.0 & 0.05& 14.2 & 960& 2.1 & 144 & 751 	& 0.27 & 1.42\\
{\footnotesize W43-MM1, 17 beams 	    }       & 6.0 & 0.05& 12.3 & 770& 1.7 & - & - & - & -\\
{\footnotesize W43-MM2, 18 beams 	     }      & 6.0 & 0.05& 14.6 & 970& 2.2 & - & - & - & -\\
\hline 
{\footnotesize All comparative studies  }      &2.9& -    & 31.8 & 331 & 3.0 & 13 & 72 & 0.1\ & 0.26\\
{\footnotesize M07, 31 sources in Cygnus~X }	       &1.4& 0.03& 18.7 & 37  & 1.5 & - & - & - & -\\
{\footnotesize LS11, 49 sources	}	 	   &3.0& 0.03& 35.9 & 290 & 2.2 & - & - & - & -\\
{\footnotesize BS07, 14 sources	}		   &3.9& 0.02& 27.0 & 240 & 1.5 & - & - & - & -\\
{\footnotesize J10, 3 beams in G035.39--00.33} &2.9& 0.006& 10.0 & 330 & 1.6 & - & - & - & -\\
{\footnotesize J12, CMZ G0.253+0.016	}		&8.0& 0.006& 70.0 &  608   & 9.5 &- & - & - & -\\
{\footnotesize J12, CMZ G1.6--0.025      }       &8.0& 0.006& 93.8 &  288   & 4.5 & - & - & - & -\\
\hline
\end{tabular}
\flushleft
$^{a}$ References: M07=\cite{motte07}; LS11=\cite{lopez-sepulcre11}; BS07=\cite{beuther07c}; J10=\cite{jimenez-serra10}, J12=\cite{jones12}. Sources types: MDCs=massive dense cores; MSFCs= star-forming cores; HMPOs= high-mass protostellar objects; CMZ= Central Molecular Zone\\
$^{b}$ Linear fit parameters measured from Figs.~\ref{fig:SiOcomparison}a-b. The units of $\alpha_\text{L}$ and $\alpha_\text{I}$ are K\,kpc$^2$ and K while those of $\beta_\text{L}$ and $\beta_\text{I}$ are K$\,\kms$\,kpc$^2$ and K$\,\kms$. We note that the measurements of CMZ are not included in the fit.\\
\end{table*}

\subsection{SiO~(2--1) luminosity \& velocity extent}
\label{sect:SiOstrength}

We measured the luminosity and the velocity extent (or Full Width at Zero Power) of the SiO~(2--1) lines detected in beam-averaged, $\sim$0.9~pc, pixels. To do so we first 
regridded the SiO spectra to $\sim$31$\arcsec$ pixels, corresponding to the beam size of our observations and $\sim$0.9~pc at 6~kpc from the Sun. 
We then selected 17 spectra above the $3\sigma$ level detection observed in W43-MM1 ridge and 18 in W43-MM2 ridge,
and thus being located within the blue contour of Fig.~\ref{fig:w43ridgeSiO}. 
The velocity extents, luminosities, and intensities of these SiO~(2--1) spectra were then compared in Table~\ref{table:SiOlum} to the 
values determined for a few selected samples \citep{lopez-sepulcre11,beuther07c,motte07,jimenez-serra10}.
The velocity extents of the SiO lines in W43-Main were defined above the 3$\sigma$, i.e. 0.05~K/kms level. This level is eight times less sensitive than G035.39-00.33 and three time less sensitive than the rest of the comparative studies (for details see Table~\ref{table:SiOlum}). However, even with higher sensitivity, G035.39-00.33 do not show wings in spectra.
We therefore expect both our velocity extents and SiO intensities to be directly comparable with those reported in the comparative studies within a factor of  $\sim$3 times. 

The sources from the \cite{lopez-sepulcre11}, \cite{beuther07c}, \cite{motte07}, and \cite{jimenez-serra10} samples have all been studied in SiO~(2--1) with the IRAM 30~m telescope and thus also with a $\sim$31$\arcsec$ beam. They span a large variety of distances: $\sim$1.4~kpc for the Cygnus~X  massive dense cores identified by \cite{motte07}, 1.7-5.7~kpc for the massive star-forming cores of \cite{lopez-sepulcre11} and 0.7-7.3~kpc for high-mass protostellar objects \citep{beuther07c} and $\sim$2.9~kpc for IRDC G035.39--00.33 \citep{jimenez-serra10}. 
In the first three studies, the SiO (2--1) emission is mainly intepreted as a tracer of protostellar outflows, in the last one the SiO emission is interpreted as associated with colliding flows.
CO outflows of high-mass protostars are known to typically be $\sim$0.05--0.8~pc long (e.g. \citealt{beuther02b}; Duarte-Cabral et al. subm.). 
The SiO emission in the three comparative studies arises from local regions where high-velocity shocks are developing and should thus be confined within a  $\sim$31$\arcsec$ beam for these $\sim$2.9~kpc sources.

\begin{figure*}[htbp]
\centering
$\begin{array}{cc}
\includegraphics[angle=0,width=9cm]{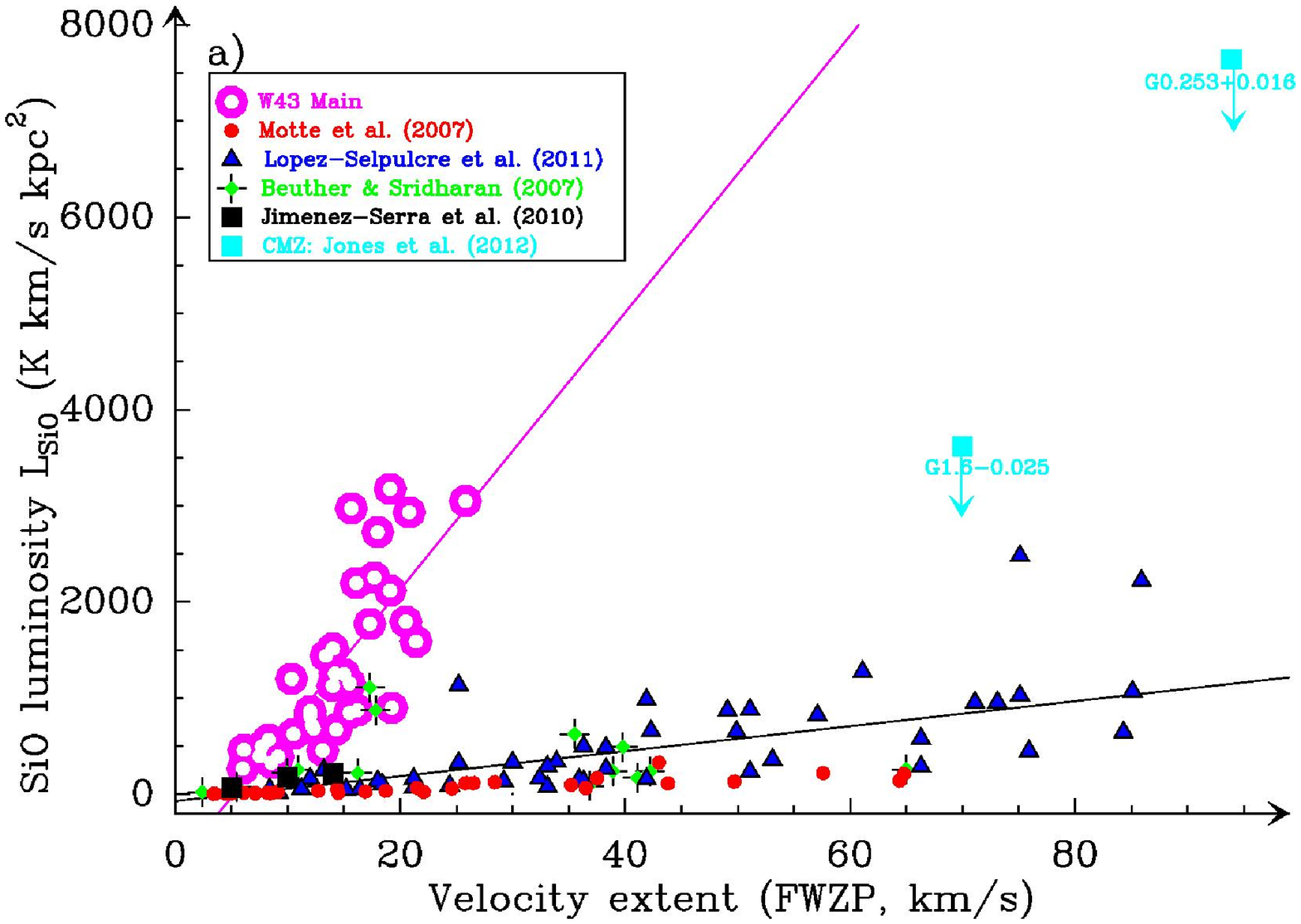} &
\hspace{0.2cm}
\includegraphics[angle=0,width=9cm]{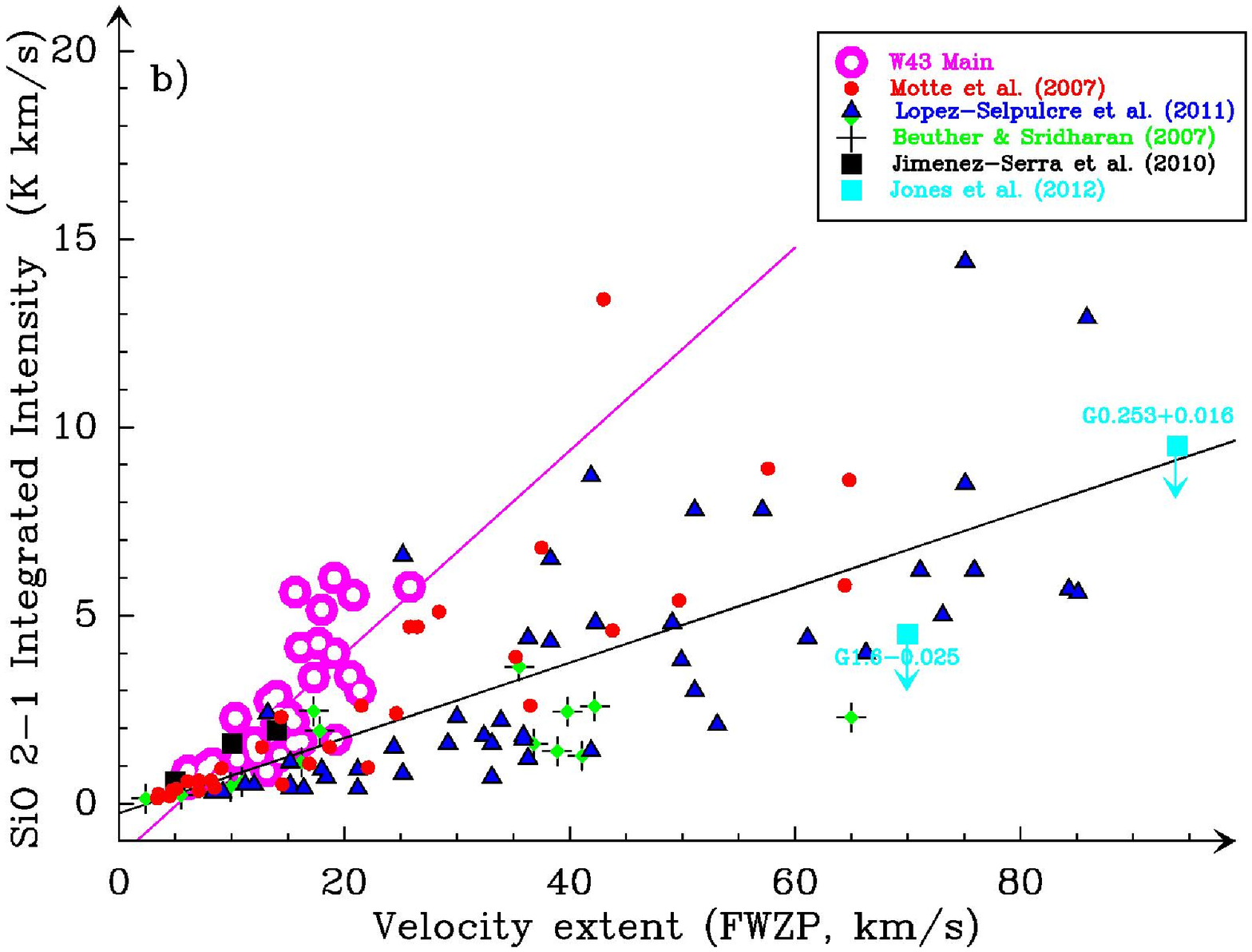} 
\end{array}$
\caption{Luminosity {\bf (a)} and intensity {\bf (b)} of the SiO lines detected in W43-Main (pink open circles) as a function of their \textit{FWZP} velocity extent. The 35 circles represent the SiO measurements of all the 31\arcsec-beam independent points measured aboved $3\sigma$ at the IRAM 30~m telescope. They are compared with those of the SiO lines measured toward the Cygnus~X massive dense cores \citep[red filled circles,][]{motte07}, massive star-forming cores \citep[blue filled triangles,][]{lopez-sepulcre11}, and high-mass protostellar objects \citep[green diamonds and plus signs,][]{beuther07c}. 
Most of the latter sources host high-mass protostars driving powerful SiO outflows. W43-Main and IRDC G035.39--00.33 \citep[black filled squares,][]{jimenez-serra10} display SiO emission that may be associated to colliding flows shocks. 
The pink solid line is the linear fit 
for W43-Main while the black solid line is the one for sources in all other samples  taken together (except the CMZ) (see Sect.\ref{sect:SiOstrength}). The CMZ sources (blue squares, \citealt{jones12}) have SiO emission whose nature is still not clearly known.
 }
\label{fig:SiOcomparison}
\end{figure*}

None of the published SiO measurements of the Central Molecular Zone (CMZ) were observed with the same transition and spatial resolution as in our W43 observations. To compare our W43 SiO observations with this well-known region, we here used the SiO (2--1) measurements of the G0.253+0.016 and G1.6--0.025 sources of the CMZ obtained with the MOPRA telescope \citep{jones12}. The larger beams (40\arcsec) of this observation over-estimate their SiO (2-1) integrated intensity and luminosity compared to ours but probably not more than 50\%. A distance of 8 kpc is assumed for Fig.~\ref{fig:SiOcomparison}b. The origin of the SiO emission from the CMZ clouds is still a matter of debate since star formation may be inhibited \citep{menten09,kruijssen13}.
For W43-Main and the outflow sources of the comparative 
samples, we estimated the SiO~(2--1) luminosity, $L_{\rm SiO~(2-1)}$, from the integrated intensity, $\int T_{\rm MB}\,dv$, via:

\begin{eqnarray}
\label{eq:lsio}
L_{\rm SiO (2-1)} & = & 4\pi \times d^2 \times \int T_{\rm MB}\, dv  \\
		   & \simeq & 450\,{\rm K\,\kms\,kpc^2} \times \left( \frac{d}{\rm 6\,kpc} \right)^2  		   \frac{\int T_{\rm MB}\,dv}{\rm 1\,K\,\kms} \nonumber \\
   		   & \simeq & 1.9\times 10^{-5}\,{\rm \lsun} \times \left( \frac{d}{\rm 6\,kpc} \right)^2  		   \frac{\int T_{\rm MB}\,dv}{\rm 1\,K\,\kms}.
		   \nonumber
\end{eqnarray}
The conversion of the SiO~2--1 luminosity unit from ${\rm K\,\kms\,kpc^2}$ to \lsun~ is discussed in Appendix~\ref{sect:conversion}.
Luminosity allows a more direct comparison of published intensities with 
those measured in W43-Main. For sources closer than W43, Eq.~\ref{eq:lsio} is a proxy of the dilution of their intensities measured within $\sim$0.1--0.9~pc beams to the largest physical scale, here 0.9~pc. This is correct for the unresolved SiO emisison of outflow sources \citep[e.g.][]{lopez-sepulcre11,beuther07c,motte07} but it is obviously wrong for IRDC G035.39--00.33 
whose emission is extended \citep{jimenez-serra10}. The comparison of intensities rather than luminosities is more appropriate in such cases but with an implicit assumption that the sources 
have a constant intensity emission over, at least, 0.9~pc.

Fig.~\ref{fig:SiOcomparison} shows the relations for both the SiO~(2--1) luminosity (panel a) and integrated intensity (panel b) as a function of their velocity extent. We called hereafter them as the $L-$\textit{FWZP} and $I_{\rm int}-$\textit{FWZP} relations.  We do not include the measurements of the CMZ in the fits since their $L-$\textit{FWZP} go off from the W43 sample or the comparative samples.
The most notable trait is that the SiO emission in W43--Main does not spread over more than 25~\kms, in contrast with that of the high-mass molecular outflows which spread up to $\sim$50~$\kms$ \citep[e.g.][]{lopez-sepulcre11}. 
The median velocity extent of W43 is much smaller than the median values in other samples: 
$\sim$14~$\kms$ compared to $\sim$32~$\kms$ (see Table~\ref{table:SiOlum}). 
Furthermore, the SiO~(2--1) luminosity and intensity of W43-Main pixels are much steeper functions of their velocity extent than the relations measured for other samples (see Figs.~\ref{fig:SiOcomparison}a-b). The parameters of the best linear fits are given in Table~\ref{table:SiOlum}.
In more details, with a linear relation between the SiO (2--1) luminosity and velocity extent such as 
\begin{eqnarray}
\label{eq:fitlum}
L_{\rm SiO (2-1)}  =  \alpha_\text{L} \times \textit{FWZP}_{\rm SiO (2-1)} - \beta_\text{L}, 
\end{eqnarray}
we measured $\alpha_\text{L}\sim140$~K\,kpc$^2$ for W43-Main and $\alpha_\text{L}\sim 13$~K\,kpc$^2$ for the comparative samples taken together.
With a similarly linear relation between the SiO (2--1) integrated intensity and velocity extent, $I_{\rm SiO (2-1)}  =  \alpha_\text{I} \times \textit{FWZP}_{\rm SiO (2-1)} - \beta_\text{I}$, the slope of the W43-Main region remains three times steeper than for the comparative samples (see Table~\ref{table:SiOlum}).

Despite a large range of distances, the SiO~(2--1) integrated intensities in in all the sources of the different samples, including W43-Main except the CMZ sources, are almost similar: $\langle I_{\rm SiO (2-1)}^{\rm int} \rangle \sim2-3$~K\,$\kms$.
In contrast, the median SiO~(2--1) luminosity in W43-Main, $\langle L_{\rm SiO (2-1)} \rangle \sim 10^3$~K\,\kms\,kpc$^2$, is three times
larger than the median value observed for comparative sources. 
The total SiO~(2--1) luminosity measured toward the W43-Main region is $L^{\rm total}_{\rm SiO (2-1)} \sim 4 \times 10^4$~K\,\kms\,kpc$^2$, with 50\% concentrated in the W43-MM1 and another 50\% in W43-MM2 ridge surroundings.

Within the three catalogs of massive young stellar objects, we found three sources that lie near the steep fit of the $L-$\textit{FWZP} relation : G34.26+0.15 in  the \cite{lopez-sepulcre11} sample and a couple of sources from \cite{beuther07c} (see Fig.~\ref{fig:SiOcomparison}a).
The G34.26+0.15 source is an ultra compact H~{\scriptsize II} region with a strong hot core SiO emission \citep{watt99} and the two  \cite{beuther07c} sources in fact correspond to W43-MM24 and another location in W43-Main. 
The SiO line emission observed toward the CMZ has also higher luminosities than the SiO lines from outflows with the same \textit{FWZP} (see Fig.~\ref{fig:SiOcomparison}a). 
If the lack of star formation activity in these regions is confirmed (see Sect.~\ref{sect:obsconstraints}) other mechanism such as violent cloud-cloud collisions could be responsible for producing SiO in the CMZ \citep{martin-pintado97,requena-torres06}.

\subsection{SiO column density}
Assuming that the SiO emission is optically thin in shocked regions ($\tau\ll1$) and that the Local Thermodynamic Equilibrium (LTE) hypothesis is valid, we can in principle calculate 
the SiO column density 
from the integrated intensity by
solving the basic radiative transfer equation \citep{rohlfs00}. The total SiO~(2--1) column density  
can be estimated with the following equation:
\begin{eqnarray}
N_\text{tot}^\text{SiO(2-1)} &= & \frac{3\,k}{8\pi^{3}S\,\mu_{s}^{2}\,\nu}  
\frac{Q_\text{rot}(T_\text{\rm ex})}{g_{u}\,g_{K}\,g_{s}}	\;\mbox{exp}\left(\frac{E_u}{k\,T_\text{ex}}\right)  \int \frac{T_\text{MB}}{f}dV \nonumber \\
& \simeq & 1.98\times 10^{10} \,\cmd
\times \frac{1}{f}\times Q_\text{rot}(T_{\rm ex}) \nonumber \\
&   & \times \mbox{exp} \left[ 0.0625 \left( \frac{T_\text{ex}} {100~\K} \right)^{-1}  \right] 
\times  \frac{\int T_\text{MB}\, dV}{1\,\K \kms}\, 
\label{eq:Ncolthin}
\end{eqnarray}
where $k$ is the Boltzmann constant and $\nu= 86.846$ GHz the emitting frequency.
The line strength ($S=2.0$), the dipole moment ($\mu_{s}=3.098$~Debye), and the upper level energy ($\frac{E_u}{k}=6.25~\K$) of the SiO~(2--1) transition were taken from the splatalogue catalog.
$Q_\text{rot}(T_{\rm ex})$ is the partition function which was extrapolated, for a particular excitation temperature, from the tabulated values of the Cologne Database for Molecular Spectroscopy (CDMS) \footnote{The CMDS catalog can be searched here: \url{http://www.astro.uni-koeln.de/cdms}} \citep{muller05}. The rotational degeneracy is given by $g_u=2J_{u}+1=5$ while the K and nuclear spin degeneracies are $g_{K}=1$ and $g_{s}=1$. 
A beam filling factor of $f=1$ is justified in our study since the SiO emission is extended in W43-Main and fills the entire beam
as confirmed by interferometric observations (Louvet et al. in prep.). We recall that the LTE calculation relies on the assumption that the gas density is larger than the critical density
of the SiO~(2--1) transition given by
\begin{eqnarray}
n_{\rm crit}  & =\frac{A_{\rm ul}}{\sigma \sqrt{
\frac{3\,k\,T_{\rm ex}}{m_{\rm H}}}}
\simeq 
2.93\times10^{5}\,\cmc
\left( \frac{T_{\rm ex}} {100\,\K}\right)^{-1/2}
\end{eqnarray}
where the Einstein coefficient ($A_{ul}=2.93\times 10^{-5}$\,s$^{-1}$), the cross-section ($\sigma=10^{-15}~\cmd$) of the SiO molecule are taken from the splatalogue catalog. $\sqrt{\frac{3\,k\,T_{\rm ex}}{m_{\rm H}}}$ is the approximate
ensemble average velocity of the collision partner, hydrogen in this case. 
Equation~\ref{eq:Ncolthin} yields a critical density of 
$n_{\rm crit}=0.3-65\times 10^4~\cmc$ for an excitation temperature ranging from 10$^{6}$\,K to 20\,K. The median density values of the W43-MM1 and W43-MM2 ridges, $n_{\rm H_{2}}\sim 6.0\times 10^4~\cmc$ and $3.4\times 10^4~\cmc$ (see Table~\ref{tab:ridgeproperties}), prove that at least there are some parts of these cloud ridges where the density surpasses the critical density. Therefore Eq.~\ref{eq:Ncolthin} can be used as a good upper limit estimation of the SiO column density.
Besides, if the SiO emission is optically thick, the SiO column density must be increased by a factor $\sim$ $\frac{\tau}{1-{\rm exp}(-\tau)}$.

\begin{figure*}[thbp]
\centering
$\begin{array}{cc}
\hspace{-.3cm}
\includegraphics[angle=0,width=8.5cm,height=7.cm]{{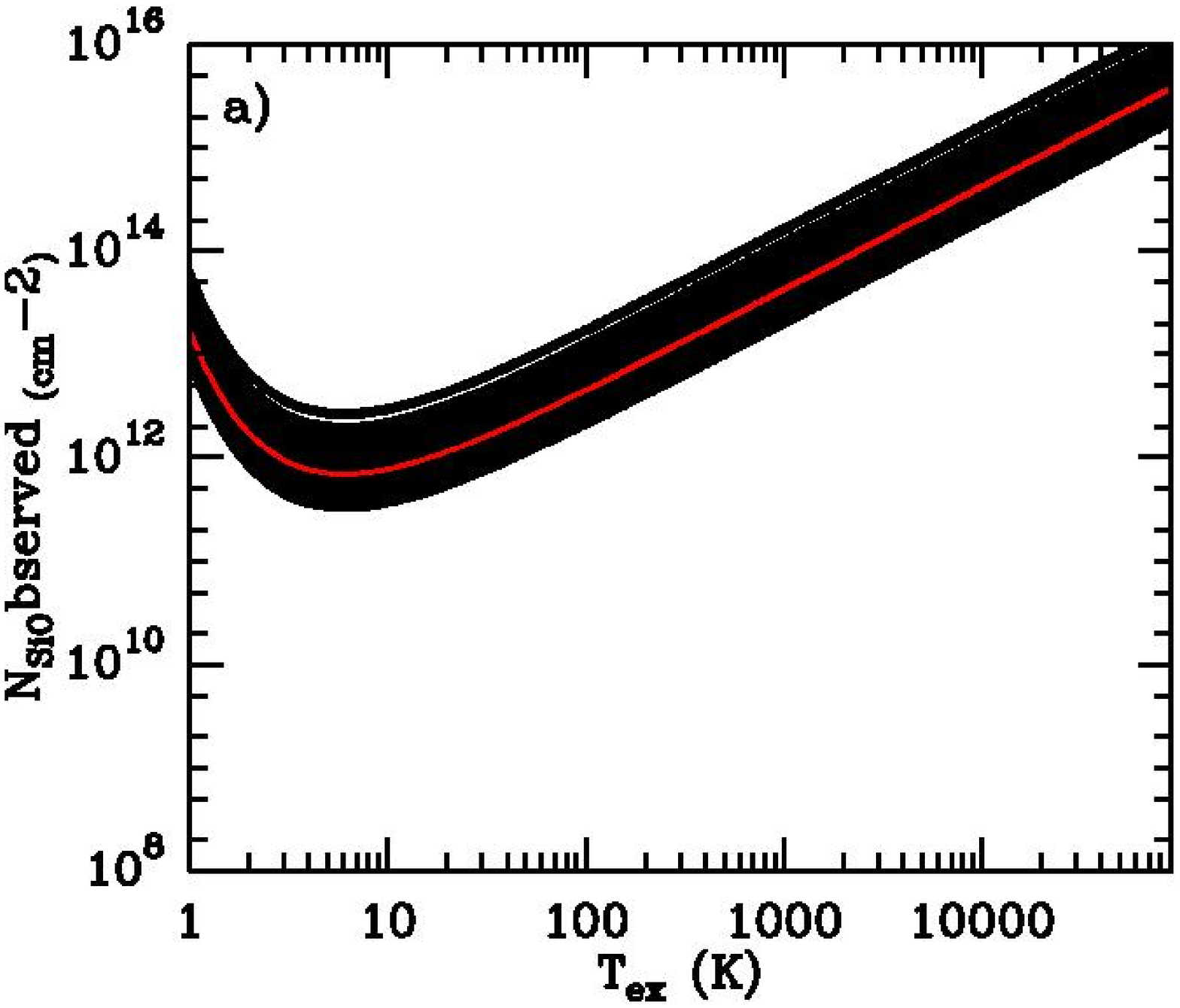}} &
\includegraphics[angle=0,width=10.cm,height=7.5cm]{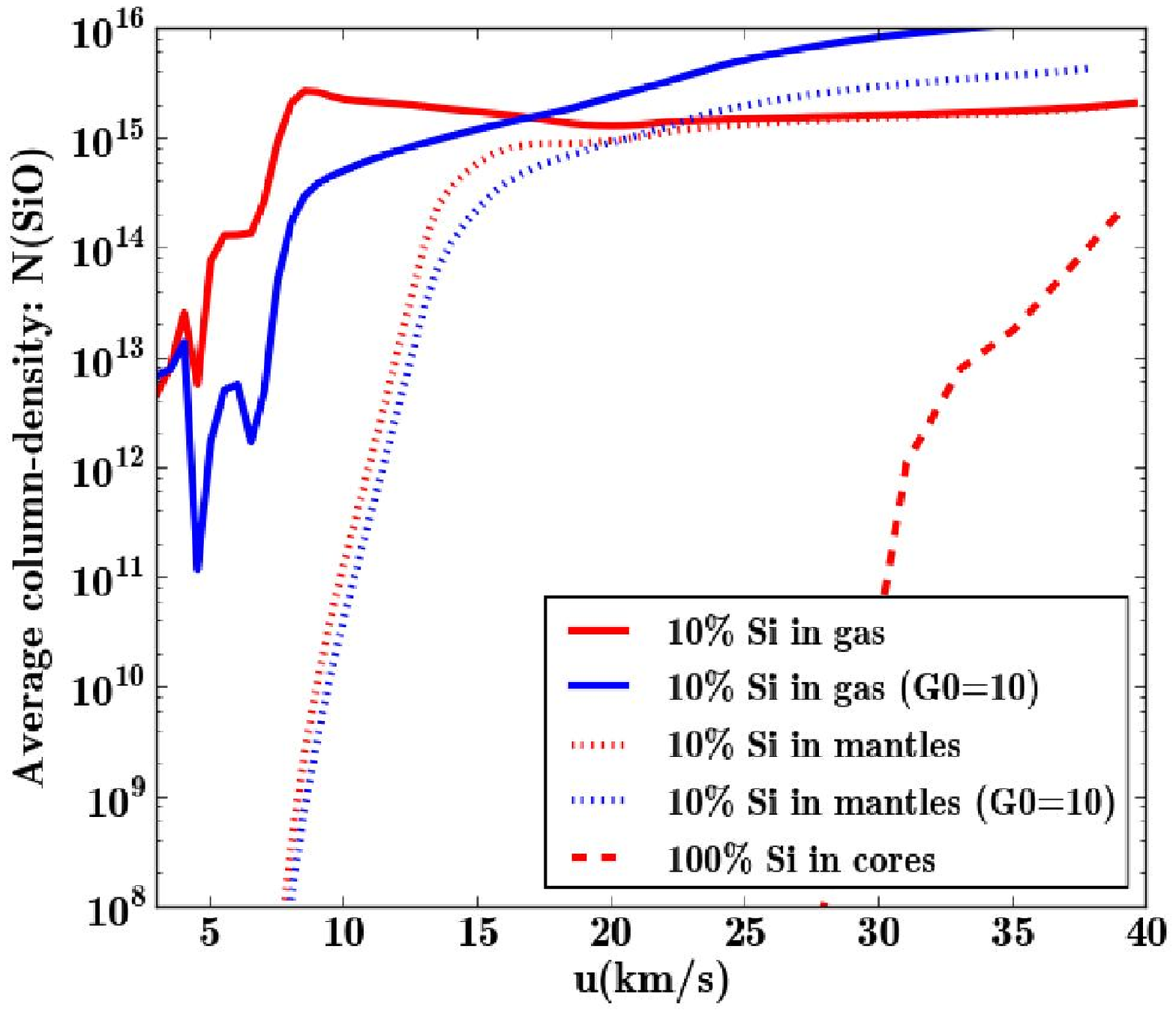} \\
\end{array}$
\caption{
{\bf a):} 
The SiO column density observed for 35 $31\arcsec$-pixels in W43-MM1 and W43-MM2 as a function of the assumed excitation temperature (see Eq.~\ref{eq:Ncolthin}). The $T_\text{ex}$ abscissa range is set to roughly correspond to shock velocities selected in {\bf b}. The red curve represents the mean column density of the 35 pixels for the assumed temperature range.
{\bf b)}: 
 The  SiO column density expected
in shocks for various initial conditions in the pre-shock gas. The pre-shock
elemental abundance of Si can be either 90\% in grain cores and 10\%
in gas (solid curves), 90\% in grain cores and 10\% in mantles as in
\cite{gusdorf08a} (dashed curves), or 100\% in grain cores as in \cite{schilke97}
(dotted curve). Parameter used: $G_0=1$ (red curve) or  $G_0=10$ (blue curve), and $n_{\rm H}=10^5~$cm$^{-3}$, magnetization parameter $b=3$ (see \citealt{lesaffre13} for a more detailed discussion on the model and the input parameters).
}
\label{fig:siocoldenmodel}
\end{figure*}

Assuming conservative shock temperatures $T_\text{ex}=20-10^6$~K in Eq.~\ref{eq:Ncolthin}, we derived the SiO column density of each of the W43-Main points of Figs.~\ref{fig:siocoldenmodel}a. 
We varied the excitation temperatures from a minimum of $T_\text{ex}=T_\text{dust}=20$~K to a maximum $T_\text{ex}=10^6$~K, which is a typical temperature range for shocks \citep{draine80,chieze98}. 
As shown in Figs.~1a and 5a of \cite{lesaffre13}, each maximum gas temperature can be related to a single shock velocity for a specific set of inputs. Therefore, the observed column density plotted in Fig.~\ref{fig:siocoldenmodel}a is almost directly comparable to the theoretical column density plotted in Fig.~\ref{fig:siocoldenmodel}b (see Sect~\ref{sect:lowvelmodel} for a fuller discussion of the models).
For instance, a C-type shock with a low-velocity of $\sim$7.5~$\kms$ can easily reach a temperature of $\sim$3000~K (see \citealt{lesaffre13} and Sect.~\ref{sect:lowvelmodel}). 
Depending on the model, the SiO molecules could be formed either in the shock front or just after it in the post-shock gas, thus arguing for excitation temperatures from a few thousands to a few hundreds Kelvin. This is coherent with the fact that
at $\sim$1900~K, Si starts to evaporate from ice mantles \citep[e.g.][]{gusdorf08b}.
The SiO column density of the 35 independent 0.9~pc-beam points, detected above our $3\sigma$ level in the W43-MM1 and W43-MM2 ridges plus surroundings, are shown in Fig.~\ref{fig:siocoldenmodel}a as a function of $T_\text{ex}$. 
Our SiO column density thus ranges from $5\times10^{11}~\cmd$ to $10^{16}~\cmd$, with a mean value of $\sim$6$~\times 10^{13}~\cmd$.
Such values are typical for shocked regions associated with the most powerful outflows driven by high-mass star-forming regions \citep[e.g. G5.89-0.39 and G34.26 by][]{acord97,hatchell01} or observed toward the Central Molecular Zone \citep{yusef-zadeh13}.

\section{Discussion}
\label{sect:discussion}

We discuss the origin of the SiO emission in Sects.~\ref{sect:obsconstraints}-\ref{sect:lowvelmodel} and their implications on the formation of ridge and subsequent the star formation processes in Sect.~\ref{sect:discussion_lowvelocity}.

\subsection{Observational constraints on the origin of the SiO emission in the W43-MM1 and W43-MM2 ridges}
\label{sect:obsconstraints}

We here review the various physical processes that could explain the bright SiO~(2--1) emission spreading along the W43-MM1 and W43-MM2 ridges. As will be explained in Sect.\ref{sect:lowvelmodel}, all SiO emission stem from shocks. We investigate the hypotheses of high-velocity shocks ($v_{\rm shock}\sim20-50~\kms$) in outflows, moderate- to high-velocity shocks ($v_{\rm shock}\sim10-20~\kms$) in photon dominated regions (PDRs) and hot cores, and low-velocity shocks ($v_{\rm shock}\le 10~\kms$) associated with clouds collision.

SiO emission coming from outflows driven by young stellar objects is usually caused by high-velocity shocks \citep[$v_{\rm shock}\sim20-50~\kms$,][]{schilke97}. This mechanism, that should translate into high-velocity spectral wings, could play a role in W43-Main since it is an active star-forming region. However, the SiO lines observed in W43-Main do not spread over more than 25~$\kms$, in contrast with lines measured toward high-mass protostellar outflows which have up to $\sim$50~$\kms$ wings \citep[e.g.][see Figs.~\ref{fig:SiOcomparison}a-b]{lopez-sepulcre11,motte07}.
As it will be shown in Louvet et al. (in prep.), from a higher-angular resolution SiO imaging, beam dilution and sensitivity limit may partly explain why the protostellar outflows of the W43-MM1, W43-MM2, and W43-MM3 massive dense cores  are hardly seen.
We cannot, however, rule out the scenario of a widespread population of low-mass protostars driving outflows along the W43-MM1 and W43-MM2 ridges. 
Higher-angular resolution and higher sensitivity SiO line and dust continuum mappings are necessary to confirm/denounce it (see Louvet et al. in prep.) 

SiO emission tracing irradiated shocks in PDRs have been observed to be associated with medium-velocity shocked gas \citep[$v_{\rm shock}\sim10-20~\kms$, e.g.][]{schilke01}. Irradiated shocks are to be expected in the W43-Main mini-starburst region since the cloud is associated with a very luminous cluster \citep[$\sim$3.5~$\times$ 10$^6~\lsun$,][]{smith78,bik05}. Fast winds and ionization from the OB/WR stars are observed to be interacting with intermediate-density clouds, which are swept up and accumulated at the periphery of the giant H~{\scriptsize II} region \citep{motte03}. We however do not detect SiO emission over the PDR ring (see Fig.~\ref{fig:w43ridgeSiO}), delineated in near-infrared and centimeter continuum emission \cite[][Carlhoff et al. subm.]{motte03}. Moreover, the process of medium-velocity shocks in irradiated PDRs is rather unlikely to develop in the higher-density and thus better-shielded cloud structures that are the W43-MM1 and W43-MM2 ridges. 
We prove this statement by making shock models for the solar or stronger radiation fields, i.e.\ref{•} $G_0=1$ and $G_0=10$. The SiO production is slightly more efficient in the case of $v_{\rm shocks}> 15\kms$ with $G_0=10$ but the weaker radiation field seems to produce more SiO for low-velocity shocks (see Fig.~\ref{fig:siocoldenmodel}b).
 
We cannot neglect the impact of the OB cluster on the ridges via its radiation pressure and UV photons. However, as shown by \cite{krumholz09b}, even for the most powerful clusters in the Galaxy, the radiation pressure effect is stalled at a radius much smaller than 5~pc. Moreover, the ridges are very dense regions with $n_{\rm H_2} > 10^4~\cmc$, the penetrating depths of UV photons from a cluster with $G_0\sim 10^5$, similar to the W43 OB cluster, will decrease rapidly and reach a value of  $G_0 < 10$ within the ridge \citep{meixner93}. 
Therefore, W43 OB cluster cannot be the main responsibility for creating the extended shocks observed in the W43 ridges although itcould still contribute.

  Hot cores and hot corinos are emitting SiO lines that trace high-velocity \citep[$v_{\rm shock}\sim50~\kms$,][]{hatchell01,jorgensen11} shocks  
associated with the strong thermal heating of young stars.
The SiO lines with hot core origin could account for part of the emission observed toward the high-mass protostars of W43-Main. However, like for the outflow hypothesis, our analysis of SiO lines did not reveal many high-velocity components (see Figs.~\ref{fig:siospectraw43} and \ref{fig:SiOcomparison}a-b) and the protostellar content of the W43-MM1 and W43-MM2 ridges must still be assessed. This scenario is thus not very likely.

Therefore, none of the physical processes usually advocated to explain SiO emission in star-forming regions, i.e. high-velocity or irradiated shocks, explain reasonably all SiO~(2--1) emission observed in W43-Main. 
Given that the observed SiO lines have Gaussian-like shapes without any high-velocity components (see, e.g., Fig.~\ref{fig:w43ridgeSiO}),  a series of low-velocity, i.e. $v_{\rm shock}\le 10~\kms$, shocks are more appropriate (see Sect.~\ref{sect:lowvelmodel}).
The very similar line shapes in terms of intensity and width over the two $\sim$10~pc$^2$ areas (see Table~\ref{tab:n2h+fitting}) moreover support the hypothesis of a large-scale mechanism homogeneously producing shocks throughout both the W43-MM1 and W43-MM2 ridges.
The similar and overlapping extended morphologies between the shocked gas and the dense gas of ridge structures (see Figs.~\ref{fig:w43ridgen2h+} and \ref{fig:w43ridgeSiO}), suggest that the formation of ridges is associated with such shocks.
The mechanism at the origin of SiO emission in W43-Main could thus be colliding flows or merging of filaments that are interpreted as the processes necessary to build up cloud ridges \citep[e.g.][]{schneider10b,hennemann12}.

\cite{jimenez-serra10} also proposed that the SiO emission observed along IRDC G035.39--00.33 could partly arise from 
the collision of gas streams that form the higher-density IRDC structure.
Their measurements however suffer from a lack of sensitivity and spatial resolution: only three independent beams are above their $3\sigma$ detection level. A detailed comparison shows that the SiO~(2--1) observed along each of the W43-MM1 and W43-MM2 ridges are $>$30 times more intense and $\sim$10 times more extended than that of IRDC G035.39--00.33.
In Fig.~\ref{fig:w43ridgeSiO}, we detected SiO~(2--1) integrated intensities of 1--11~K$\,\kms$ over $\sim$12~pc$^2$ and $\sim$16~pc$^2$ areas along the W43-MM1 and W43-MM2 ridges and their immediate surroundings, respectively. These values are to be compared with the 0.02--0.05~K$\,\kms$ emission observed along the $\sim$1~pc$^2$ IRDC G035.39--00.33 filament \citep[][see also Fig.~\ref{fig:SiOcomparison}b]{jimenez-serra10,henshaw13}. 
The W43-MM1 and W43-MM2 ridges could therefore be two more regions, as IRDC G035.39--00.33, where extended SiO emission is detected and interpreted as arising from low-velocity shocks associated with the formation of high-density cloud structures.
From the analysis of Sect.~\ref{sect:SiOstrength}, we propose to use a linear relation in the luminosity versus velocity extent diagram of Fig.~\ref{fig:SiOcomparison}a 
to distinguish the low-velocity ($\le$10~\kms) shocks with steep slopes ($\alpha_\text{L}\simeq140$~K\,kpc$^2$ defined in Eq.~\ref{eq:fitlum}) possibly associated with cloud collision from more classical high-velocity ($\sim$20--100~\kms) shocks with flatter slopes  ($\alpha_\text{L}=13\pm7$~K\,kpc$^2$). The SiO emission in W43 is as luminous as the SiO emission associated with extremely high-velocity outflows despite having narrower velocity range. This may suggest that the low-velocity shocks in W43 are highly efficient in forming/releasing SiO over an extended region.

Very widespread SiO emission, with velocity dispersion
exceeding those observed toward W43-Main, has been reported toward the entire Central Molecular Zone of our Galaxy \citep[CMZ covering $\sim$ $10^4$~pc$^2$,][]{martin-pintado97,jones12}. The CMZ is in general characterized by much higher temperatures ($\tdust\ge50~$K)
and more turbulence and shears, resulting in larger velocity dispersion \citep[$\sigma$~$\ge10-30\kms$; see the recent review by][]{gusten04}. 
Two CMZ clouds with strong and extended SiO emission, G1.6--0.025 and G0.253+0.016, have been observed to be devoid of signs of massive star formation \citep{menten09,kauffmann13,kruijssen13} but see however \cite{rodriguez13}.
Recent studies by \cite{yusef-zadeh13} still interpreted the very broad SiO lines observed close to Sgr~A$^*$ as arising from highly embedded protostellar outflows.
We propose that
the process explaining the SiO emission measured in the CMZ is partly, like in the case of W43-Main, velocity shears arising from cloud collision. Given the much higher shock velocities in CMZ ($v_{\rm shock}$ is up to $80~\kms$) and interstellar radiation field, its SiO emission could be explained either in the frame of highly-irradiated and highly magnetized shock models, or in that of fast J-type shock models (see Section~\ref{sect:lowvelmodel} for a short review on shock models).

This is in marked contrast with the strong SiO emission of W43-Main detected at the ridges' standard velocities at rest. 
In W43, the $2-5~\kms$ velocity dispersion observed for SiO~(2--1) lines and the $<$5~$\kms$ variation found for the ridge material suggest the collision of gas flows separated by $\sim$5--10~$\kms$ only. Note that SiO molecules could form in the post-shock gas and thus display a factor of two lower velocities than the initial shock velocity
(see Sect.~\ref{sect:lowvelmodel}).

We therefore conclude that most of the SiO emission in W43-Main mainly comes from low-velocity ($\le$10~\kms) shocks.
This is a convincing evidence that SiO molecules being formed in shock front or post-shock gas under the influence of low-velocity shocks.

\subsection{Theoretical ability to create SiO molecules through low-velocity shocks}
\label{sect:lowvelmodel}

SiO emission has been first demonstrated to possibly arise from magnetized, stationnary C-type shock models, with $v_{\rm shock}\ge25~\kms$ \citep{schilke97,gusdorf08a,guillet07,guillet11}.
 In bipolar outflows associated to low-mass star-forming regions, for instance, \cite{gusdorf08b} showed that young, non-irradiated, non-stationary intermediate-to-high velocity C-shock models can also account for the SiO emission. On the other hand, low-velocity C-type shocks with magnetized, ``irradiated'' models is also proposed to explain the origin of SiO emission in irradiated environment \citep{lesaffre13}. \cite{guillet09} have shown that stationnary, J-type shock models also constitute reasonable candidates to generate significant amounts of SiO emission in molecular outflows and jets.

We examine here a few dedicated shock models, similar to models in \cite{lesaffre13}, which 
demonstrate that SiO could be produced in low-velocity, 
$v_{\rm shock}\le10~\kms$,
shocks in W43 ridges. 
Figure~\ref{fig:siocoldenmodel}b is comparing six models with different radiation fields and different fraction of Si in the gas phase, grain mantles, or grain cores \cite[e.g.][]{lesaffre13}. Their $N_{\rm SiO}$ predictions as a function of the shock velocity can directly be compared with our observations (see Figs.~\ref{fig:siocoldenmodel}a-b).
Silicon in the interstellar medium resides mainly in the cores of
silicate grains \citep{flower96}. Impinging particles or photons may however
release some of the Si in the gas phase where it can potentially be
oxidized by either O${_2}$ or OH to form the SiO molecule (see \citealt{lepicard01} and references therein). In even mildly irradiated
media, like e.g. with a far-UV field of $1\,G_0$ in units of the average interstellar radiation field  
 and with a column density of $\sim$2~$\times10^{20}$~\cmd, all these SiO molecules photo-dissociate and Si remains mostly in atomic form in the
gas \citep{lepicard01}. 
As shown in \cite{lesaffre13}, magnetized shocks, even
at low-velocity, can bring enough
energy to overcome the reaction barriers for OH and O${_2}$ formation
 and balance the photo-dissociation rates. These
molecules are then available to oxidize Si and form SiO.

The amount of SiO a shock can produce depends on the phase where the element Si is present in
the pre-shock medium: grain cores, grain mantles, or gas phase. 
If some Si is already in gas phase, SiO can be formed
at very low velocity, i.e. already around $5~\kms$ (see e.g. solid lines in Fig.~\ref{fig:siocoldenmodel}b). If Si is present in the
form of ice mantles on the grains, one needs to 
reach at least $\sim$10~$\kms$
to sputter Si in the gas phase \citep[see e.g. dashed lines in Fig.~\ref{fig:siocoldenmodel}b,][]{gusdorf08b}. If, on the other hand all the
Si is locked in the core of grains, one needs to spend more energy to
detach Si and a velocity of at least 25~$\kms$ is needed \citep[see e.g. dotted lines in Fig.~\ref{fig:siocoldenmodel}b,][]{schilke97,gusdorf08a}.
Figure~\ref{fig:siocoldenmodel}b illustrates in more details these three scenarii for a range of C-type shock velocities using the models by \cite{lesaffre13}. 
We used far-UV fields of $1~G_O$ (red curves) to $10~G_O$ (blue curves) and a density for the pre-shock gas of  $n_{\rm H}=10^5~$cm$^{-3}$ (see below), close to that of the W43-MM1 and W43-MM2 ridges. The magnetic field perpendicular to the shock front is another important parameter because it will determine the velocity limit at which C shocks cannot exist \citep{lesaffre13}. It is expressed through the dimensionless magnetization parameter $b=\frac{B}{1{\,\rm\mu G}} \sqrt{\frac{n_{\rm H}}{1\, \text{cm}^3}}$. 
We assumed a constant magnetic field strength over the shocked regions, equal to the one measured toward 
the W43-MM1 massive dense core: $B\sim855~\mu$G \citep{cortes10}. Given the mean density of the W43-MM1 and W43-MM2 ridges listed in Table~\ref{tab:ridgeproperties}, we estimated $n_{\rm H}=2.33\times n_{\rm H_{2}}\simeq 3-6\times 10^{4}\,\cmc$ and thus $b\simeq2.7-4.8$.
We therefore set $b=3$ for our models shown in Fig.~\ref{fig:siocoldenmodel}b.

As seen in Figs.~\ref{fig:siocoldenmodel}, the models with Si initially deposited in the gas or in mantles and shock velocities of $\sim$5--10~$\kms$ can explain the SiO column density observed in the W43-MM1 and W43-MM2 ridges. 
However, the distinction between these two types of Si reservoirs and the use of more detailed input parameters would require a multi-transition study of SiO emission lines incorporating other shock tracers.

In order to check the effect of the mean interstellar radiation field on the SiO production, we increased the far UV field strength from $G_0=1$ to $G_0=10$. We showed that the SiO abundance is not changed much (see Fig.~\ref{fig:siocoldenmodel}b). It proves that the observed SiO emission does not necessarily favor shocks in a medium highly irradiated by the UV photons of the cluster.

\subsection{Implications for the formation of ridge and young  massive cluster}
\label{sect:discussion_lowvelocity}

Our current knowledge of high-mass star-formation points toward a dynamical picture.  Accretion flows from a surrounding filamentary network \citep{schneider10,hennemann12} were observed and suggested to be necessary to create ridges that form the majority of high-mass stars.
In principle, this process could be called ``colliding flows" since they represent the collision of several gas streams. 
The collision observed here happens at high density ($n_{\rm H_2}> 10^3~$\cmc) and could be a prolongation of 
the converging flows process building up molecular gas from pockets of H{\scriptsize I} gas.
At these densities, the cloud structures considered are gravitationally bound and even collapsing (see Table~\ref{tab:ridgeproperties} and \citealt{motte05} for W43). In colliding flows models, gravity is the main force that drive the colliding flows, sometimes called ``gravitational focussing" \citep{burkert04}. Since the initial gravitational potential should anyway be explained, a converging flows plus colliding streams scenario is attractive as it self-consistently forms first the cloud and then ridges. It also agrees with observations suggesting that the W43 molecular complex has formed through large-scale colliding flows \citep{nguyenluong11} while harboring ridges with a coherent but complex velocity structure (see Sect.~\ref{sect:ridgeresult}).
The special location of W43 at the meeting point of the Galactic Bar and spiral arm certainly helps to generate such dynamics (\citealt{nguyenluong11}, Carlhoff et al. subm.).

Theoretically, colliding flows are an efficient way to gather mass in dense cloud structures \citep[e.g.][]{bergin04,heitsch08b,koyama00}. 
With the simple case of two streams frontally colliding, collected sheets would form and rapidly fragment into filaments \citep{heitsch11}. In the practical case of streams which are turbulent and/or colliding with non-zero impact angles, shears are expected to develop.
Series of shocks in compressed layers should thus be natural consequences of realistic colliding flows \citep{bergin01,koyama00,vanloo07}.  
In W43-Main, the high-density ridges coincide with extended SiO emission, thus delineating the shocked and compressed cloud structures (see Sect.~\ref{sect:ridgeresult}). 
The low-velocity ($\le$10\,\kms) shocks observed along the ridges (see Sect.~\ref{sect:sioresult}) tend to confirm that cloud ridges are formed through the gradual merging of several gas flows, as previously proposed for the DR21 ridge by \cite{schneider10,hennemann12}. 
The W43-MM1 and W43-MM2 immediate surroundings have $\sim$1.5 more mass than the ridges themselves (i.e. $M_\text{ridges}\sim 1.5\times 10^4~\msun$) and could feed them for the next $10^6$~yrs with an infall rate close to the shock velocity found here, a few \kms.

Within the shocked filaments created by colliding flows, higher-density cloud structures/cores are expected to form, possibly leading to the formation of high-mass stars \citep{whitworth94b,whitworth94,vanloo07}. 
Ridges are large-scale, a few pc$^2$, elongated structures with a mean density typical of protostars ($n_{\rm H_2}\sim 10^4-10^5$~\cmc) and forming clusters of high-mass stars \citep{hill11,nguyenluong11b,hennemann12}. The W43-MM1 and W43-MM2 ridges are exceptionally massive and can be qualified as candidate precursors of starburst clusters \citep{motte03} or or precursors of young massive clusters \citep{ginsburg12}.
The massive dense cores identified by \cite{motte03} were assumed to be hosting high-mass protostars and used to estimate a star formation rate reminiscent of extragalactic starbursts: $\Sigma_\text{SFR}~\sim 1000~\msun$yr$^{-1}$kpc$^{-2}$). 
While the three most massive dense cores (W43-MM1, W43-MM2, and W43-MM3 with $M=1000-4000~\msun$ within $\sim$0.2~pc sizes) are clearly forming one/a few high-mass stars, the exact nature of the others is still unclear.
One needs higher-angular observations to search for high-mass protostars in the W43-MM1 and W43-MM2 ridges and confirm that the star formation rates of these ridges are high (Louvet et al. in prep.).
Such a result is consistent with the high star formation activity measured toward other ridges studied with \emph{Herschel} (e.g. \citealt {nguyenluong11}, Hennemann et al. in prep.). Note that, in contrast, the CMZ presents high-density clouds that may be devoid of high-mass star formation (e.g. \citealt{menten09}). If confirmed, such a low star formation activity could be related to the too-large strength of the velocity shears in the CMZ. The W43-MM1 and W43-MM2 ridges are thus high-density cloud structures forming from constructive low-velocity colliding flows while other regions, like CMZ, might suffer from destructive colliding flows or strong radiation fields.

\section{Conclusion \& Summary}
\label{sect:conclusion}
We report the discovery of two ridge structures and extended SiO emission in the W43-Main mini-starburst region. We used a combination of \emph{Herschel} continuum data in the 160--350~$\micron$ wavebands and IRAM 30~m telescope data of the SiO~(2--1) and N$_2$H$^+$~(1--0) emission lines (see Tables~\ref{table:obs}-\ref{table:SiOobs}). Our main findings can be summarized as follows:
\begin{itemize}
\item 
On the \emph{Herschel} column density image,
W43-Main harbors two high-column density (\Nhtwo~$>10^{23}~\cmd$)  structures, which we call the W43-MM1 and W43-MM2 ridges (see Fig.~\ref{fig:w43coldens}a). 
\item Above a background level of $4\times 10^{22}~\cmd$, the W43-MM1 and W43-MM2 ridges have masses of  $21000~\msun$ and $35000~\msun$ within 6 and 14~pc$^2$ areas, respectively (see Table~\ref{tab:ridgeproperties}). 
The corresponding volume and surface densities are $n_{\rm H_2}\sim 6 \times 10^4$~cm$^{-3}$ and $\Sigma_S = 3500~\msun\,$pc$^{-2}$ for W43-MM1 ridge or $n_{\rm H_2}\sim 3.4 \times 10^4$~cm$^{-3}$ and $\Sigma_S = 2300~\msun\,$pc$^{-2}$ for W43-MM2 ridge.

\item The immediate surroundings of the W43-MM1 and W43-MM2 ridges, defined as cloud structures above $4\times10^{22}~\cmd$, have a mean density of $n_{\rm H_2}\sim 5 \times 10^3$~cm$^{-3}$ (see Fig.~\ref{fig:w43coldens}a and Table~\ref{tab:ridgeproperties}). They provide a mass reservoir of $32000~\msun$ for W43-MM1 ridge and $53000~\msun$ for W43-MM2 ridge that could be used to further build up the ridges and enhance its subsequent star formation.
\item The N$_2$H$^+$~(1--0) spectra measured toward the W43-MM1 and W43-MM2 ridges (see Fig.~\ref{fig:w43ridgen2h+} and Table~\ref{tab:n2h+fitting}) suggest that these dense cloud structures are complex, with velocities spanning the $\vlsr\sim94-97~\kms$ and $88-95~\kms$ ranges, respectively. The W43-MM1 and W43-MM2 ridges however are coherent velocity structures and 
despite their large N$_2$H$^+$ velocity dispersion, $\sigma\sim2~\kms$, 
they are gravitationally bound.
\item In W43-Main, the SiO (2--1) emission spreads over two large areas: $\sim$12~pc$^2$ and $\sim$16~pc$^2$  (see Fig.~\ref{fig:w43ridgeSiO}). They cover the W43-MM1 and W43-MM2 ridges and part of their immediate surroundings. This SiO emission is, with a few clouds in the CMZ, among the most extended observed up to now in our Galaxy.
\item The SiO~(2--1) spectra observed along the ridges and integrated over the entire W43-Main (see Figs.~\ref{fig:siospectraw43}-\ref{fig:w43ridgeSiO})
can correctly be described by single Gaussian velocity dispersions ($\sigma~\sim2.5-5.2~\kms$) without any obvious wings (\textit{FWZP}~$<25~\kms$, see Table~\ref{table:SiOlum}). 
SiO (2--1) lines are also peaking at the same velocities as N$_2$H$^+$~(1--0) and $^{13}$CO~(1--0) lines.
The absence of high-velocity SiO wing emission in W43 is consistent with not having SiO arising from outflow shocks, as observed in other star-forming regions.
\item Despite the $\sim$0.9~pc resolution of our SiO~(2--1) map, the lines are bright (1--11~K$\,\kms$, see Fig.~\ref{fig:w43ridgeSiO}), roughly corresponding to a SiO column density of $N_{\rm SiO}\sim 6 \times 10^{13}~\cmd$.
The luminosities of $31\arcsec$ beam-pixels in the W43-MM1 and W43-MM2 regions are $L_{\rm SiO~(2-1)} \sim 
200-2700~$K$\,\kms\,$kpc$^2$, summing up to a total luminosity of $L^{\rm total}_{\rm SiO (2-1)} \sim$4$\times 10^4$ or $\sim$2$\times 10^4$ K$\,\kms\,$kpc$^2$ for each of the ridges. 

\item The $L-$\textit{FWZP} relation is steeper for the SiO lines arising from W43-Main than for most SiO lines detected toward comparative samples of massive young stellar objects (see Fig.~\ref{fig:w43ridgeSiO} and Table~\ref{table:SiOlum}). We propose to use such diagrams and slopes of the linear fits to distinguish between low-velocity ($\le10~\kms$) and high-velocity ($20-100~\kms$) shocks, for both observational studies and synthetic observations of shock models.
\item Small offsets in velocity and positions ($\sim$0.8~$\kms$ and $\le$0.5~pc, see Table~\ref{tab:n2h+fitting} and Fig.~\ref{fig:w43ridgeSiO}) are measured between the SiO and N$_2$H$^+$ lines observed along the W43-MM1 ridge. They recall the relative motions expected between post- and pre-shocked gases.
\item Dedicated shock models prove that low-velocity  ($5-10~\kms$) shocks could give rise to bright and extended SiO emission, provided that 10\% Si is initially deposited in gas phase or in the grain ice mantles (see Fig.~\ref{fig:siocoldenmodel}). 
The exact process responsible for SiO emission in the W43-MM1 and W43-MM2 ridges 
is surely \textit{not} grain core destruction 
as advocated for high-velocity ($>$20~$\kms$) shocks developing along protostellar outflows. 
Among the main open questions are how Si could exist in the gas phase or in grain mantles and with which abundance.
\item The structure and kinematics of the W43-MM1 and W43-MM2 ridges are tightly linked to those of the shocked gas observed in SiO. Low-velocity ($\le$10\,\kms) shocks may thus indicate that shears have developed when collisionning several gas flows arising from the ridge immediate surroundings. This interpretation is consistent with the idea that cloud ridges (and their embedded massive dense cores) are forming through colliding flows or merging of filaments.
\end{itemize}

\begin{acknowledgements}
Part of this work was supported by the ANR (\emph{Agence Nationale pour la Recherche}) projects ``PROBeS" (number: ANR-08-BLAN-0241) and  ``STARFICH'' (number: ANR-11-BS56-010). A. Gusdorf acknowledges support by the grant ANR-09-BLAN-0231-01 from the French Agence Nationale de la Recherche as pasrt of the SCHISM project.
LB acknowledges support from CONICYT project PFB-06. We would like to thank Sergio Molinari for giving us the permission to use the {\it Herschel} images and the IRAM staff for their wonderful hospitality and observing services. This work has benefited from research funding from the European Community's Seventh Framework Programme.
\end{acknowledgements}

\bibliographystyle{apj}  \bibliography{quangreference.bib}

\newpage
\appendix
\section{Appendix A: Online images}
\begin{figure*}[tbhp]
\centering
$\begin{array}{c}
\hspace{-0.1cm}
 \includegraphics[scale =0.5,angle=0]{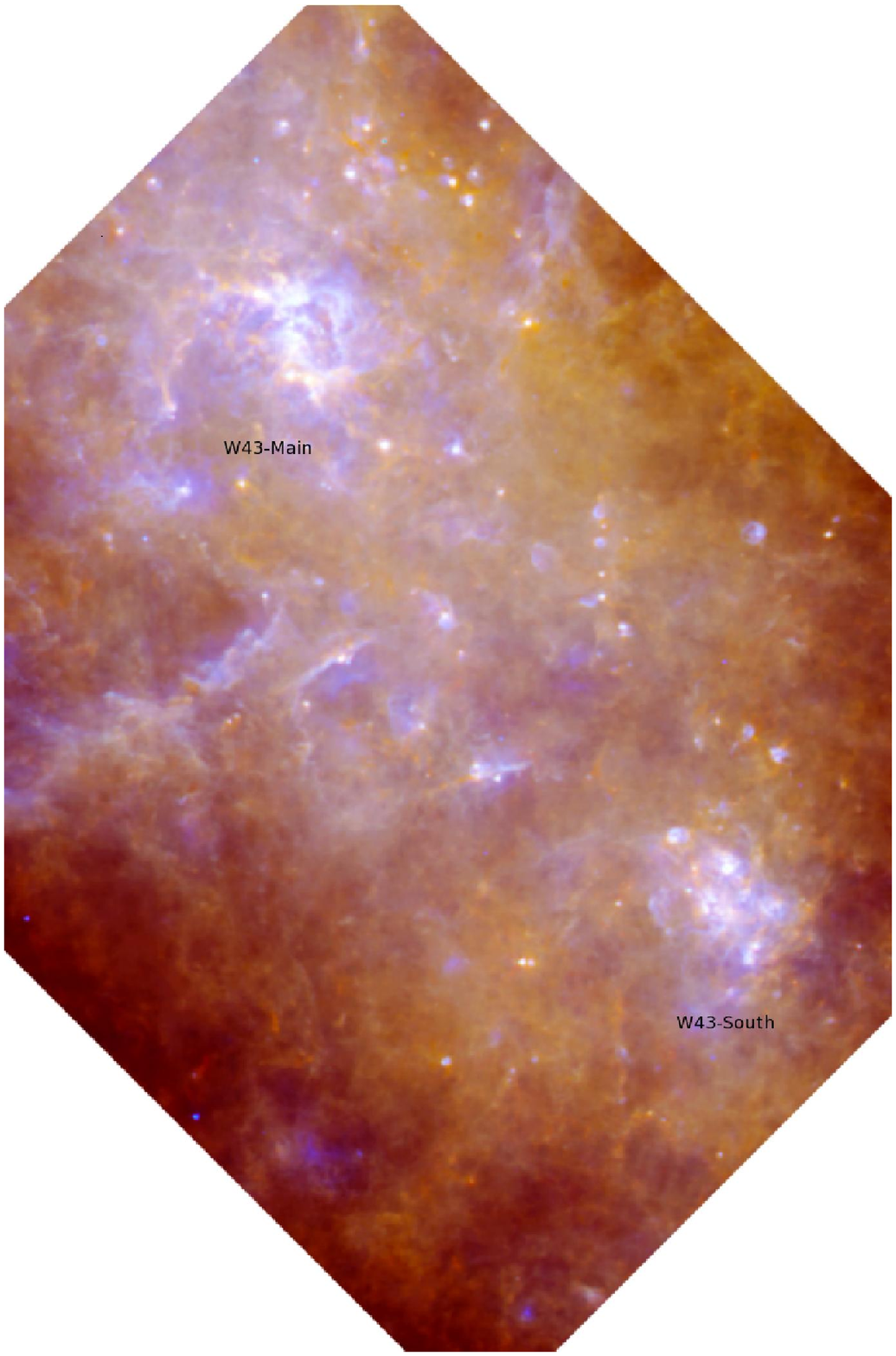} 

\\
\end{array}$
\caption{Composite 3-color \emph{Herschel} image of the W43 molecular cloud complex  (70\,\micron: blue, 160\,\micron: green, 250\,\micron: red). The image is oriented in Right Ascension and Declination directions while the complex was defined by a rectangle in Galactic coordinates \citep[see][]{nguyenluong11b}. The blue component traces H{ \scriptsize II} and photon-dominated regions while earlier stage star-forming sites such as cores and filaments are traced by the red component.}
\label{fig:w43_3colors}
\end{figure*}

 In Fig.~\ref{fig:w43_3colors}, we show the composite 3-color \emph{Herschel} image of the W43 molecular cloud complex  (70\,\micron: blue, 160\,\micron: green, 250\,\micron: red).

\begin{figure*}[tbhp]
$\begin{array}{cc}
\hspace{-0.1cm}
\includegraphics[angle=0,height=15cm]{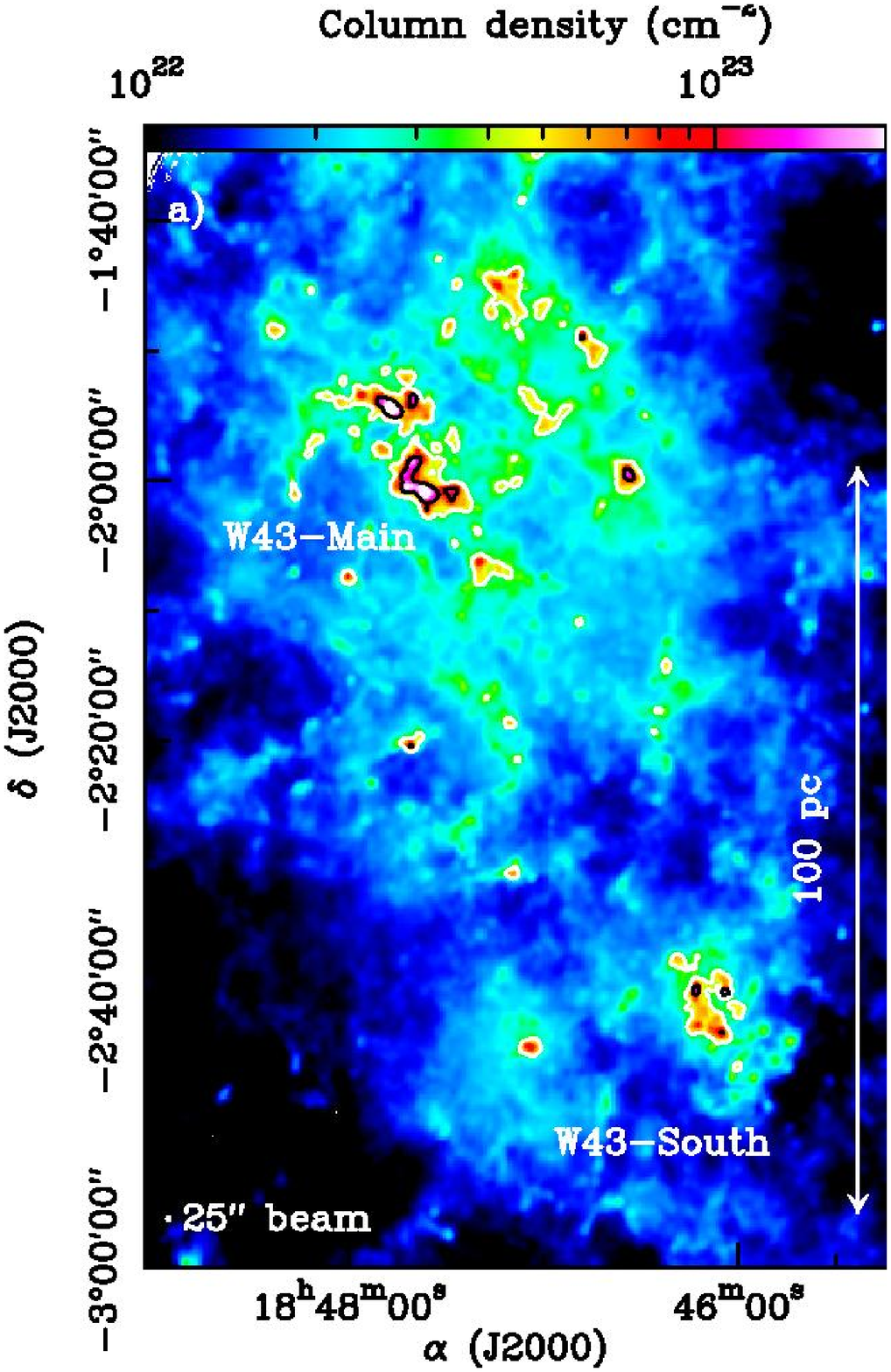} &
\includegraphics[angle=0,height=15cm]{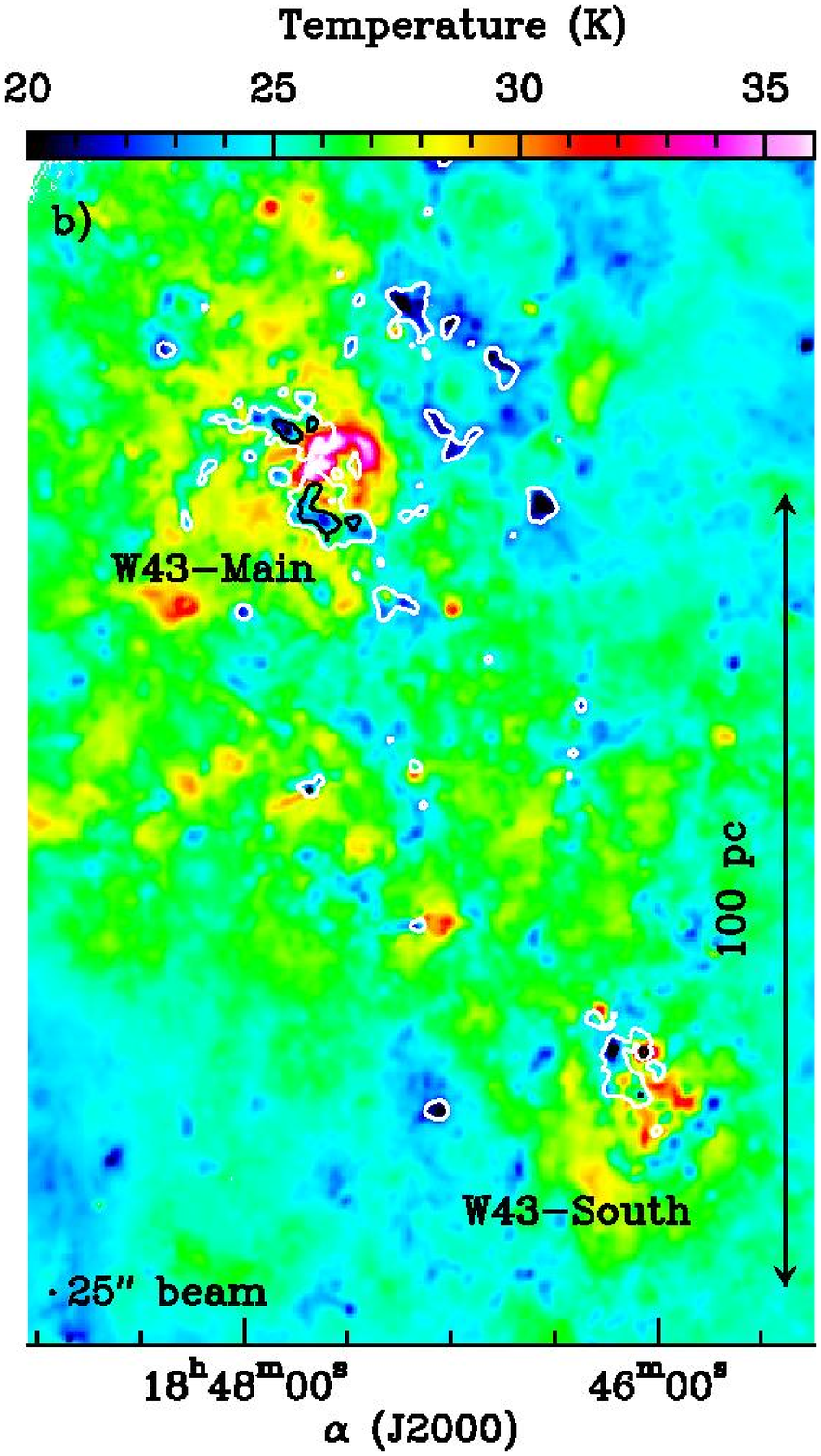} 

\\
\end{array}$
\caption{The W43 molecular complex in: {\bf (a)} \emph{Herschel} column density 
  (color and contours), and {\bf (b)} dust temperature (color) and
  column density (contours). The white and black contours, corresponding to the $4\times10^{22}~\cmd$ and $1\times10^{23}~\cmd$ levels, are outlining the W43-MM1 and W43-MM2 ridges and their immediate surroundings boundary.  }
\label{fig:w43coldenslarge}
\end{figure*}

\newpage
\section{Saturation correction of \emph{Herschel}/SPIRE images}
\label{sect:appendixC}

\begin{figure*}[tbhp]
$\begin{array}{ccc}
\hspace{-0.1cm}
\includegraphics[angle=0,height=8.1cm]{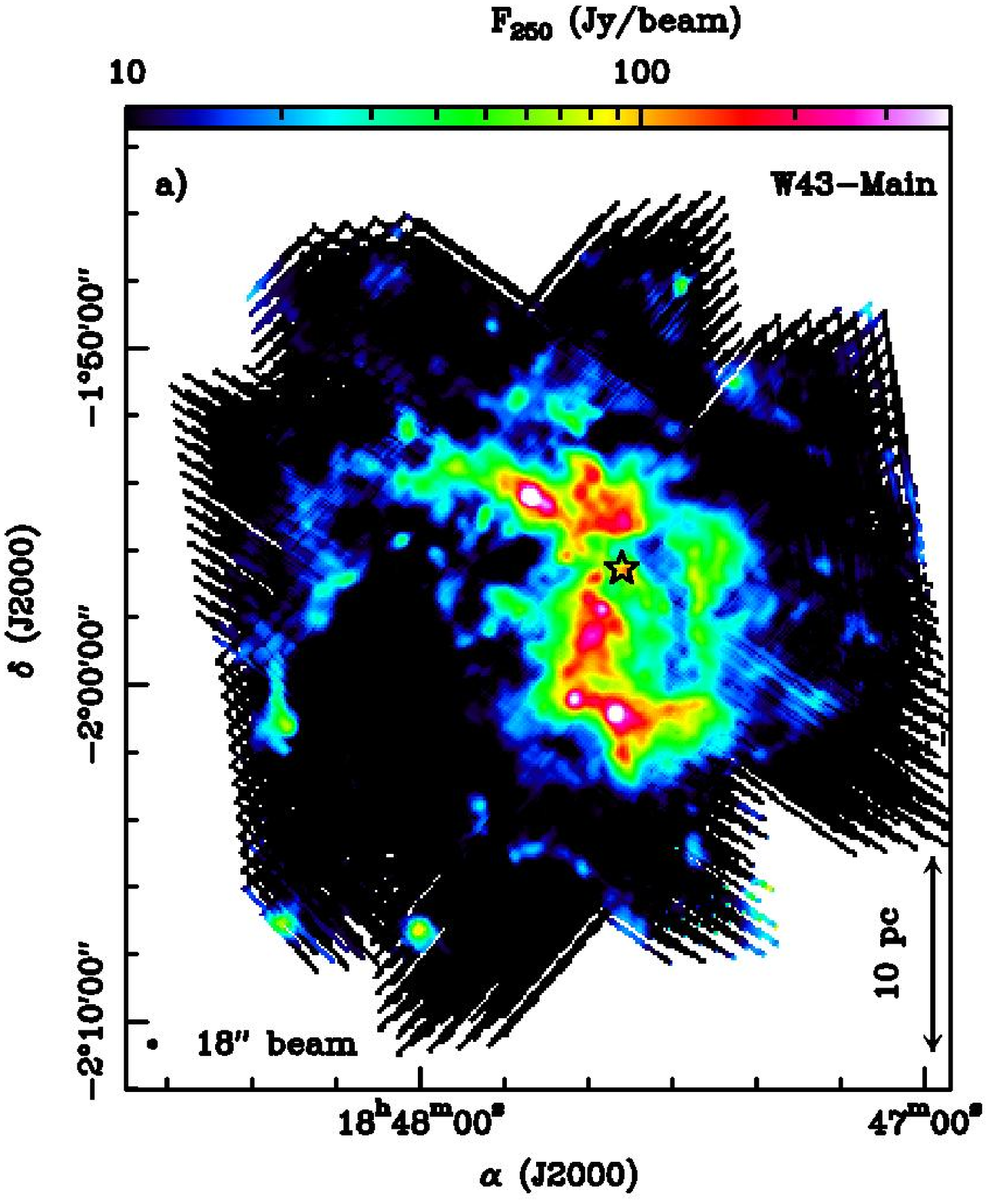} &
\hspace{-1.35cm}
\includegraphics[angle=0,height=8.1cm]{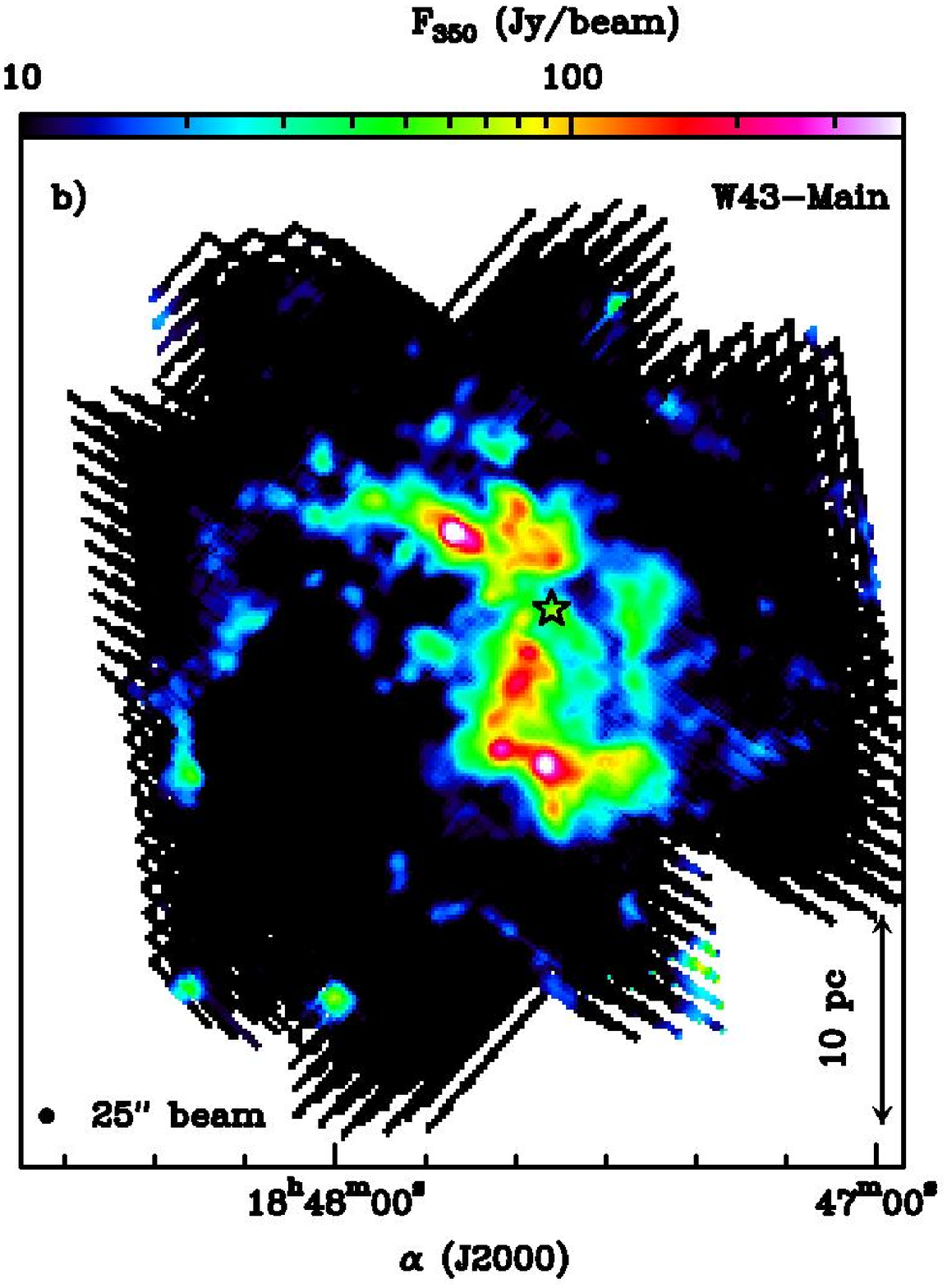} & 
\hspace{-1.35cm}
\includegraphics[angle=0,height=8.1cm]{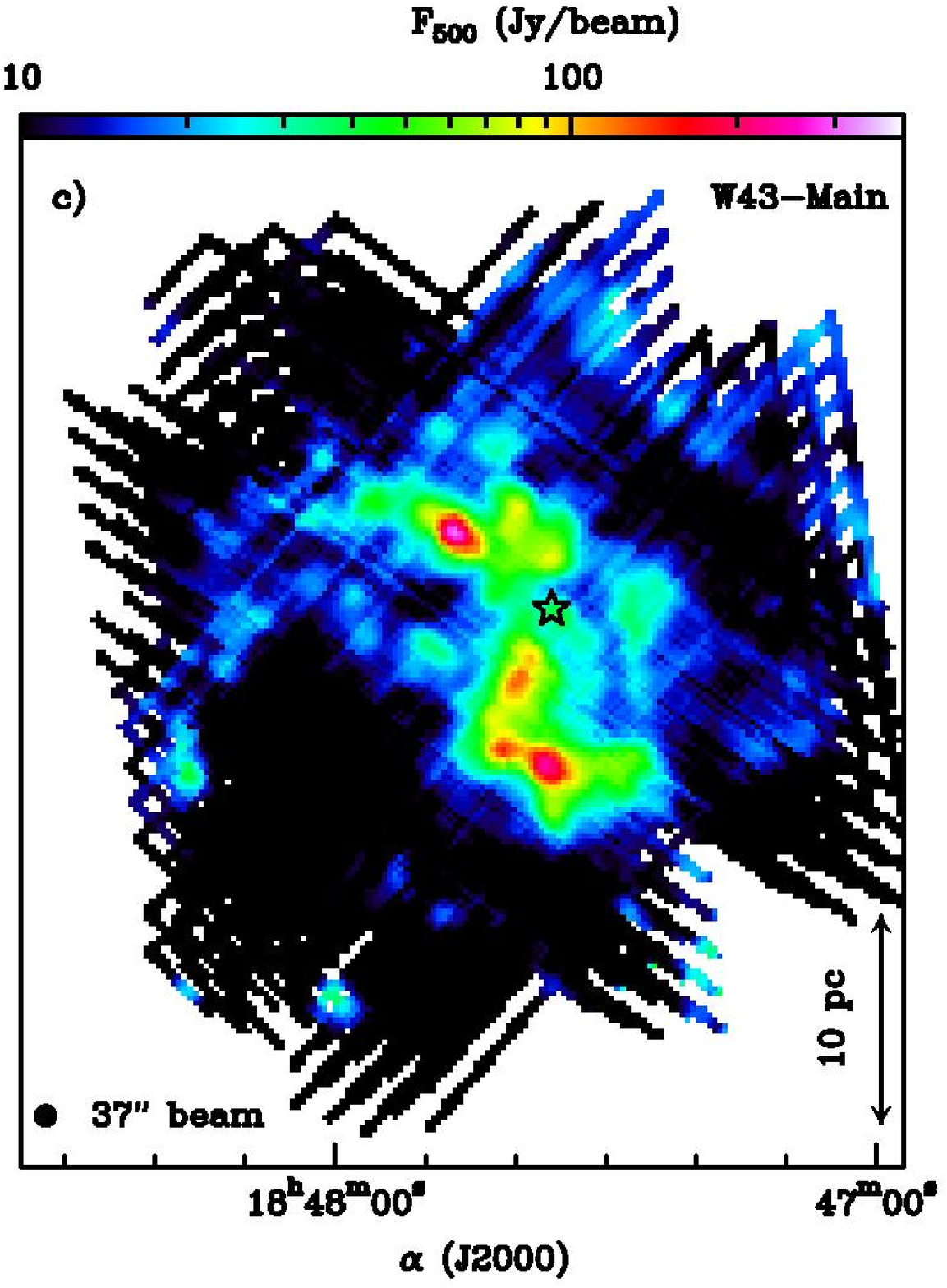} \\
\hspace{-2cm}
\includegraphics[angle=0,height=8.1cm]{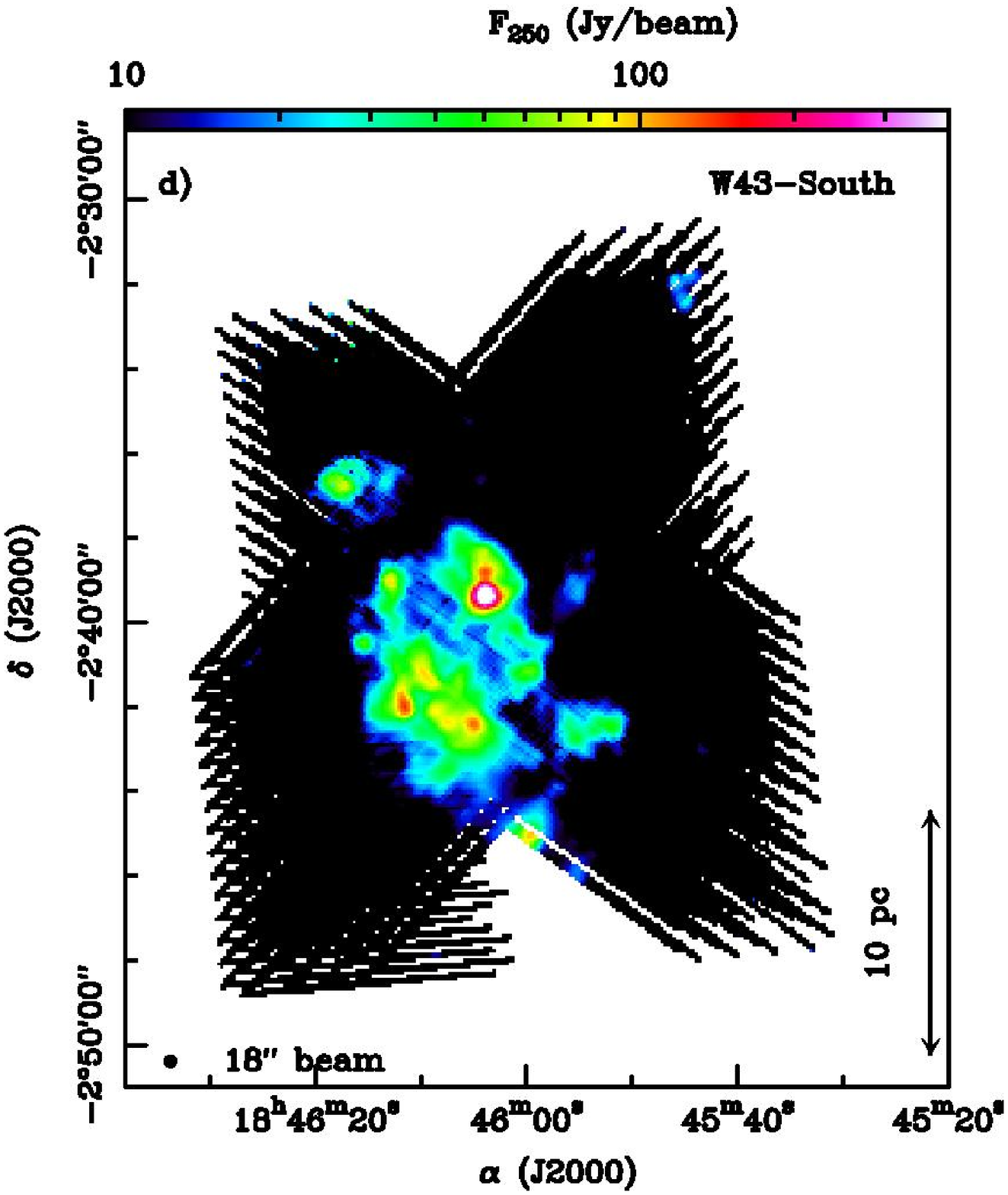} &
\hspace{-3.4cm}
\includegraphics[angle=0,height=8.1cm]{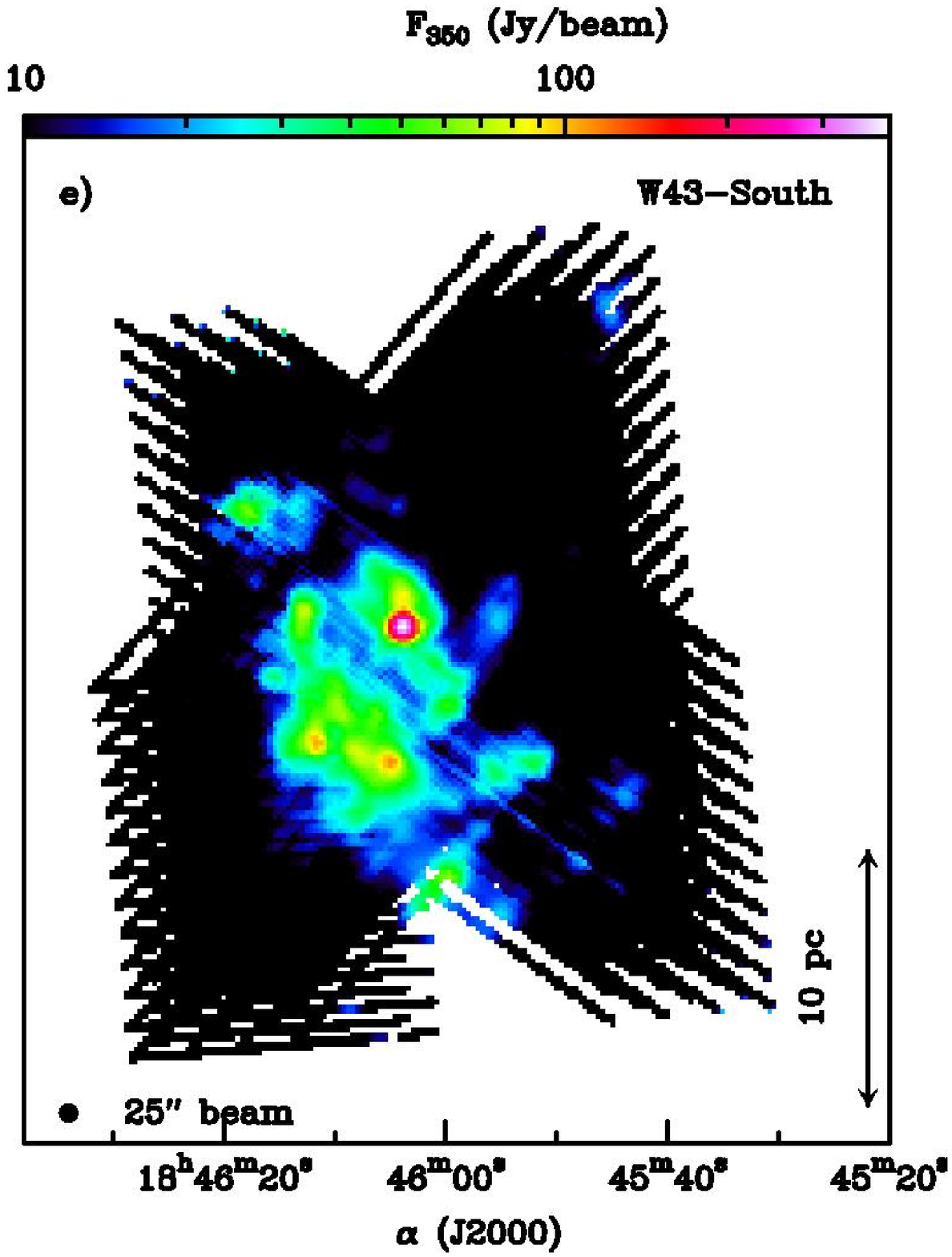} & 
\hspace{-2.9cm}
\includegraphics[angle=0,height=8.1cm]{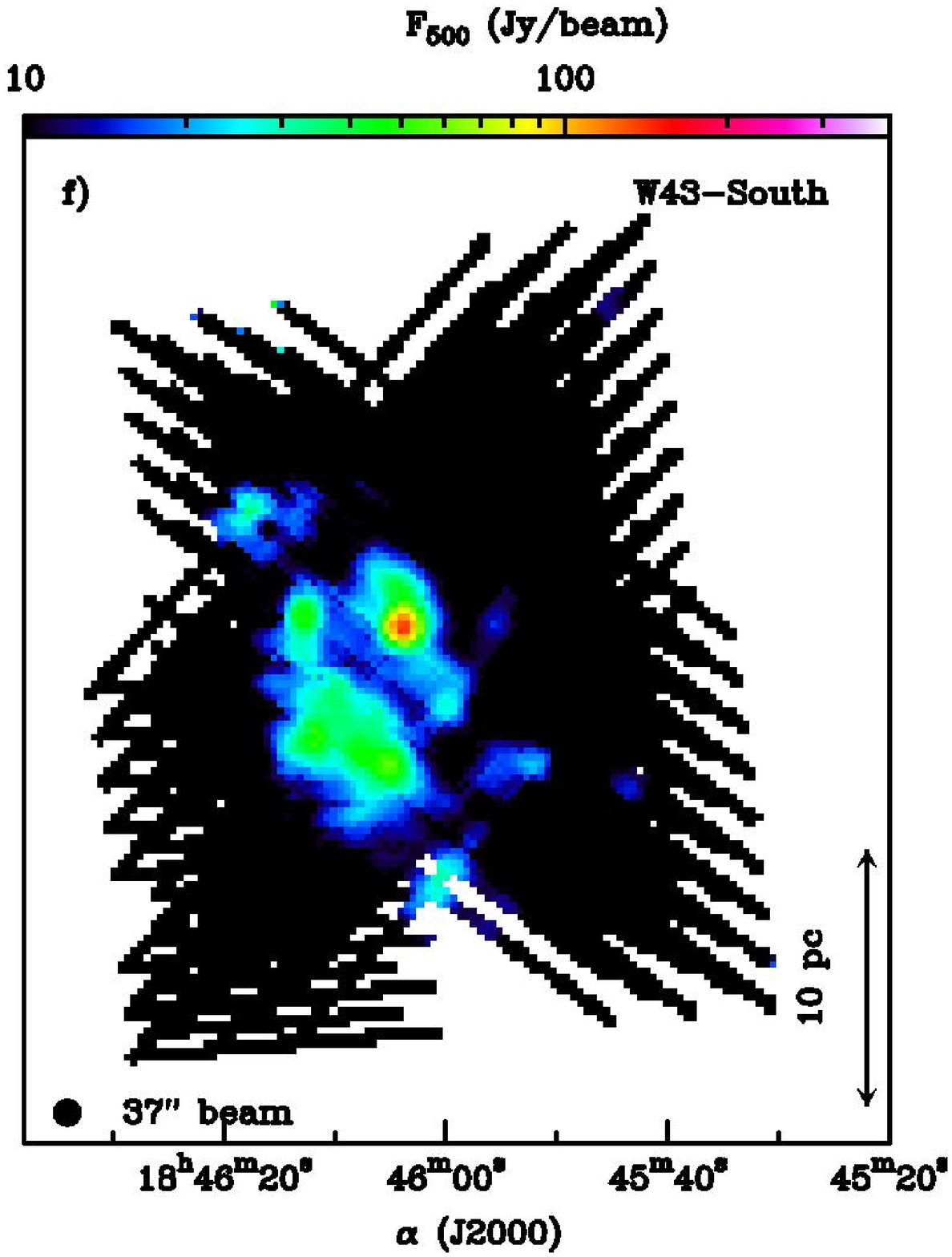} 
\end{array}$ 

\caption{The W43-Main {\bf (upper row)} and W43-South {\bf (lower row)} regions as observed by the HOBYS key program at 250, 350 and $500\,\micron$ \emph{Herschel} wavelengths. }
\label{fig:w43hobys}
\end{figure*}

We explain here the process developed to correct for the saturation of SPIRE images, presented in Sect.~\ref{sect:obsherschel} and used to build the column density and dust temperature images (see Sect.~\ref{sect:coltem}). This process has already been successfully used to correct the bright-mode SPIRE images of the HOBYS Key program (e.g. \cite{rivera-ingraham13}; Fallscheer et al. subm.). In the case of W43, 
the Hi-GAL observations were performed in the nominal voltage bias mode, and as a result, the SPIRE detectors were saturating on bright sources, i.e. the inner parts of W43-Main and W43-South. We used SPIRE data taken in the bright-source voltage bias mode, as part of the HOBYS project (see Sect.~\ref{sect:obsherschel}), to replace the Hi-GAL map pixels suffering from soft saturation (see definition in Sect.~\ref{sect:softsat}) or which are just missing. To this end, the HOBYS data were processed through the exact same steps as those used for the Hi-GAL data, including mapping into the same coordinate system using Scanamorphos (see Sect.~\ref{sect:obsherschel} and Figs.~\ref{fig:w43hobys}a-f.). 

\subsection{Determination of the ``soft''-saturation limit}
\label{sect:softsat}
\begin{figure*}[tbhp]
$\begin{array}{cc}
\hspace{-0.1cm}
\includegraphics[angle=0,height=11.5cm]{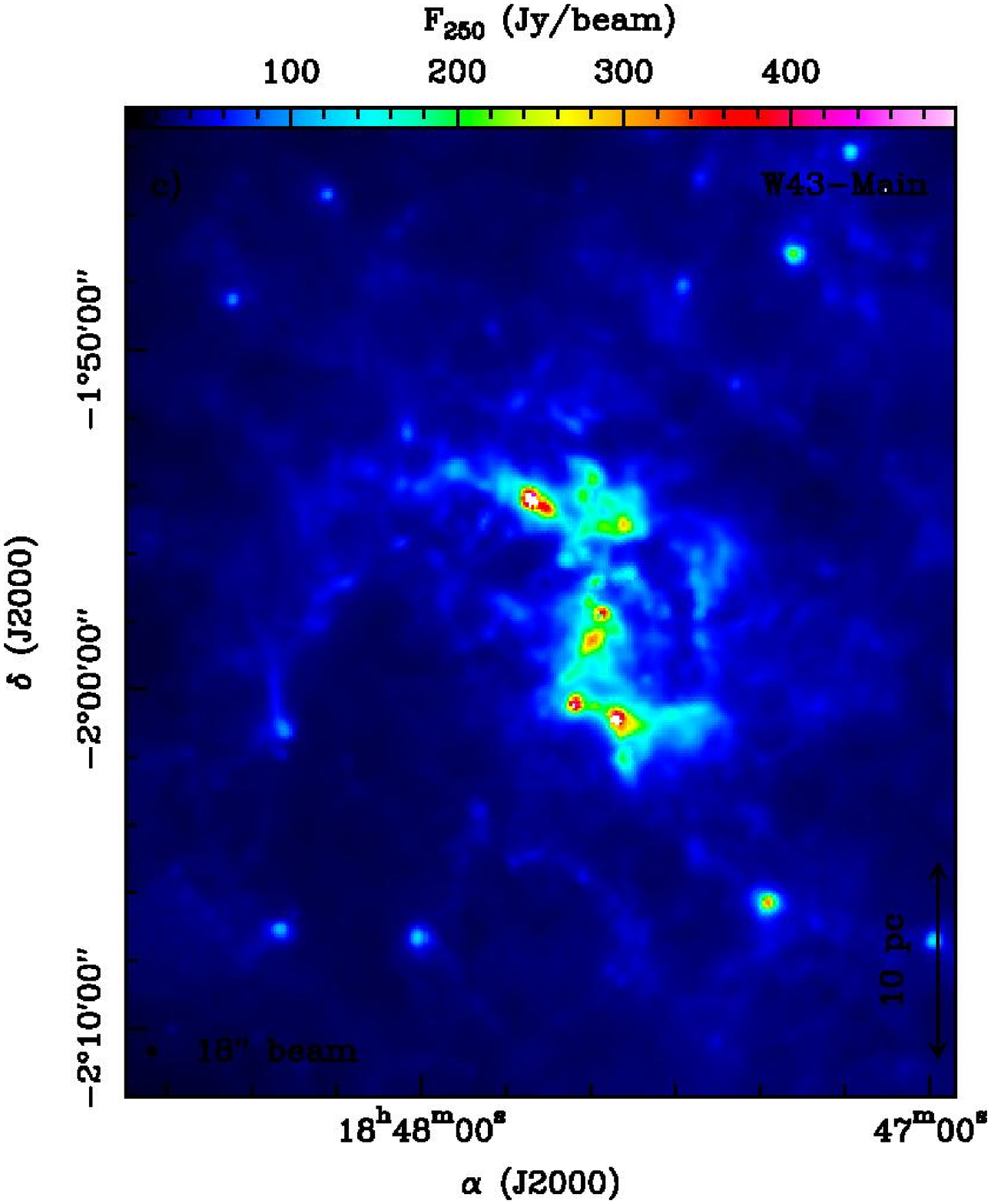} &
\hspace{-0.5cm}
\includegraphics[angle=0,height=11.5cm]{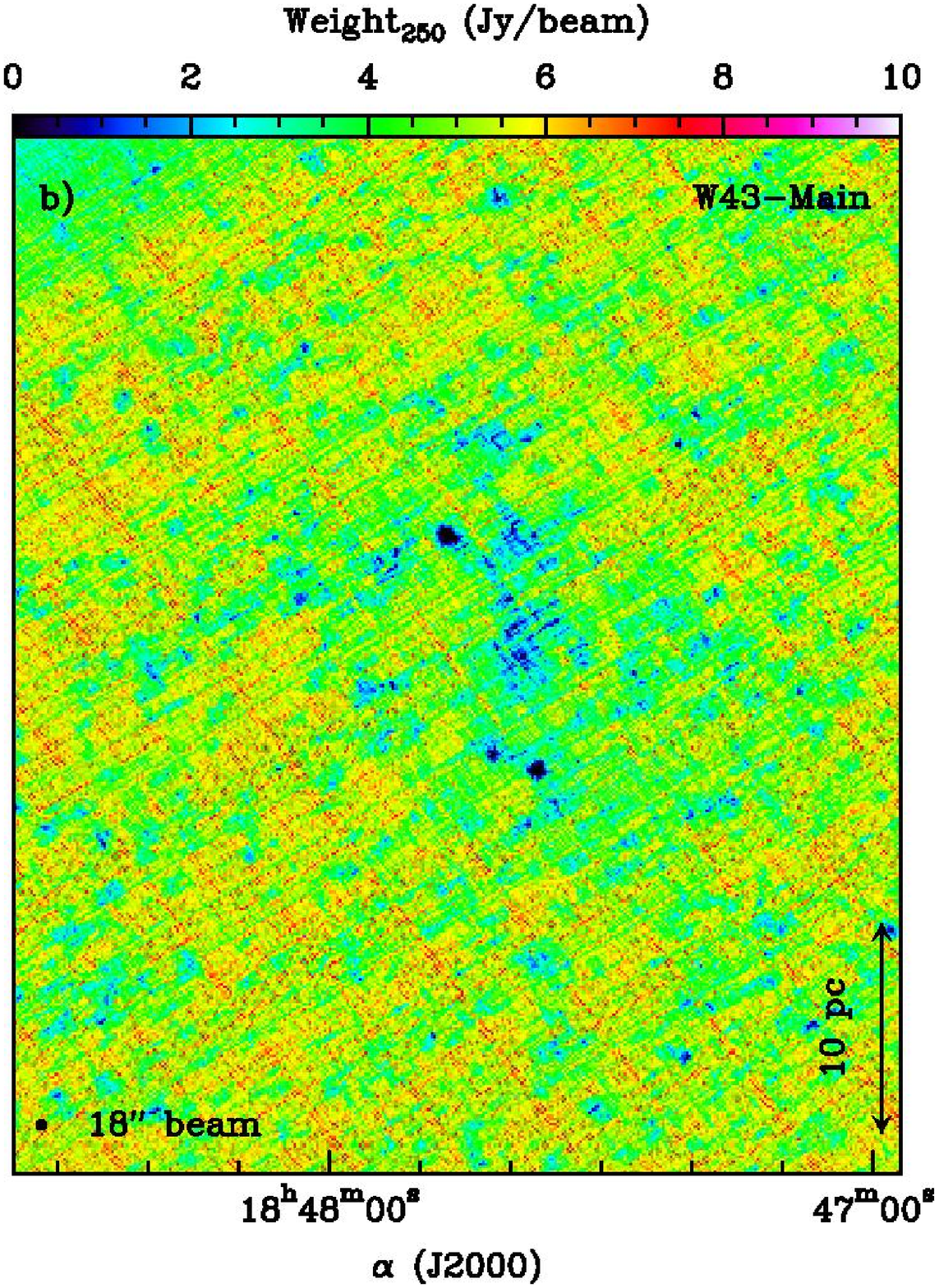} \\
\includegraphics[angle=0,height=11.5cm]{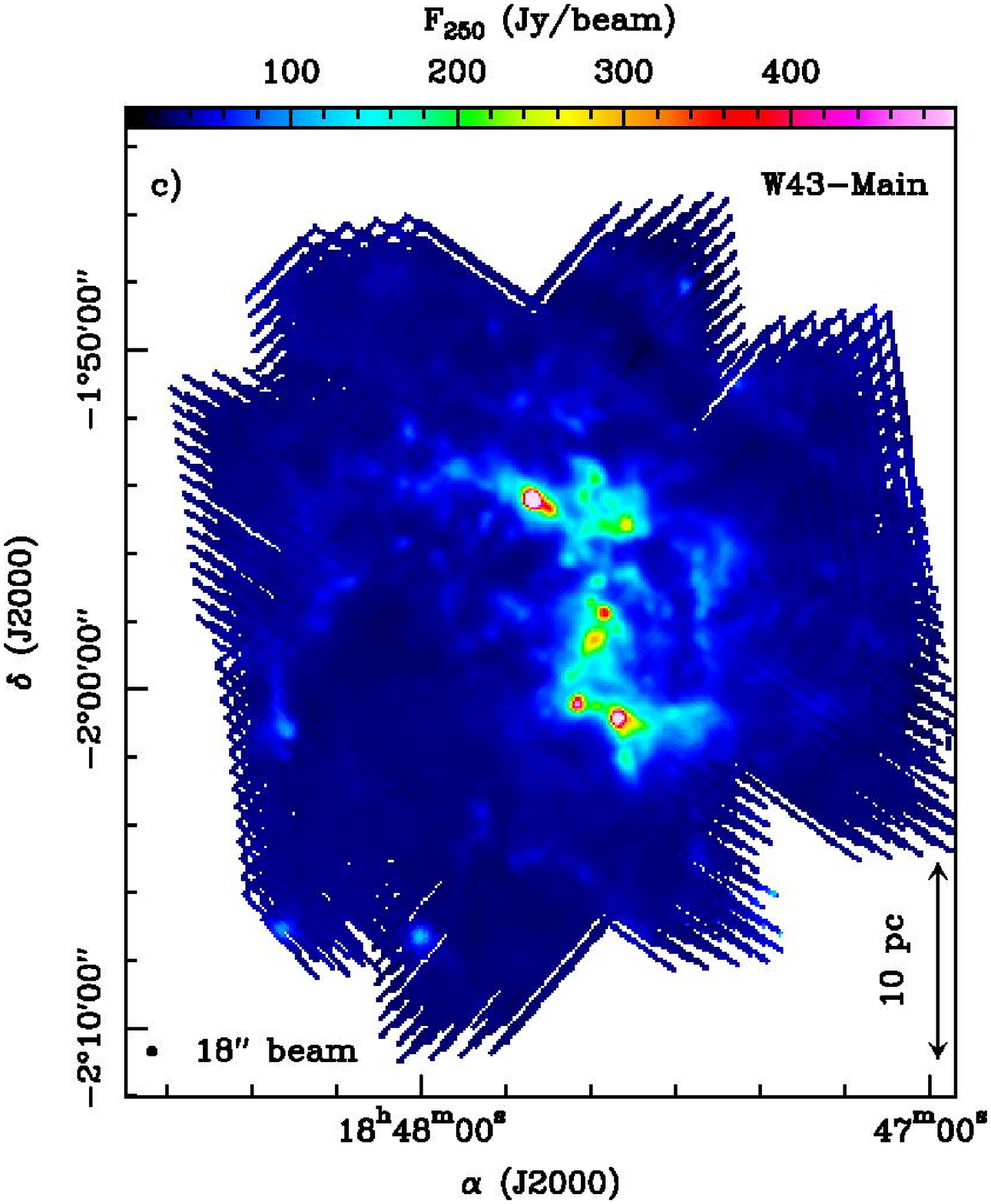} &
\hspace{-0.5cm}
\includegraphics[angle=0,height=11.5cm]{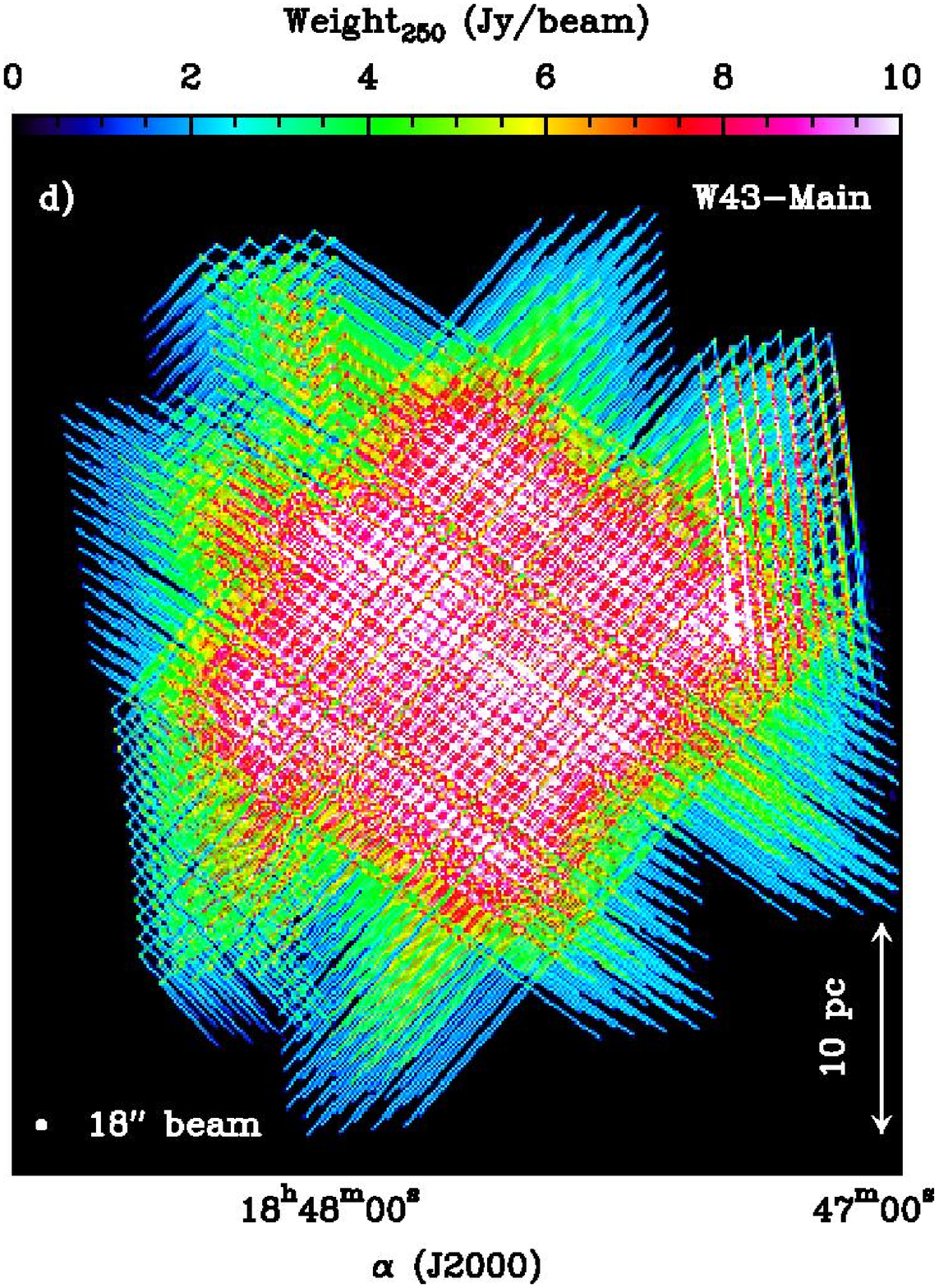} 
\end{array}$ 
\caption{\emph{Herschel} $250\,\micron$ images of W43-Main taken from the Hi-GAL in {\bf a)} and HOBYS in {\bf c)} key programs. Their weight maps are plotted in {\bf b)} and {\bf d)}, respectively.}
\label{fig:w43sat}
\end{figure*}

In Fig.~\ref{fig:w43sat}, we plotted the signal and weight maps of the
region around W43-Main as observed by the Hi-GAL and HOBYS surveys (top and bottom rows respectively). Saturated
map pixels were flagged as NaN when all bolometers passing over the
regions produced signals that fell out the range of the electronics.
These pixels are shown in white and are located over the W43-MM1, W43-MM2, and W43-MM3 massive dense cores.
The astronomical surface brightness at which this occurs varies
from bolometer to bolometer as it depends on how the voltage offset
for the detector signal was set at the beginning of the observation.  In practice, a
bolometer with a baseline signal near the high signal limit 
saturates more easily.

Furthermore, the map pixels
near the bright saturated regions
have a non-linear response to the incoming flux.
Bolometers travelling up or down the bright 
sources
produced usable data only at the faint beginning and end of their measurements.
In this
case, the signal values of the map pixels were weighted down 
with increasing
surface brightness,
making signals close to bright sources no longer reliable.  
This is illustrated by the sudden
decrease in the weight map toward the centers of the W43-MM1, W43-MM2, and W43-MM3 massive dense cores (see 
Fig.~\ref{fig:w43sat} upper right).
Therefore, it is also necessary to replace the signal in these
``softly''-saturated pixels.  We visually checked both the weight
and signal maps around the NaN pixels and determined that map pixels
with fluxes larger than $\sim$150~Jy/beam, for all \emph{Herschel} wavebands, are significantly affected by saturation effects.
This threshold value is also consistent with the surface brightness above which individual
bolometers of the SPIRE arrays were saturated in flux calibration data \citep{bendo13}.

\subsection{Offset addition to the bright-source-mode maps \& replacement of saturated pixels in the nominal-mode maps}

\begin{figure*}[tbhp]
$\begin{array}{ccc}
\hspace{-0.1cm}
\includegraphics[angle=0,height=5cm]{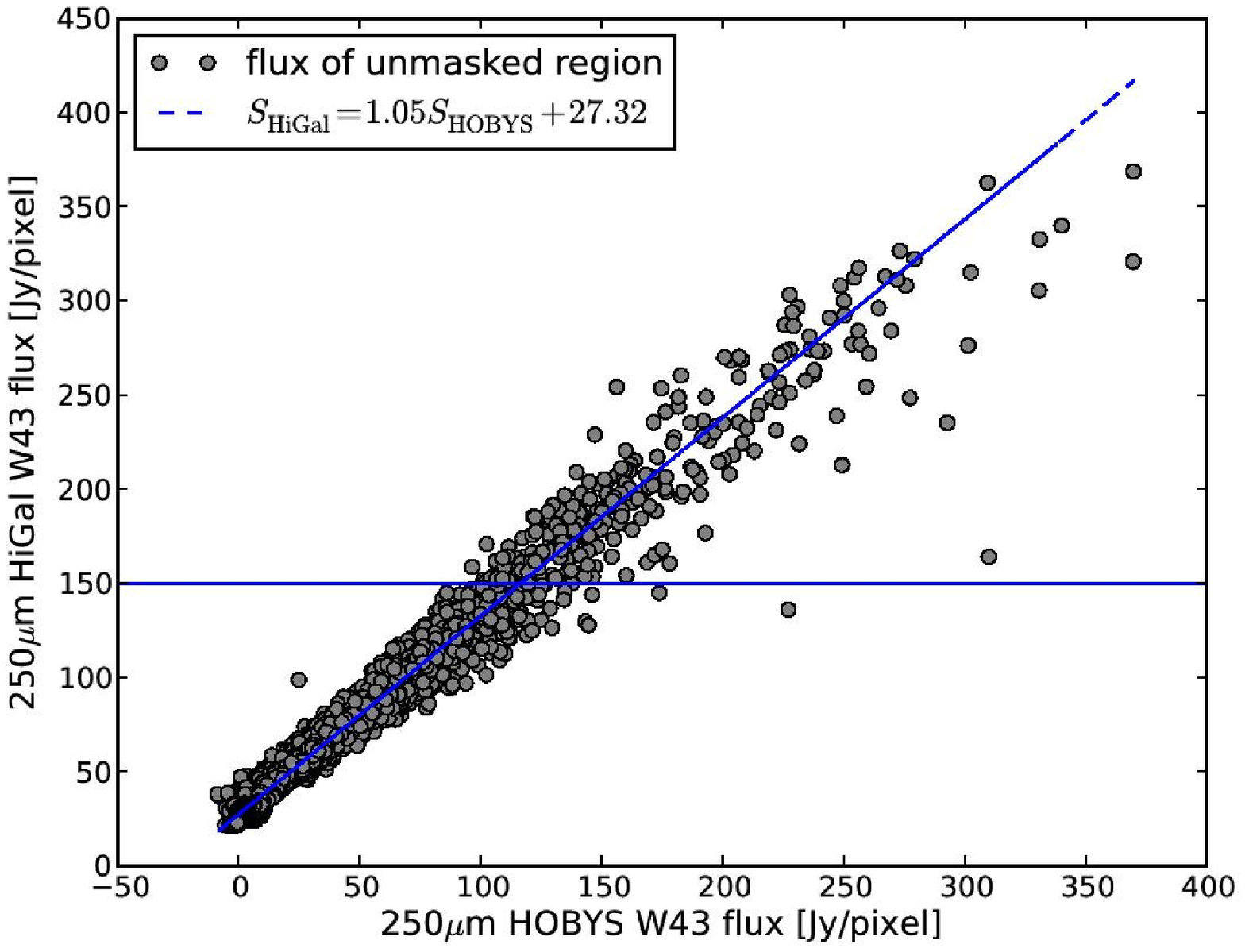} &
\hspace{-0.8cm}
\includegraphics[angle=0,height=5cm]{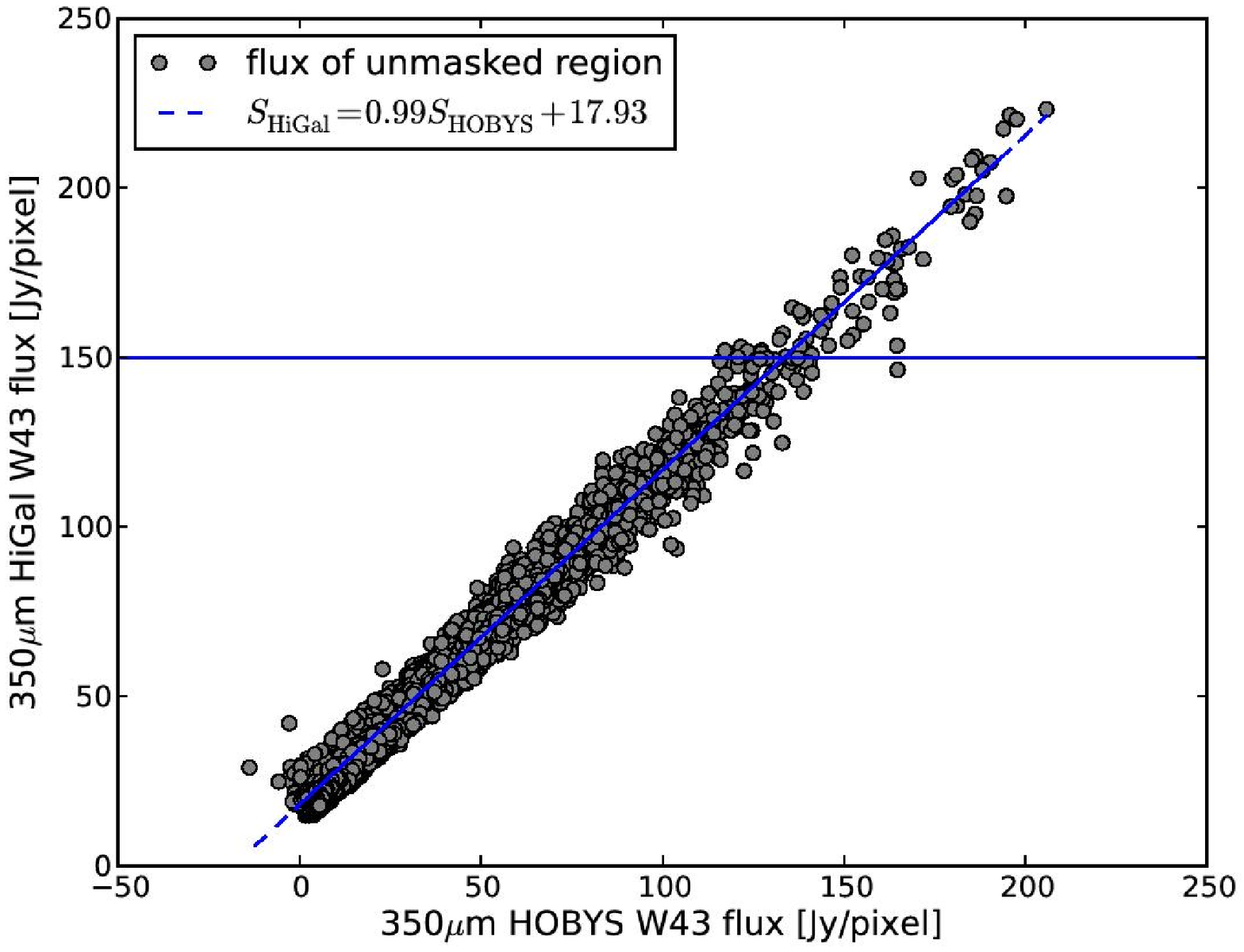} &
\hspace{-0.8cm}
\includegraphics[angle=0,height=5cm]{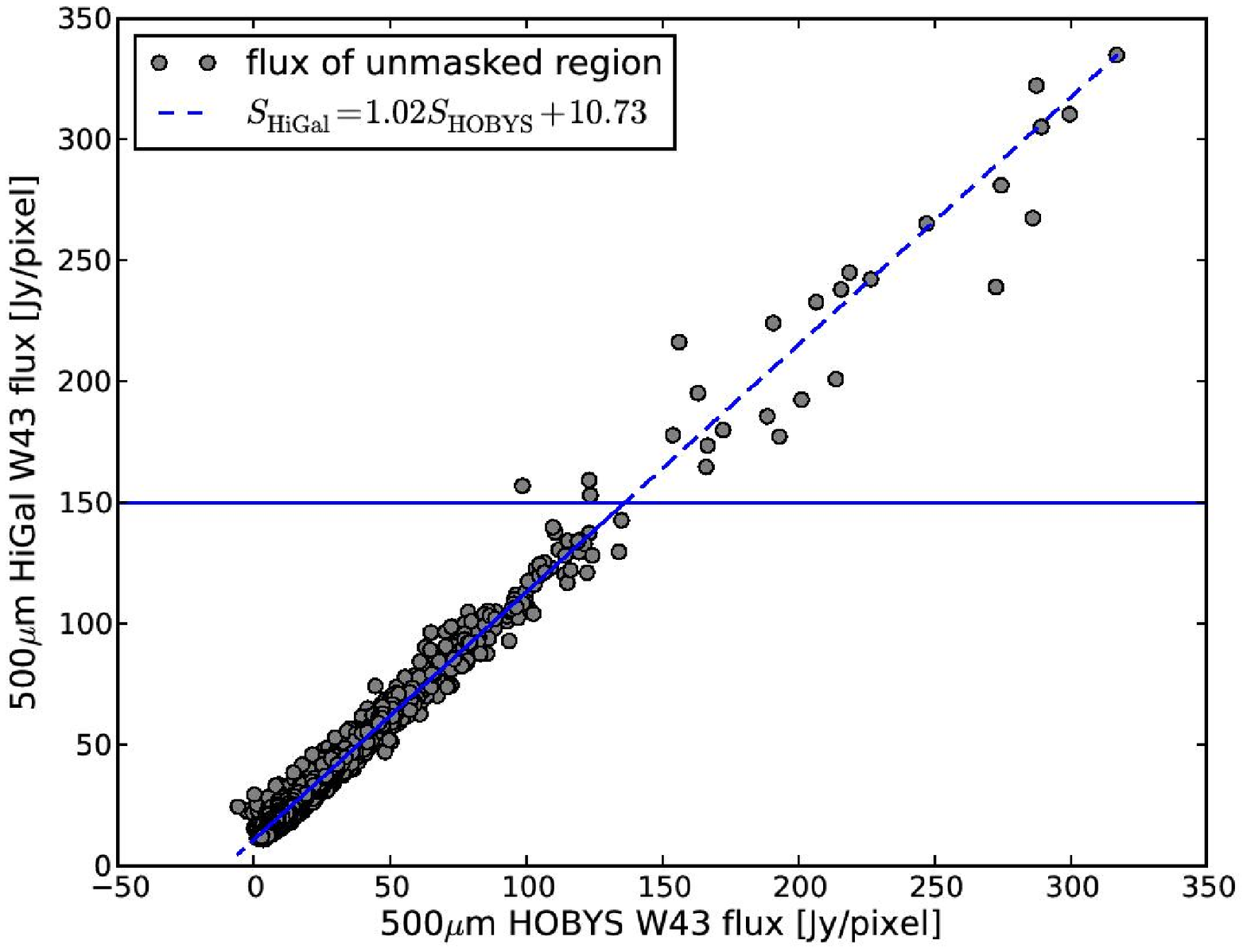} 
\end{array}$
\caption{The pixel-by-pixel correlation between initial Hi-GAL and HOBYS datasets over 
    the unsaturated common
        regions.}
\label{fig:w43offset}
\end{figure*}

\begin{figure*}[!tbhp]
$\begin{array}{ccc}
\hspace{-0.1cm}
\includegraphics[angle=0,height=5cm]{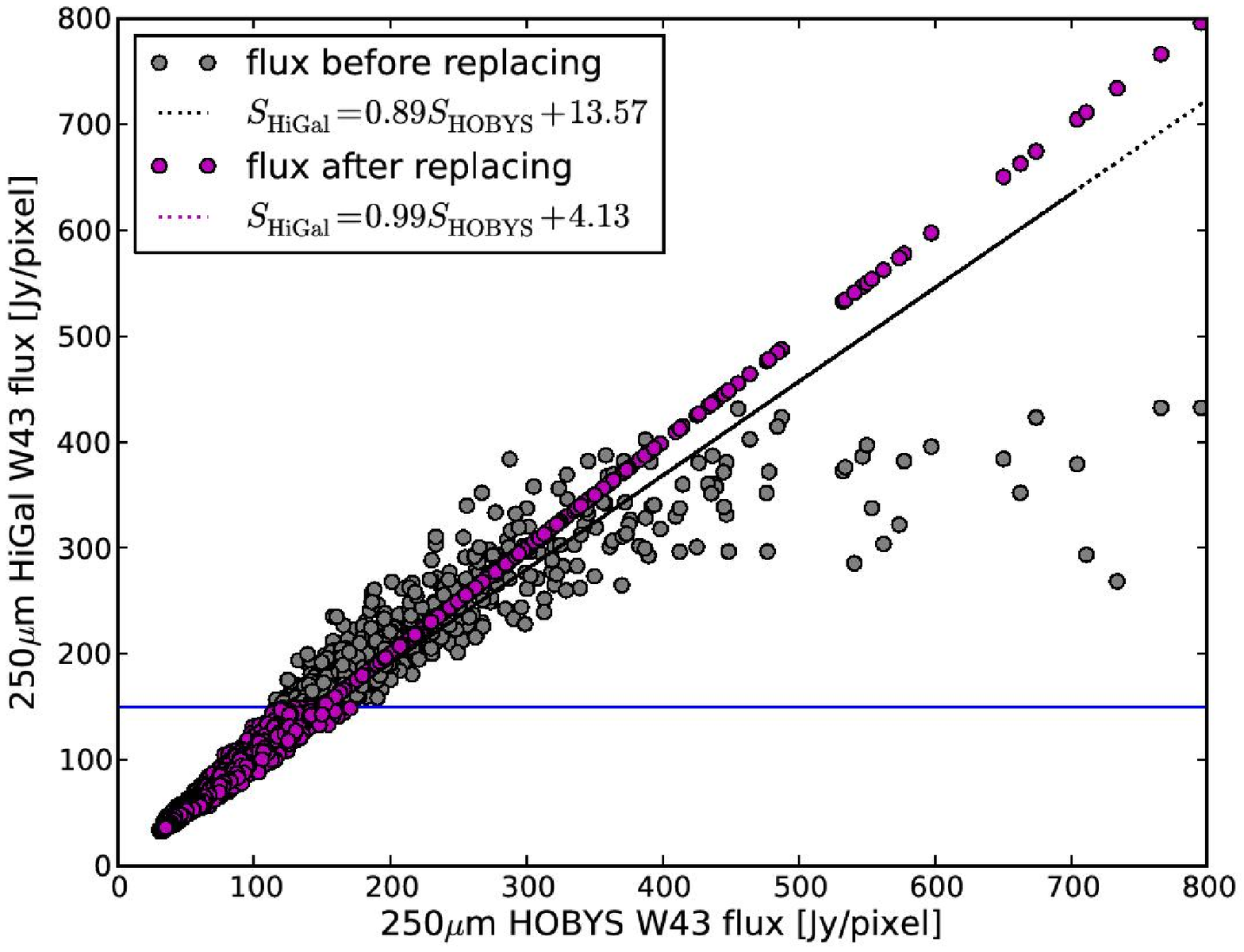} &
\hspace{-0.8cm}
\includegraphics[angle=0,height=5cm]{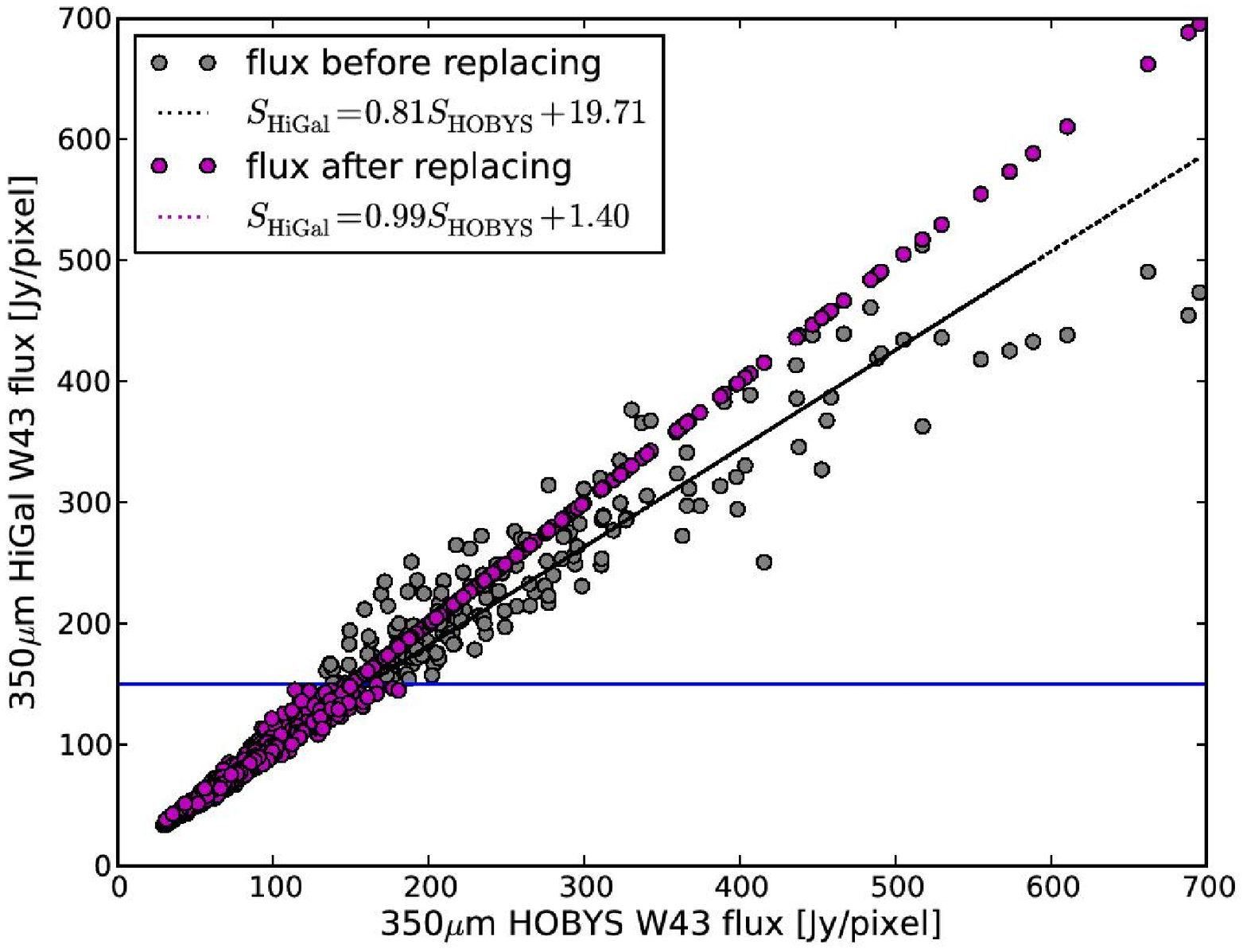} & 
\hspace{-0.8cm}
\includegraphics[angle=0,height=5cm]{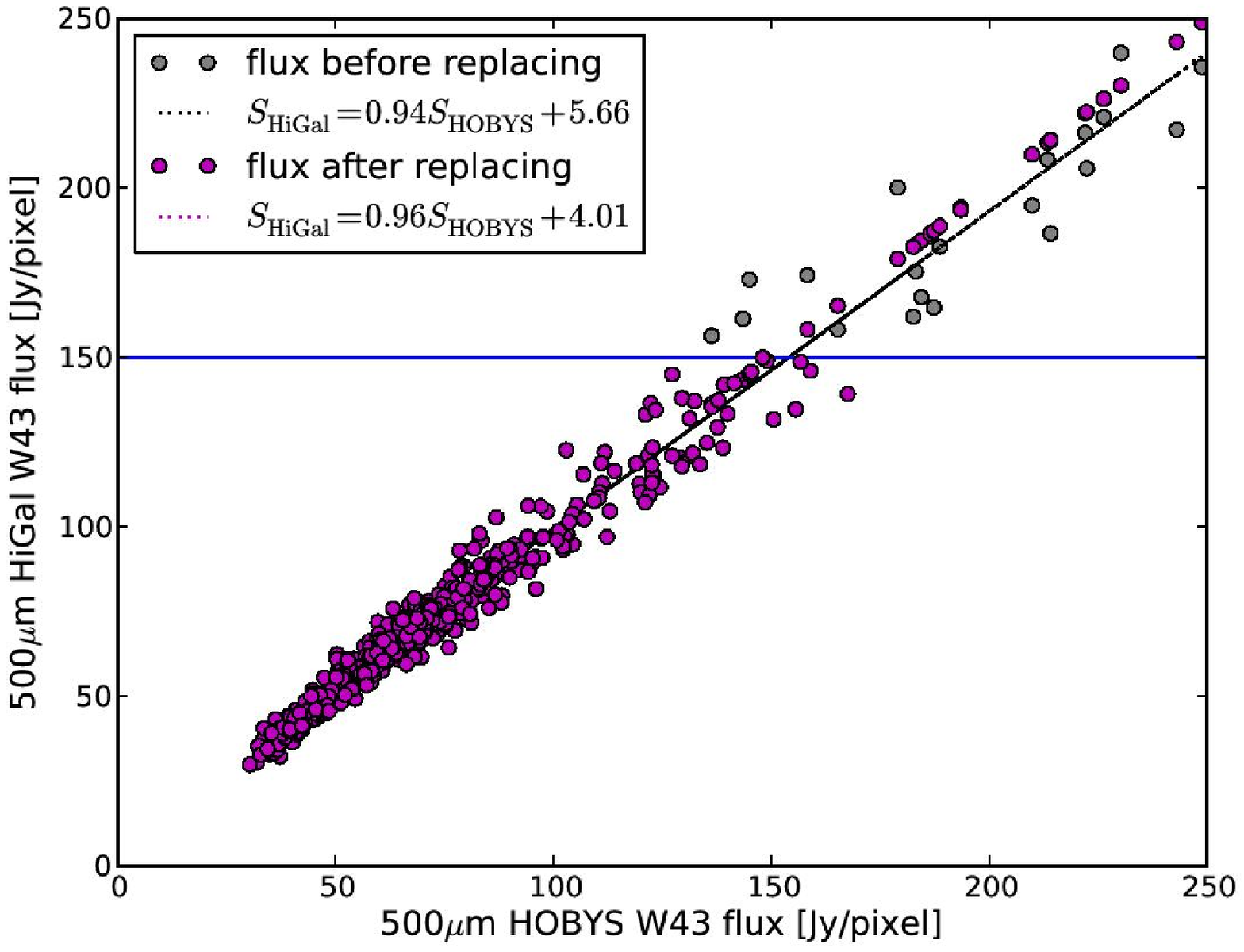} 
\end{array}$
\caption{The pixel-by-pixel correlation between Hi-GAL and HOBYS+offset datasets before (green circles) and after saturation correction (pink circles).}
\label{fig:w43correlation}
\end{figure*}

Before replacing the saturated pixels in the nominal voltage bias mode (Hi-GAL) SPIRE maps with those from
the bright-source voltage bias mode (HOBYS) maps, we added offsets to the bright-mode map pixels. To derive such
offsets, 
we first masked the region surrounding the NaN pixels, then
correlated surface brightness pixel by pixel over the 
common areas where Hi-GAL pixels are not saturated
(see Figs.~\ref{fig:w43offset}a-c). The correlation weakens as the surface
brightness increases, especially above the ``soft''-saturation limit
of 150~Jy/beam. Elsewhere, both datasets are linearly correlated, as
shown in Figs.~\ref{fig:w43offset}a-c.
The pixels with fluxes $>$150~Jy/beam in the Hi-GAL maps, i.e. the saturated pixels and those with non-linear response, were then replaced by the
ones from the HOBYS maps with offsets. As can be seen in
Figs.~\ref{fig:w43correlation}a-c, the repaired maps better correlate
with the bright-mode maps.
The resulting image presents a smooth transition from the pixels initially coming from the Hi-GAL dataset and the HOBYS dataset with offsets added.

\section{Conversion the unit of line luminosity from K~\kms~\kpc$^{2}$ into solar luminosity $\lsun$}
\label{sect:conversion}
The unit K~\kms~\kpc$^{2}$ of the molecular line luminosity calculated from Equation~\ref{eq:lsio} can be expressed also in the unit of the sun bolometric luminosity \lsun. One \lsun~equals $3.846 \times 10^{26}$~W, which can be converted into the Jy unit of flux density as $1~\lsun = 3.846 \times 10^{26}~{\rm W} = 3.846 \times 10^{52}$ Jy~Hz~m$^{2}$ = $4 \times 10^{13}$ Jy~Hz~kpc$^{2}$. The frequency-integrated flux density unit Jy~Hz is equivalent to the velocity-integrated main-beam temperature K~\kms, the conversion between the two units is as following
\begin{equation}
\frac{\int F_{\nu} d\nu}{Jy~Hz}=8.17\times 10^{-7}  \frac{\nu_{0}^2}{\rm GHz^2} \frac{\theta^2}{\rm \arcsec^2} \frac{\int T_{\rm mb} dv}{\rm K \kms} \frac{\nu_{0}}{Hz}\frac{c^{-1}}{(\kms)^{-1}}
\end{equation}
Thus, for our SiO~2--1 observation, the conversion between the line luminosity in unit of K~\kms~\kpc$^{2}$ to the unit of the sun bolometric luminosity \lsun~ is
\begin{equation}
\frac{L_{\rm SiO~2-1}}{\lsun}=  8.17\times 10^{-7} \times \frac{L_{\rm SiO~2-1}}{\rm K~\kms~\kpc^{2}} \frac{\nu_0^2}{\rm GHz^2} \frac{\theta^2}{\rm \arcsec^2} \frac{\nu_{0}}{Hz}\frac{c^{-1}}{(\kms)^{-1}} \times \frac{1}{4\times 10^{13}}\\
                             =  4.28\times 10^{-8} \times \frac{L_{\rm SiO~2-1}}{\rm K~\kms~\kpc^{2}} 
\end{equation}

\end{document}